\documentclass[runningheads]{llncs}
\usepackage[T1]{fontenc}
\usepackage{graphicx}
\usepackage{tabularx}
\usepackage{array}
\usepackage[style=numeric-comp, sorting=none]{biblatex}
\usepackage{placeins} 
\usepackage{setspace}

\usepackage{amsmath,mathrsfs, bm}
\usepackage{amsfonts} 
\usepackage{mathtools}
\usepackage{amssymb}
\usepackage{physics}
\usepackage{booktabs}
\usepackage{multirow}
\usepackage{array}
\usepackage{tikz}

\usepackage[ruled,linesnumbered]{algorithm2e}

\usepackage{geometry}
\usepackage{caption}
\usepackage{newfloat}
\usepackage{float}
\usepackage{booktabs}
\usepackage{makecell}

\PassOptionsToPackage{hyphens}{url}
\usepackage[colorlinks=true, linkcolor=blue, citecolor=blue, urlcolor=blue]{hyperref}
\usepackage{cleveref}

\newcommand{\IP}{{\rm I}\kern-0.18em{\rm P}}
\newcommand{\1}{{\rm 1}\kern-0.24em{\rm I}}
\newcommand{\E}{{\rm I}\kern-0.18em{\rm E}}
\newcommand{\Prob}{{\rm I}\kern-0.18em{\rm P}}
\newcommand{\R}{{\rm I}\kern-0.18em{\rm R}}

\newcommand{\cS}{\mathcal{S}}

\newcommand{\bT}{\mathbf{T}}

\newcommand{\Ex}{\mathbb{E}}
\newcommand{\pr}{\mathbb{P}}

\usepackage[bottom]{footmisc}
\newtheorem{simsetting}{Simulation Setting}
\newtheorem{assumption}{Assumption}

\usepackage{etoolbox}
\AtBeginEnvironment{equation}{%
  \setlength\abovedisplayskip{6pt}%
  \setlength\belowdisplayskip{6pt}%
}
\AtBeginEnvironment{equation*}{%
  \setlength\abovedisplayskip{6pt}%
  \setlength\belowdisplayskip{6pt}%
}
\AtBeginEnvironment{align*}{%
  \setlength\abovedisplayskip{6pt}%
  \setlength\belowdisplayskip{6pt}%
}
\AtBeginEnvironment{align}{%
  \setlength\abovedisplayskip{6pt}%
  \setlength\belowdisplayskip{6pt}%
}

\begin{document}

\title{Nullstrap-DE: A General Framework for Calibrating FDR and Preserving Power in Differential Expression Methods, with Adaptivity to DESeq2 and edgeR}
\titlerunning{Nullstrap-DE: A General Framework for Calibrating FDR and Preserving Power in DE Methods}
%
\author{Chenxin Flora Jiang\inst{1}$^\dagger$\orcidID{0009-0005-7369-4116} \and
Changhu Wang\inst{2}$^\dagger$\orcidID{0009-0002-8567-0961} \and
Jingyi Jessica Li\inst{1,2}\textsuperscript{*}\orcidID{0000-0002-9288-5648}}
\authorrunning{C. F. Jiang et al.}
%
\institute{Department of Statistics and Data Science, University of California, Los Angeles, CA 90095, USA \and
Biostatistics Program, Public Health Science Division, Fred Hutchinson Cancer Center, Seattle, WA 98109, USA 
\\[6pt]
$^\dagger$\ These authors contributed equally to this work.\\[3pt]
\textsuperscript{*}\ Contact: \email{lijy03@fredhutch.org}; \email{jli@stat.ucla.edu}
}
\maketitle              
\begin{abstract}
Differential expression (DE) analysis aims to identify genes whose expression differs across biological conditions, with a central challenge being the balance between false discovery rate (FDR) and statistical power. Popular parametric methods, DESeq2 and edgeR, achieve high power by modeling gene counts using negative binomial generalized linear models, but can exhibit FDR inflation because their parameter estimates are not maximum likelihood estimates (MLEs) while $p$-values are computed using asymptotic null distributions of MLEs. In contrast, the nonparametric Wilcoxon rank-sum test provides more robust FDR control but often suffers from low power and does not support covariate adjustment. We propose Nullstrap-DE to combine the power of parametric DE methods with data-driven FDR calibration. Designed as an add-on to an existing parametric “parent” method, Nullstrap-DE generates synthetic null data under gene-specific null hypotheses, applies the same estimation procedure to the original and null data, and calibrates significance thresholds to control FDR while accounting for estimation uncertainty and bias not captured by asymptotic theory. We show that Nullstrap-DE asymptotically controls the FDR and achieves power consistency. Simulations and analyses of bulk and single-cell RNA-seq datasets demonstrate that, compared with existing approaches, Nullstrap-DE reliably controls the FDR while identifying biologically meaningful genes.
\end{abstract}

\newpage

\begin{refsection}

\section{Introduction}

Differential expression (DE) analysis is central to transcriptomic studies, particularly RNA sequencing (RNA-seq), for interrogating the functions of thousands of genes. Its main goal is to identify genes whose expression levels vary significantly between biological conditions, such as treatment and control groups. These differentially expressed (DE) genes can provide valuable insights into underlying biological pathways, disease mechanisms, and potential therapeutic targets. A critical challenge in DE analysis is balancing false discovery rate (FDR) control with statistical power. On one hand, stringent FDR control is essential in DE analysis because thousands of genes are tested simultaneously, and even modest per-gene type I error rates can result in hundreds of false positives, potentially compromising downstream analyses such as pathway enrichment and biomarker discovery~\cite{reiner2003identifying}. On the other hand, maintaining high power is crucial to detect genes exhibiting meaningful expression changes. 
This creates a well-known tradeoff within a given method: tightening FDR control often reduces power, but boosting power may increase false discoveries. However, this tradeoff does not necessarily hold when comparing different methods. Some methods may achieve both better FDR control and higher power, motivating the continuous development of statistical methods for DE analysis.

To date, numerous statistical methods have been developed for DE analysis, ranging from parametric methods that assume specific distributions for gene expression to non-parametric methods that avoid such assumptions. In practice, parametric methods are more widely used, particularly in settings with limited sample sizes or complex experimental designs, where they offer higher power and facilitate covariate adjustment. 
In parametric methods, each gene's expression levels are modeled using a specified distribution and linked to experimental conditions through (generalized) linear models. For example, limma-voom~\cite{ritchie2015limma} assumes normality on log-transformed counts and applies linear models with precision weights to account for the mean–variance relationship. Another common approach assumes a negative binomial (NB) distribution for gene counts and uses generalized linear models (GLMs) to associate gene expression with experimental conditions. This NB-GLM framework underlies methods such as DESeq2~\cite{love2014moderated} and edgeR~\cite{robinson2010edger}, both of which incorporate an offset term (called \textit{size-factor normalization}) in the GLM to adjust for different sequencing depths and apply empirical Bayes shrinkage to stabilize NB dispersion estimates. While the modeling framework is shared, the two methods differ in their approaches to size-factor normalization, dispersion parameter estimation, and the resulting statistical tests. 
In addition to the methods discussed above, numerous other statistical approaches have been developed for DE analysis, each with its own assumptions, strengths, and limitations. For detailed comparisons, see the reviews by~\cite{costa2017rna, rosati2024differential}.

Among the available DE methods, DESeq2 and edgeR remain the most widely used~\cite{rosati2024differential} due to their strong power, flexibility, and integration into standard bioinformatics workflows. However, empirical studies have shown that both methods may exhibit inflated FDR, particularly in large-sample settings~\cite{li2022exaggerated}.
The inflation arises from the size-factor normalization and shrinkage procedures used in parameter estimation by DESeq2 and edgeR. While these procedures reduce the variance of gene-level mean estimates by borrowing information across genes, they deviate from the standard maximum likelihood estimation (MLE), yielding parameter estimates that are no longer exact maximum likelihood estimates (MLEs). Because $p$-values are computed using the asymptotic null distributions of the MLEs, without accounting for the additional uncertainty introduced by normalization or the bias induced by shrinkage, the null distribution of the test statistic can be misspecified. This misspecification may lead to inaccurate $p$-values and an inflated FDR, even when the underlying negative binomial generalized linear model (NB-GLM) is correctly specified.

To mitigate this FDR inflation, Li et al. recommended non-parametric alternatives such as the Wilcoxon rank-sum test, which better control the FDR in large samples without relying on parametric assumptions~\cite{li2022exaggerated}. This robustness, however, comes at the cost of reduced power, especially when sample sizes are small. Moreover, the Wilcoxon rank-sum test does not accommodate covariate adjustment, which can lead to confounding in the presence of known batch effects or other sample-level covariates. Given the vulnerability of parametric methods to FDR inflation and the limitations of non-parametric methods in power and covariate adjustment, there is a clear need for approaches that can achieve reliable FDR control while maintaining statistical power and practical usability.

To address this gap, we propose Nullstrap-DE, a general framework that combines the robustness of non-parametric methods with the power and modeling flexibility of parametric approaches. 
Designed as an \textit{add-on} to widely used DE methods such as DESeq2 and edgeR (referred to as parent methods), Nullstrap-DE improves FDR control without sacrificing power, while preserving the modeling structure and estimation procedures of the parent method. As a result, it retains key parametric estimates such as dispersions and effect sizes, and, when using the same test statistic, preserves the original gene ranking, modifying only the decision threshold for declaring significance. This design enables Nullstrap-DE to seamlessly integrate into existing DESeq2 or edgeR workflows and ensures full compatibility with downstream analysis and biological interpretation. Importantly, Nullstrap-DE offers a user-friendly extension for researchers already familiar with DESeq2 or edgeR, while improving the FDR control with power preservation. 
For ease of reference, we refer to the general framework as Nullstrap-DE, and denote its application to a specific DE method as Nullstrap-[method] (e.g., Nullstrap-DESeq2 or Nullstrap-edgeR), enabling direct integration with the parent method’s implementation.

The core idea of Nullstrap-DE builds on Nullstrap~\cite{wang2025nullstrapsimplehighpowerfast}, a recently proposed framework that leverages synthetic null data for FDR control. Nullstrap-DE fits a per-gene parametric null model based on the parametric model and the null hypothesis of the parent method. It then generates synthetic null data from the null models and compares the parent method's test statistics on the original data to those from the synthetic null data, enabling false discovery proportion (FDP) estimation and FDR calibration without sacrificing the statistical power of the parent method. By mimicking the entire estimation procedure of the parent method, including normalization and shrinkage, Nullstrap-DE captures the additional uncertainty and bias introduced by these steps. This enables Nullstrap-DE to recalibrate the significance threshold on the resulting test statistics or $p$-values, thereby correcting FDR inflation while preserving statistical power.
A recent preprint introduces a computationally efficient strategy for type I error control in NB-GLM–based DE analysis~\cite{barry2025permuted}, though no software is currently available for comparison. Unlike Nullstrap-DE, which serves as a general add-on framework applicable to existing parametric DE methods, not limited to DESeq2 and edgeR, this preprint offers an independent method implementing its own NB-GLM and score-based testing procedure, focusing on type I error control rather than explicit FDR calibration. Beyond bulk RNA-seq, Nullstrap-DE is also well-suited for DE analysis in single-cell RNA-seq (scRNA-seq) under the pseudobulk framework, where gene counts from individual cells are aggregated at the sample level before inter-sample variability is modelled~\cite{squair2021confronting}.

We introduce the Nullstrap-DE framework, algorithm, and theory in Section~\ref{method}. Extensive simulations (Section~\ref{simulation}) demonstrate that Nullstrap-DESeq2 and Nullstrap-edgeR consistently maintain valid FDR control across diverse settings, including varying sample sizes, effect sizes, true DE gene proportions, target FDR levels, and the presence of confounding covariates, even when DESeq2 or edgeR fail. 
Real data analyses (Sections~\ref{real1}--\ref{real4}) further show that Nullstrap-DE eliminates false positives in negative-control datasets and enhances the detection of biologically relevant DE genes across multiple bulk and scRNA-seq studies.

\section{Methods}\label{method}

\subsection{NB-GLM for DE Analysis}

We begin by introducing the NB-GLM, a widely adopted model for DE analysis on RNA-seq count data. Both DESeq2 and edgeR rely on NB-GLM, but with method-specific estimation and testing procedures. Let
$
\mathbf{Y} = [Y_{ij}] \in \mathbb{N}^{n \times p}
$
denote the observed RNA-seq count matrix, where \(n\) is the number of samples, \(p\) is the number of genes, and \(Y_{ij}\) is the count for gene \(j\) in sample \(i\). Suppose the dataset contains \(K\) biological conditions (e.g., treatment groups, cell types, time points), with the $K$th condition treated as the reference condition. To encode the condition for each sample $i$, we construct \(K-1\) dummy variables. Specifically, for \(k = 1, \ldots, K-1\), we define
$x_{ik} = \mathbb{I}(\text{sample } i \text{ belongs to condition } k)$, 
and let
$
\mathbf{x}_i = (x_{i1}, \ldots, x_{i(K-1)})^\top \in \{0,1\}^{K-1}
$
denote the condition vector for sample \(i\).
Additional sample-level covariates (e.g., batch, age, sex, etc.) are denoted by
$
\mathbf{z}_i = (z_{i1}, \ldots, z_{id})^\top \in \mathbb{R}^d,
$
where \(d\) is the number of covariates.

For each gene $j$ and sample $i$, the observed count is modeled as 
\[
Y_{ij} \overset{\mathrm{ind}}{\sim} \mathrm{NB}\left(\, \mu_{ij}, \phi_j\right),
\]
where \(\mu_{ij} > 0\) is the mean and \(\phi_j > 0\) is the dispersion parameter of the NB distribution. The mean is linked to the condition via a log-linear model:
\begin{equation*}
\log(\mu_{ij}) = \log(s_i) + \alpha_j + \mathbf{x}_i^\top \bm{\beta}_j + \mathbf{z}_i^\top \bm{\gamma}_j,
\label{eq:mean}
\end{equation*}
where \(s_i > 0\) is a sample-specific size factor that accounts for sequencing depth (i.e., library size) and enters the model as an offset; \(\alpha_j \in \mathbb{R}\) is the gene-specific intercept; \(\bm{\beta}_j \in \mathbb{R}^{K-1}\) contains the effects of the \(K-1\) non-reference conditions; and \(\bm{\gamma}_j \in \mathbb{R}^{d}\) represents the effects of additional sample-level covariates.

The goal of DE analysis is to identify genes whose expression levels differ significantly across experimental conditions. For gene~\(j\), the DE test can be formulated as
\[
H_{0,j}\!:\;\bm{\beta}_j = \mathbf{0} 
\quad\text{vs.}\quad 
H_{1,j}\!:\;\bm{\beta}_j \neq \mathbf{0}.
\]
In cases where interest lies in a specific condition \(k\in\{1,\cdots,K\}\), the test becomes
\[
H_{0,jk}\!:\;\beta_{jk}=0 
\quad\text{vs.}\quad 
H_{1,jk}\!:\;\beta_{jk}\neq 0,
\]
where \(\beta_{jk}\) is the \(k\)-th enrty of \(\bm{\beta}_j\).
To express the model compactly across all samples, we define the design matrix 
\(\mathbf{X} = (\mathbf{x}_1, \ldots, \mathbf{x}_n)^\top \in \mathbb{R}^{n \times (K-1)}\) 
and the covariate matrix 
\(\mathbf{Z} = (\mathbf{z}_1, \ldots, \mathbf{z}_n)^\top \in \mathbb{R}^{n \times d}\), 
where each row corresponds to a sample. These matrices capture the experimental design and sample-level covariates used in modeling gene expression.

\subsection{Nullstrap-DE Framework}
Motivated by the idea of Nullstrap~\cite{wang2025nullstrapsimplehighpowerfast}, we propose a synthetic null data-based framework for DE analysis, denoted as Nullstrap-DE. The core of Nullstrap-DE has three main components: 
(1) \textbf{synthetic null data generation}, based on the fitted models under gene-specific null hypotheses;
(2) \textbf{computation of test statistics} $\mathcal{E}(\cdot, \cdot, \cdot):\mathbb{N}^{n \times p} \times \{0,1\}^{n \times (K-1)} \times \mathbb{R}^{n \times d} \to \mathbb{R}^{p}$, applied to calculate test statistics, one per gene, from the original data and synthetic null data in parallel, and 
(3) \textbf{FDR calibration}, where test statistics from the synthetic null data are used to calibrate the significance threshold for those from the original data.

For each gene $j$, the synthetic null data for DE analysis is generated based on the fitted model under the null hypothesis, which assumes gene $j$ is not DE.  

\begin{definition}[Synthetic null data generation for DE analysis]\label{def:synde}
The synthetic null data matrix \(\widetilde{\mathbf{Y}} = [\widetilde{Y}_{ij}] \in \mathbb{N}^{n \times p}\) is generated from the fitted model under the null hypothesis as follows:
\begin{equation*}
\widetilde{Y}_{ij} \overset{\mathrm{ind}}{\sim} \mathrm{NB} \left({\mu}_{ij}^0,\, \hat{\phi}_j \right),
\end{equation*}
where the mean parameter \({\mu}_{ij}^0\) is defined by
\begin{equation*}
{\mu}_{ij}^0 =\exp\left( \log(\hat{s}_i) + \hat{\alpha}_j + \mathbf{x}_i^\top \bm{\beta}_0 + \mathbf{z}_i^\top \hat{\bm{\gamma}}_j  \right)=  \exp\left( \log(\hat{s}_i) + \hat{\alpha}_j + \mathbf{z}_i^\top \hat{\bm{\gamma}}_j \right),
\end{equation*}
with the null coefficient vector \(\bm{\beta}_0 = \mathbf{0}\). 
Here, \(\hat{s}_i\) is the sample-specific size factor, \(\hat{\alpha}_j\) is the intercept, \(\hat{\bm{\gamma}}_j\) is the covariate coefficient, and \(\hat{\phi}_j\) is the dispersion parameter, each estimated from the original data \((\mathbf{Y}, \mathbf{X}, \mathbf{Z}) \).
\end{definition}

With the synthetic null data \(\widetilde{\mathbf{Y}}\) generated according to Definition~\ref{def:synde}, we write the test statistics for the real data and synthetic null data as
\begin{equation*}
\widehat{\bT} = (\widehat{T}_{1},\dots, \widehat{T}_{p})^\top = \mathcal{E}(\mathbf{Y}, \mathbf{X}, \mathbf{Z}), \quad \widetilde{\bT} = (\widetilde{T}_{1},\dots, \widetilde{T}_p)^\top = 
\mathcal{E}(\widetilde{\mathbf{Y}}, \mathbf{X}, \mathbf{Z}).
\end{equation*}
Several choices are available for the test statistics. An example is the Wald test statistic for each gene \(j\), 
\begin{equation*}\label{eq:wald}
    \widehat{T}_j = \hat{\bm{\beta}}^\top_j \widehat{\operatorname{Cov}}(\hat{\bm{\beta}}_j)^{-1} \hat{\bm{\beta}}_j, \quad \widetilde{T}_j = \tilde{\bm{\beta}}^\top_j \widehat{\operatorname{Cov}}(\tilde{\bm{\beta}}_j)^{-1} \tilde{\bm{\beta}}_j, 
\end{equation*}
where \(\hat{\bm{\beta}}_j\) and \(\tilde{\bm{\beta}}_j\) are the estimated coefficients from the original data
\(\{\mathbf Y, \mathbf X, \mathbf Z\}\) and the synthetic null data
\(\{\widetilde{\mathbf Y}, \mathbf X, \mathbf Z\}\), respectively.
The operator \(\widehat{\operatorname{Cov}}(\cdot)\) refers to the estimator of covariance.

In our results of Nullstrap-DE for a two-condition scenario (where $\hat{\beta}_j$ is scalar), DESeq2 uses the Wald statistic $\widehat{T}_j = |\hat{\beta}_j|/\mathrm{se}(\hat{\beta}_j)$; using the same statistic ensures that Nullstrap-DE preserves DESeq2’s gene ranking and differs only through its recalibrated significance threshold. For edgeR, whose native test is a likelihood-ratio statistic, we used the coefficient magnitude $\widehat{T}_j = |\hat{\beta}_j|$ (edgeR does not provide $\mathrm{se}(\hat{\beta}_j)$), making the statistic related to DESeq2’s but allowing Nullstrap-edgeR to differ from edgeR in both ranking and DE calls because both the statistic and threshold change. The Nullstrap-DE R package also supports an alternative test statistic $\widehat{T}_j = -\log p_j$, matching the parent method’s ranking while doing FDR calibration.

With the negative control $\tilde{\bT}$, we can estimate the FDP as
\begin{align*}\label{eq:FDP}
\widehat{\text{FDP}}(t) = \frac{\#\{j : |\widetilde{T}_{j}| \ge t\}}{\max\{\#\{j : |\widehat{T}_{j}| \ge t\},\,1\}}.
\end{align*}
Then, given a target FDR level \( q \in (0,1) \), the set of selected DE genes \( \widehat{\mathcal{S}}(\tau_q) \) depends on the threshold \( \tau_q \):
\begin{equation*}\label{eq:tau_q}
\tau_q = \min \left\{ t > 0 : \widehat{\operatorname{FDP}}(t) \le q \right\}, \quad
\widehat{\mathcal{S}}(\tau_q) = \{ j : |\widehat{T}_{j}| \ge \tau_q \}. 
\end{equation*}


We outline the Nullstrap-DE procedure for DESeq2 and edgeR in Algorithm~\ref{alg:Nullstrap-DE}. The theoretical guarantees, with full proofs, are provided in Supplementary Methods~\ref{supp:theory}, \ref{supp:proof1} and~\ref{supp:proof2}. Assuming the parent method’s parameter estimates are consistent, we show that Nullstrap-DE asymptotically controls the FDR at the nominal level and attains power converging to one in probability as the sample size grows. 

\begin{algorithm}[H]
\caption{Nullstrap-DE: An \textit{add-on} framework for NB-GLM--based DE methods}
\label{alg:Nullstrap-DE}
\SetKwInOut{Input}{Input}\SetKwInOut{Output}{Output}

\Input{Count matrix $\mathbf{Y}\in\mathbb{N}^{n\times p}$;\\
  Treatment design matrix $\mathbf{X}\in\{0,1\}^{n\times (K-1)}$; 
  Covariate matrix $\mathbf{Z}\in\mathbb{R}^{n\times d}$;\\
  Target FDR level $q\in(0,1)$;\\
}
\Output{Set of genes $\widehat{\mathcal{S}}(\tau_q) \subset \{1,\ldots,p\}$ declared DE;\\}

\BlankLine
\textbf{Step 1: Fit NB-GLMs to original data}\;
Fit DESeq2 or edgeR NB-GLMs to $(\mathbf{Y},\mathbf{X},\mathbf{Z})$ to obtain size factors $\hat{s}_i$ for $i=1,\ldots,n$; gene-specific dispersions $\hat{\phi}_j$, and parameter estimates $(\hat{\alpha}_j,\hat{\bm{\beta}}_j,\hat{\bm{\gamma}}_j)$ for \(j=1,\dots,p\)\;
 Compute test statistics $\widehat{\mathbf{T}}=\mathcal{E}(\mathbf{Y},\mathbf{X},\mathbf{Z})$ based on the parameter estimates\;

\BlankLine
\textbf{Step 2: Generate synthetic null data}\;
\For{$j=1$ \KwTo $p$}{
  \For{$i=1$ \KwTo $n$}{
  Set $\tilde{s}_i = \hat{s}_i$, or sample $\tilde{s}_i$ from the empirical distribution of $\{\hat{s}_1, \ldots, \hat{s}_n\}$\;
Set $\mu_{ij}^{0}\gets\exp\bigl(\log\tilde{s}_i+\hat{\alpha}_j+\mathbf{z}_i^{\top}\hat{\bm{\gamma}}_j\bigr)$\;
    Draw $\widetilde{Y}_{ij} \overset{\mathrm{ind}}{\sim} \mathrm{NB} \left({\mu}_{ij}^0,\, \hat{\phi}_j \right)$\;
  }
}
Denote the synthetic null data matrix by $\widetilde{\mathbf{Y}}$\;

\BlankLine
\textbf{Step 3: Fit NB-GLMs to synthetic null data}\;
 Fit DESeq2 or edgeR NB-GLMs to $(\widetilde{\mathbf{Y}},\mathbf{X},\mathbf{Z})$ with
 $\tilde{s}_i$ and $\hat{\phi}_j$ fixed to obtain parameter estimates $(\tilde{\alpha}_j,\tilde{\bm{\beta}}_j)$ for \(j=1,\dots,p\)\;
 Compute test statistics 
 $\widetilde{\mathbf{T}}=\mathcal{E}(\widetilde{\mathbf{Y}},\mathbf{X},\mathbf{Z})$ based on the parameter estimates\;

\BlankLine
\textbf{Step 4: Compute the data-driven threshold and declare discoveries (DE genes)}\;
 Compute
$\widehat{\text{FDP}}(t)=
  \dfrac{\#\{j:|\,\widetilde{T}_j|\ge t\}}
        {\max\{\#\{j:|\,\widehat{T}_j|\ge t\},1\}}$; Set
$\displaystyle
  \tau_q=\min\bigl\{t>0 : \widehat{\text{FDP}}(t)\le q\bigr\}$\;
%
\Return  $\widehat{\mathcal{S}}(\tau_q)=\{j:|\,\widehat{T}_j|>\tau_q\}$\;

\end{algorithm}

\section{Results}

\subsection{Simulations Demonstrate Nullstrap-DE’s Reliable FDR Control and High Power}\label{simulation}

We evaluated the performance of Nullstrap-DE \textit{add-on} for both DESeq2 and edgeR (referred to as Nullstrap-DESeq2 and Nullstrap-edgeR) through comprehensive simulations. For comparison, we also include the original DESeq2 and edgeR methods, as well as the Wilcoxon rank-sum test applied to normalized data (Wilcoxon\_norm; TMM normalized counts from edgeR) and raw counts (Wilcoxon\_raw; no normalization).

Following the data-generating scheme of the DESeq2 paper~\cite{love2014moderated}, we simulated semi-synthetic RNA-seq datasets under an NB-GLM. 
Two main settings were considered: one without covariates and one with additional confounding covariates. For each setting, 50 replicate datasets were generated, and empirical FDR and power were calculated based on these replicates. Simulation details and additional settings with model mis-specification (Poisson-GLM and zero-inflated NB-GLM) are provided in Supplementary Method~\ref{supp:simu}.

Fig.~\ref{fig:simu1} presents representative results, with additional findings shown in Supplementary Figs.~\ref{fig:simu1_de=0.2}--\ref{fig:simu2_de=0.1}. Across all scenarios, both without covariates (Fig.~\ref{fig:simu1}a,b) and with covariates (Fig.~\ref{fig:simu1}c,d), Nullstrap-DESeq2 
and Nullstrap-edgeR maintain effective FDR control at the target level while achieving comparable or higher statistical power than alternative approaches, including the original DESeq2, edgeR, and two 
Wilcoxon rank-sum tests. In contrast, DESeq2 and edgeR exhibit increasing FDR inflation as sample size, fold change, or target FDR level increases, consistent with previous findings~\cite{li2022exaggerated}. 
The Wilcoxon\_norm method fails to control the FDR in most settings (consistent with \cite{ge2024response}), and both Wilcoxon rank-sum tests show low power with small samples and cannot accommodate additional covariates. Notably, these results remain to hold under model mis-specification (data generated from Poisson-GLM and zero-inflated NB-GLM; Supplementary Fig.~\ref{fig:simu3_de=0.2}, \ref{fig:simu4_de=0.2}). Overall, Nullstrap-DESeq2 and Nullstrap-edgeR provide reliable FDR calibration and strong power even in small-sample or model-misspecified scenarios, whereas their parent methods (DESeq2 and edgeR) and nonparametric methods (Wilcoxon\_norm and Wilcoxon\_raw) often exhibit inflated FDR or low power.

\subsection{Permutation Negative Controls Confirm Nullstrap-DE’s Robust FDR Control}\label{real1}

We benchmarked Nullstrap-DESeq2 and Nullstrap-edgeR against DESeq2, edgeR, and Wilcoxon\_norm for FDR control using an RNA-seq dataset of classical and non-classical human monocytes~\cite{williams2017empirical}. 
The dataset includes 34 samples (17 classical CD14\(^+\)CD16\(^-\) vs. 17 non-classical CD14\(^\text{low}\)CD16\(^+\)) spanning 52{,}376 genes derived from peripheral blood mononuclear cells of Ugandan children. 
Following~\cite{li2022exaggerated}, we generated 1{,}000 negative-control datasets by permuting condition labels across samples, providing an empirical basis to assess FDR control since no truly DE genes are expected under permutation.

Fig.~\ref{fig:case1} summarizes the number of DE genes at the target FDR level of $0.05$ across the permuted datasets. DESeq2 and edgeR detect a substantial number of spurious DE genes across the permutations, whereas Nullstrap-DESeq2 and Nullstrap-edgeR consistently report near-zero DE genes, demonstrating strong empirical FDR control (Fig.~\ref{fig:case1}a). Wilcoxon\_norm yields fewer false positives than DESeq2 and edgeR, but still identifies a non-negligible number of spurious DE genes. The full distribution of DE gene counts across the permuted datasets (Fig.~\ref{fig:case1}b) further highlights the contrast: DESeq2 and edgeR show heavy tails with frequent false discoveries, while Nullstrap-DE methods are sharply centered at zero. These results underscore the robustness of Nullstrap-DE in the presence of spurious variation introduced by permutation, and their reliability in maintaining FDR control where DESeq2 and edgeR tend to be anti-conservative.

\subsection{Bulk RNA-seq: Classical vs. Non-Classical Human Monocytes}\label{real2}

In this subsection, we compared Nullstrap-DE with DESeq2, edgeR, and Wilcoxon\_norm using the same dataset as in Subsection~\ref{real1}. At a target FDR of 0.05, Nullstrap-DESeq2 and Nullstrap-edgeR identified fewer DE genes than DESeq2, edgeR, and Wilcoxon\_norm (Fig.~\ref{fig:case2}a). 
As an \textit{add-on} to DESeq2 or edgeR, Nullstrap-DE preserves the parent method’s fold-change estimates while improving FDR control, allowing direct comparison using the same MA plots (Fig.~\ref{fig:case2}b). Nullstrap-DE methods tended to select genes with larger fold changes and higher expression levels, and their top-ranked DE genes largely overlapped with those from DESeq2 and edgeR. Heatmaps of scaled expression (Fig.~\ref{fig:case2_supp1}a) further show that DE genes identified by Nullstrap-DE exhibit clearer separation between the two monocyte subsets.

To investigate whether DESeq2, edgeR, and Wilcoxon\_norm yielded excess false discoveries, and whether the Nullstrap-DE \textit{add-on} effectively screened for biologically meaningful DE genes from its parent method, we performed gene ontology (GO) enrichment analysis on the sets of DE genes from each method (Fig.~\ref{fig:case2_supp1}b,c,d). The results indicate that DE genes identified by Nullstrap-DESeq2 and Nullstrap-edgeR are mainly associated with immune-related GO terms, whereas those found by DESeq2, edgeR, and Wilcoxon\_norm are associated with broader and less specific functions. 
We further compared the ranks of four representative GO terms in the enrichment analysis (Fig.~\ref{fig:case2}c). The immune-related GO terms (e.g., GO:0001819 ``positive regulation of cytokine production'') are among the highest-ranked terms for Nullstrap-DESeq2 and Nullstrap-edgeR, but are ranked much lower by DESeq2, edgeR, and Wilcoxon\_norm. Conversely, GO terms related to general cellular processes (e.g., GO:0042254 ``ribosome biogenesis'') are prioritized by DESeq2, edgeR, and Wilcoxon\_norm, but are ranked lower or not detected by Nullstrap-DE. Together, these results indicate that although DESeq2, edgeR, and Wilcoxon\_norm identify more DE genes, many are less biologically relevant, whereas Nullstrap-DESeq2 and Nullstrap-edgeR more effectively enrich for meaningful DE genes.

Building on these findings, we performed separate GO enrichment analyses for genes upregulated in non-classical versus classical monocytes (Fig.~\ref{fig:case2}d), and vice versa (Fig.~\ref{fig:case2}e). Marked differences in biological specificity emerged between Nullstrap-edgeR and edgeR. 
For non-classical (CD14\textsuperscript{low}CD16\textsuperscript{+}) monocytes, Nullstrap-edgeR highlighted immune processes such as leukocyte activation, T cell proliferation, and NK cell–mediated immunity, consistent with their roles in immune surveillance and coordination of adaptive responses~\cite{auffray2007monitoring, cros2010human}. For classical (CD14\textsuperscript{+}CD16\textsuperscript{–}) monocytes, Nullstrap-edgeR enriched for leukocyte migration, response to lipopolysaccharide, and cell chemotaxis, consistent with their well-established roles in pathogen sensing, migration, chemotaxis, and inflammatory initiation~\cite{geissmann2003blood,ziegler2007cd14+}. 
In contrast, GO terms identified by edgeR were a mixture of immune-related and general cellular processes, and edgeR-only DE genes were mainly enriched for broad housekeeping functions. These results indicate that Nullstrap-edgeR substantially improves the biological specificity of edgeR in identifying DE genes that distinguish monocyte subsets.

\subsection{Bulk RNA-seq: Dexamethasone Treatment in Airway Smooth Muscle}\label{real3}

In this subsection, we applied Nullstrap-DE, DESeq2, edgeR, and Wilcoxon\_norm to an RNA-seq dataset from a dexamethasone treatment study in airway smooth muscle cells~\cite{himes2014rna}, consisting of 4 untreated and 4 treated samples and 63{,}677 genes. At a target FDR of 0.05, Nullstrap-DESeq2 and Nullstrap-edgeR identifies fewer DE genes than DESeq2 and edgeR, while Wilcoxon\_norm does not identify any DE genes, reflecting its low power in small-sample settings (Fig.~\ref{fig:case3}a). 
The MA plots show that Nullstrap-DE methods preferentially retain genes with larger fold changes and higher expression levels compared to their parent methods (Fig.~\ref{fig:case3}b). 
Heatmaps of normalized expression (Fig.~\ref{fig:case3}c) further demonstrate that DE genes selected by Nullstrap-edgeR exhibit clearer separation between treated and untreated samples compared to those identified by edgeR or edgeR only (Fig.~\ref{fig:case3}c). Additionally, hierarchical clustering based on Nullstrap-edgeR or edgeR DE genes groups samples correctly by treatment status, whereas clustering with edgeR-only DE genes fails to do so, indicating that Nullstrap-DE more effectively captures treatment-associated signals.

To assess the biological relevance of identified DE genes, we performed KEGG enrichment analysis on DE genes upregulated under dexamethasone treatment for each method. The DE genes identified by Nullstrap-DESeq2 and Nullstrap-edgeR are significantly enriched in biological pathways that are directly relevant to the expected effects of dexamethasone on airway smooth muscle cells (Fig.~\ref{fig:case3}d), including the calcium signaling pathway, cytokine–cytokine receptor interaction, and the cAMP signaling pathway, all of which are known to mediate glucocorticoid signaling and regulate airway smooth muscle function~\cite{newton2000molecular, panettieri2019non}. 
In contrast, DESeq2 and edgeR identify many pathways with limited relevance to dexamethasone response, including human papillomavirus infection and various cancer-related pathways. Moreover, DE genes uniquely detected by DESeq2 or edgeR showed little enrichment for pathways consistent with known dexamethasone biology, suggesting that these additional genes likely represent spurious signals filtered out by Nullstrap-DE.

\subsection{Single-cell RNA-seq: Monocytes from COVID-19 Patients with Varying Severity}\label{real4}

In this subsection, we analyzed scRNA-seq COVID-19 data~\cite{ren2021covid} under the pseudobulk framework, where cell-level counts are aggregated by sample. We used 31 PBMC samples, focusing on monocytes, yielding 24,134 genes after removing mitochondrial genes. The DE analysis was performed between patients with mild/moderate versus severe/critical symptoms. At a target FDR of 0.1, Nullstrap-DESeq2 and Nullstrap-edgeR identify fewer DE genes than DESeq2 and edgeR (Fig.~\ref{fig:case4}a,b). GO enrichment analysis shows that Nullstrap-DESeq2 DE genes are strongly enriched for monocyte-specific immune processes, such as antigen processing and MHC class II presentation (Fig.~\ref{fig:case4}c, left), closely linked to COVID-19 severity~\cite{vanderbeke2021monocyte, saichi2021single, leon2022virus}. DESeq2 also produces meaningful pathways but prioritizes broader processes (Fig.~\ref{fig:case4}c, middle). In contrast, DESeq2 only genes are enriched for nonspecific terms such as transferase activity and metabolic processes (Fig.~\ref{fig:case4}c, right), suggesting that Nullstrap-DESeq2 effectively filters out lower-relevance or spurious signals.

To visualize how enriched biological processes relate to their underlying genes, we constructed bipartite gene–concept networks for Nullstrap-DESeq2 and DESeq2 (Fig.~\ref{fig:case4}d). Each edge links a DE gene to an enriched GO term. The Nullstrap-DESeq2 network forms a coherent module centered on antigen presentation, driven by core genes such as \textit{HLA-DRA}, \textit{HLA-DPB1}, and \textit{HLA-DQA1}, all well-established markers of monocyte immune activity~\cite{kim2010differential, joshi2023utility} and highly relevant in COVID-19~\cite{liu2023monocytic, henao2024classical}. 
In contrast, the DESeq2 network is more diffuse, spanning broader functional categories. Although it includes some antigen-related processes, its top GO term, ``positive regulation of cell adhesion,'' was driven by genes that are not central to monocyte immune function in COVID-19. These results show that Nullstrap-DESeq2 improves biological interpretability by filtering spurious signals while preserving core immune pathways.

\section{Discussion}
In this paper, we introduce Nullstrap-DE, a statistical framework that enhances FDR control while retaining statistical power in DE analysis. As an add-on to popular parametric DE methods such as DESeq2 and edgeR, Nullstrap-DE preserves the parent method’s modeling structure, parameter estimation, and workflow, yet provides more reliable FDR calibration when the parent method exhibits inflated FDR. Additionally, compared with nonparametric methods such as the Wilcoxon rank-sum test, Nullstrap-DE maintains high power in small-sample settings and supports covariate adjustment. These features make Nullstrap-DE a reliable, powerful, and user-friendly enhancement to existing DE workflows for both bulk and pseudobulk scRNA-seq data, offering improved statistical rigor without compromising interpretability and usability across diverse transcriptomic settings.

Nullstrap-DE builds on a synthetic-null-data-based framework for DE testing, which enables empirical FDR control by comparing the original data to synthetic null data generated under the null hypotheses. The synthetic null data preserve the structural features of the original data while removing the DE effects of interest. For DESeq2 or edgeR, this is achieved by generating data from the fitted NB-GLMs with all samples assigned to a common reference condition while retaining other parameter estimates. Unlike conventional methods that rely on theoretical $p$-values, Nullstrap-DE’s synthetic null approach requires only accurate estimation of the null model parameters, rather than accurate specification of the null distribution of test statistics, making it well-suited for complex or noisy datasets where parameter estimation involves multiple steps and deriving accurate test-statistic distributions is challenging.

The Nullstrap-DE framework is broadly extensible and offers several promising directions for future research. Although we focus on DESeq2 and edgeR, the synthetic-null-data approach can be applied to many other statistical models. For example, Nullstrap-DE can be adapted to generalized linear mixed models to accommodate random effects or repeated measures, or to nonparametric methods that support null-based resampling. 
Another direction is extending Nullstrap-DE to scRNA-seq counts without pseudobulking, where noise, sparsity, and high dimensionality pose challenges for FDR control. Adapting the Nullstrap-DE framework to these settings would further expand its utility in modern transcriptomic analysis. More broadly, Nullstrap-DE illustrates the potential of the synthetic-null-data-based framework as a general strategy for improving the reliability of inference in high-throughput biological data.

\section{Code and Data Availability}

The \texttt{NullstrapDE} R package, tutorial, and code for reproducing the results are publicly available at \url{https://github.com/chexjiang/NullstrapDE}. The human monocyte and airway smooth muscle cell RNA-seq datasets are available at \url{https://doi.org/10.5281/zenodo.18297224}. The scRNA-seq COVID-19 dataset is available from the Gene Expression Omnibus (GEO) under accession number GSE158055 (\url{https://www.ncbi.nlm.nih.gov/geo/query/acc.cgi?acc=GSE158055}).

\section{Figures}

\section{Supplementary Material}

The supplementary material includes the theory, simulation details, and additional figures.

\FloatBarrier 

{\captionsetup{font=small} 
\begin{figure}[H]
    \centering
    \includegraphics[width=0.9\textwidth]{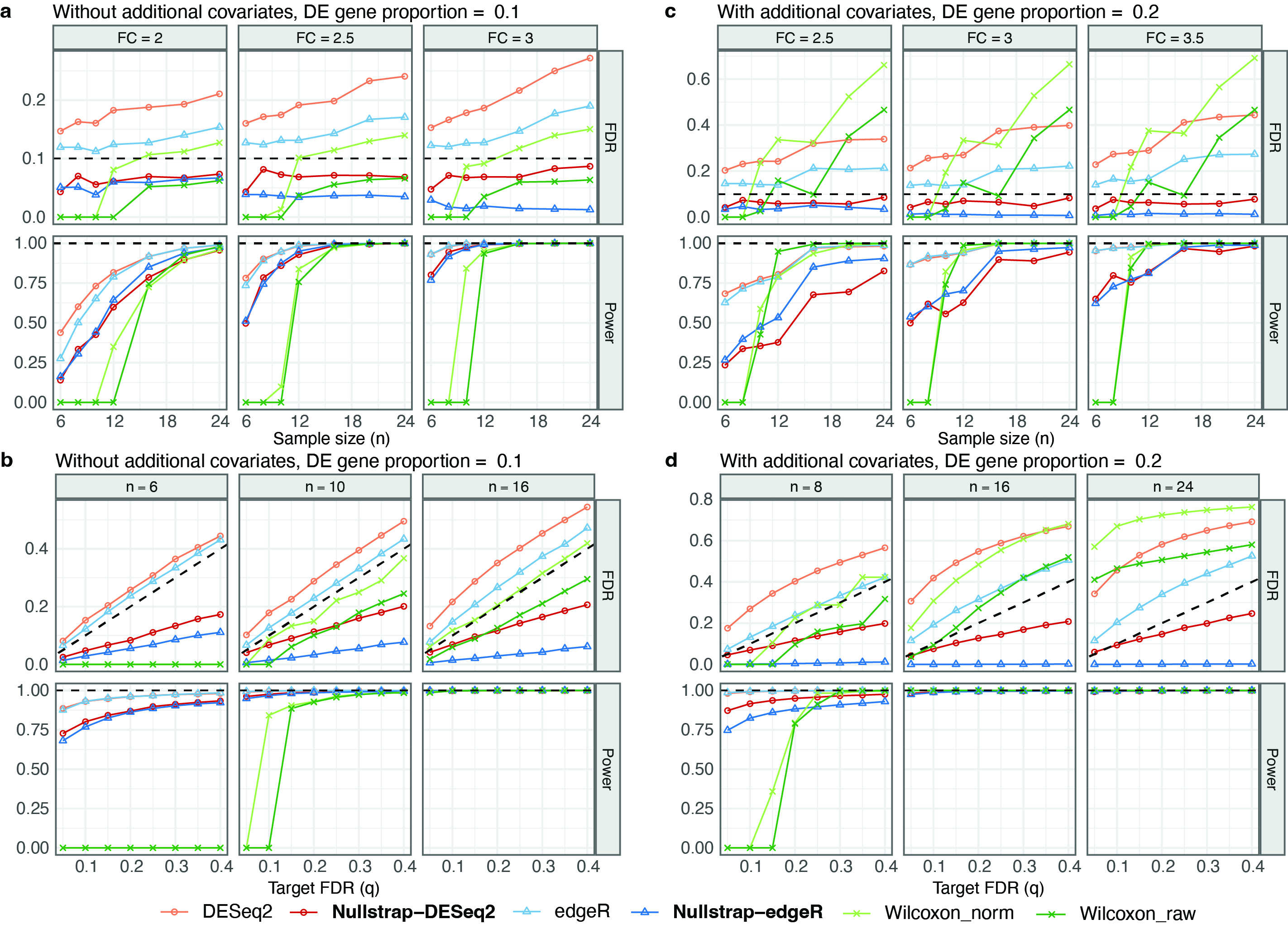}

    \caption{
    \textbf{Simulation results under two settings.} 
    \textbf{(a, b)} Results without covariates: empirical FDR (top) and power (bottom) across sample sizes ($n$) and FDR targets ($q$), with DE proportion $=0.1$.
    \textbf{(c, d)} Results with covariates, with DE proportion $=0.2$.
    Across all scenarios, Nullstrap-DESeq2 and Nullstrap-edgeR maintain effective FDR control while achieving high power, outperforming DESeq2, edgeR, and Wilcoxon rank-sum tests.}
    \label{fig:simu1}
\end{figure}
}
\vspace{-1em}

{\captionsetup{font=small} 
\begin{figure}[H]
    \centering
    \includegraphics[width=0.85\textwidth]{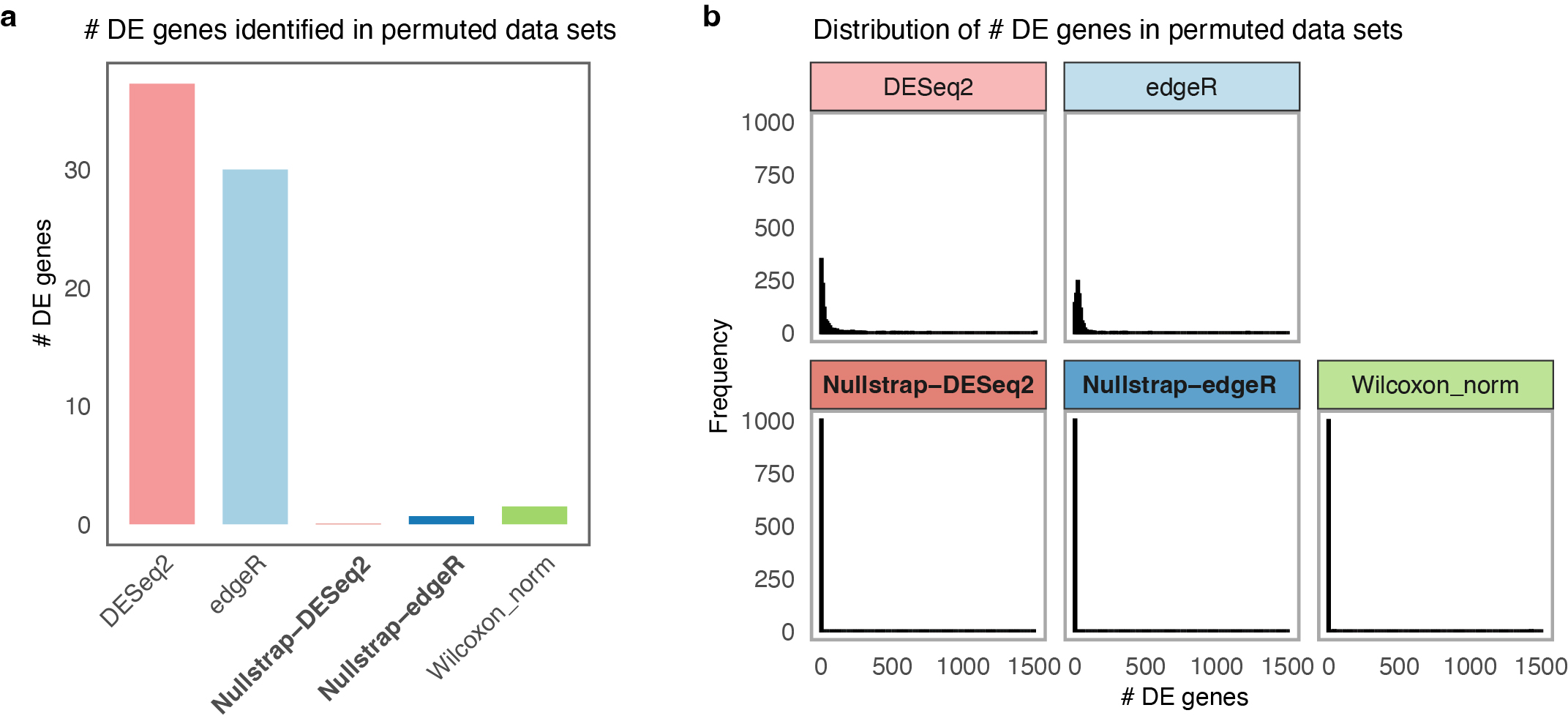}
    \vspace{-1em}
    \caption{
    \textbf{Empirical evaluation of FDR control using permuted negative-control datasets.}
    (\textbf{a,b}) Average and distribution of number of DE genes at FDR $=0.05$ across 1,000 permutations of a real RNA-seq dataset of classical and non-classical human monocytes. DESeq2 and edgeR produce many false positives, whereas Nullstrap-DESeq2 and Nullstrap-edgeR report near-zero DE genes.
}
        \label{fig:case1}
    \end{figure}
}
\FloatBarrier

{\captionsetup{font=small} 
\begin{figure}[htbp]
    \centering
    \includegraphics[width=\textwidth]{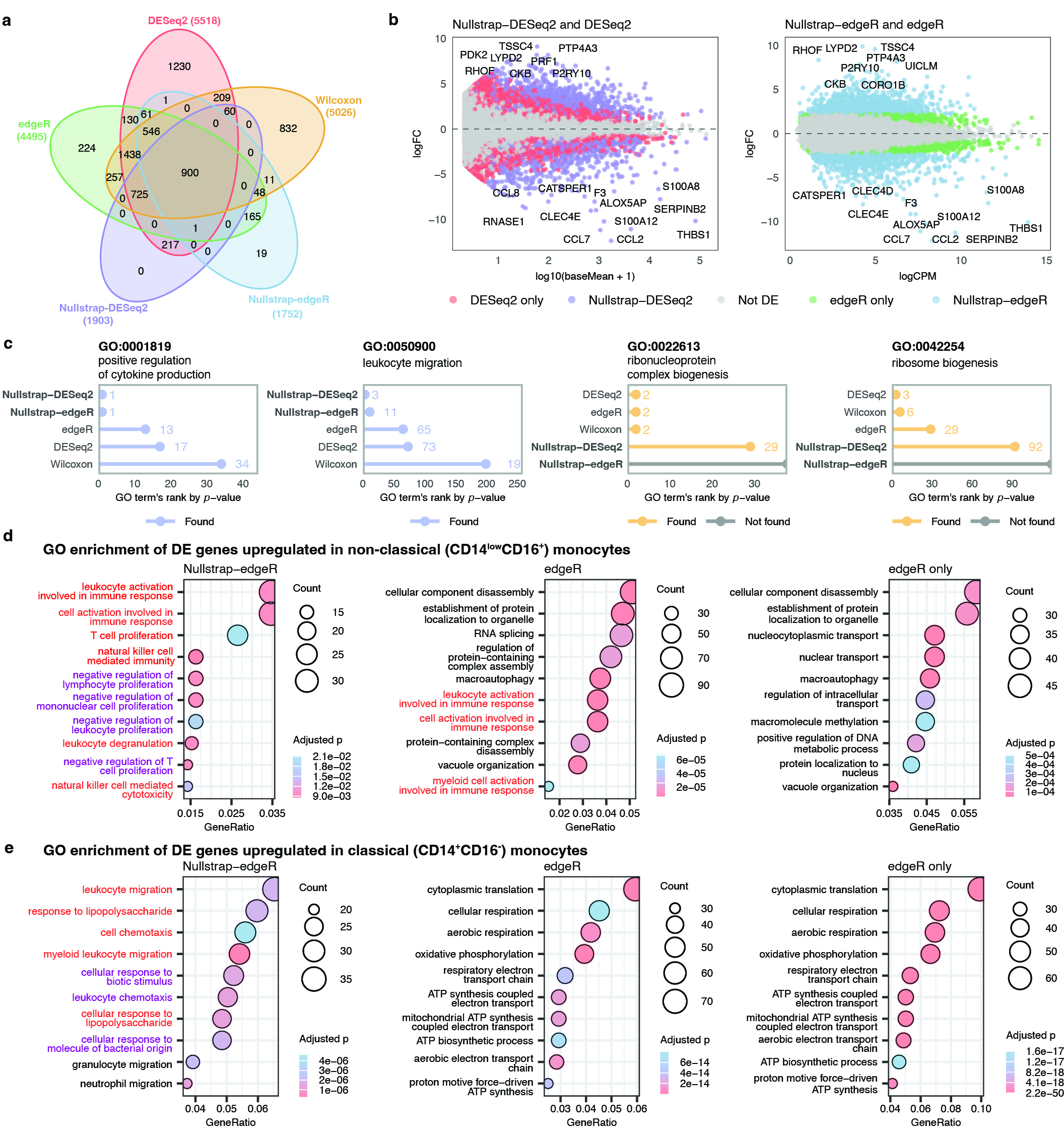}
    \caption{
    \textbf{Comparison of DE methods on RNA-seq data from classical and non-classical human monocytes.}
    (\textbf{a}) Venn diagram showing overlaps of DE genes (FDR $<$ 0.05) identified by DESeq2, edgeR, Wilcoxon, Nullstrap-DESeq2, and Nullstrap-edgeR. 
    (\textbf{b}) MA plots comparing DE genes from Nullstrap-DESeq2 vs. DESeq2 (left) and Nullstrap-edgeR vs. edgeR (right); Nullstrap-DE methods prioritize DE genes with stronger fold changes and higher expression. 
    (\textbf{c}) Ranks of four representative GO terms by enrichment $p$-value for each method, including two immune-related terms (left) and two general cellular functions (right). ``Found'' indicates that the GO term is enriched (hypergeometric test, BH-adjusted $p$-value $<0.05$), while ``Not found'' indicates that the GO term is not enriched.
    Nullstrap-DE methods prioritize immune-specific terms but deprioritize general cellular functions, highlighting their ability to enrich for biologically meaningful signals. 
    (\textbf{d}) GO terms enriched among DE genes upregulated in non-classical (CD14\textsuperscript{low}CD16\textsuperscript{+}) monocytes. GO term colors indicate their biological relevance to non-classical monocytes, with red, purple, and black denoting highly relevant, relevant, and not relevant, respectively. Nullstrap-edgeR predominantly recovers GO terms that are highly relevant or relevant to non-classical monocyte biology. 
    (\textbf{e}) GO terms enriched among DE genes upregulated in classical (CD14\textsuperscript{+}CD16\textsuperscript{–}) monocytes. GO term colors follow the same scheme as in (e). Nullstrap-edgeR predominantly highlights functions specific to classical monocyte biology.
    }
    \label{fig:case2}
\end{figure}
}

{\captionsetup{font=small} 
\begin{figure}[htbp]
    \centering
    \includegraphics[width=\textwidth]{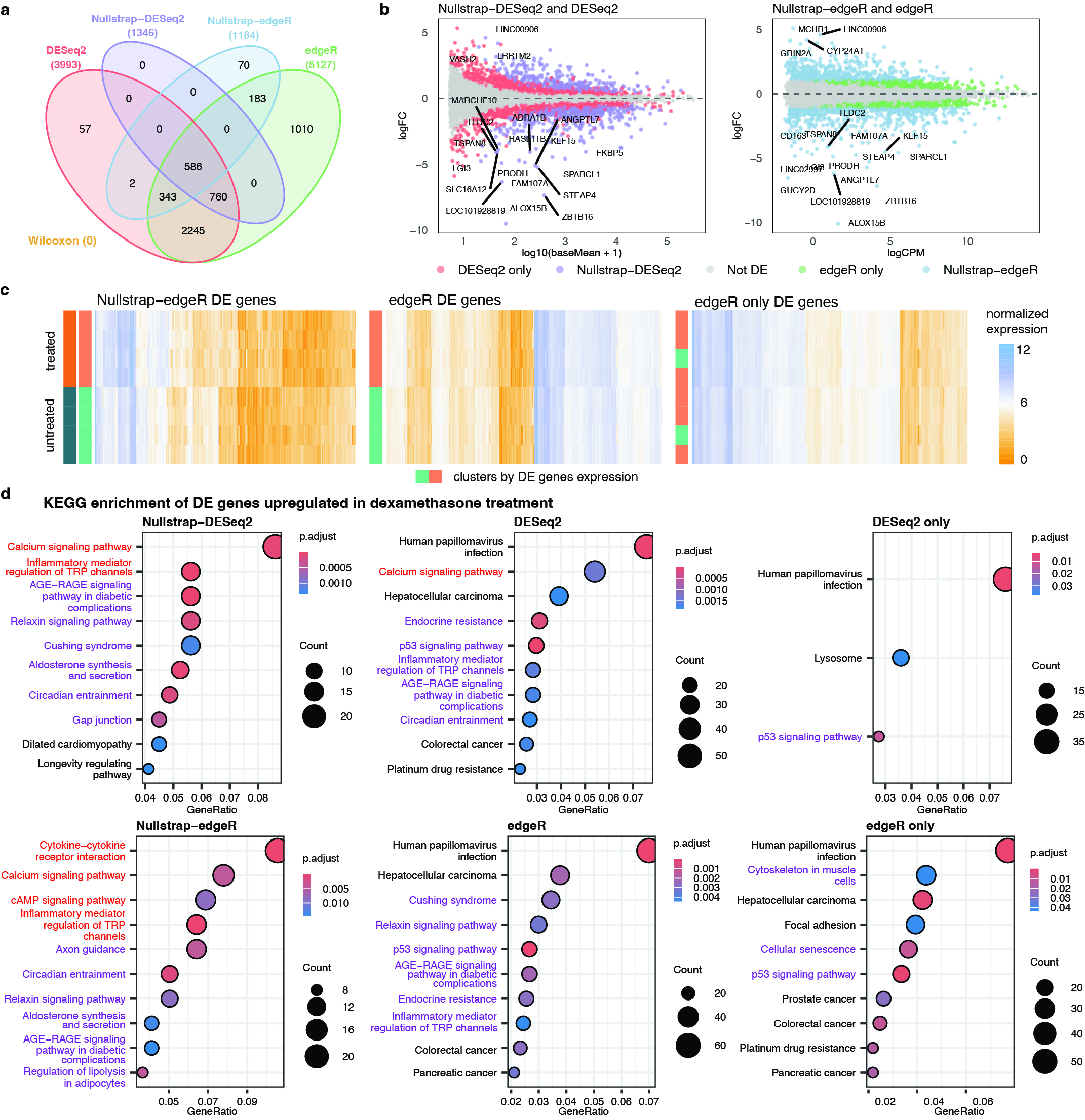}
    \caption{
        \textbf{Comparison of DE methods on RNA-seq data from dexamethasone-treated airway smooth muscle cells.}
    (\textbf{a}) Venn diagram of DE genes (FDR $<$ 0.05) identified by five methods. Nullstrap-DESeq2 and Nullstrap-edgeR methods yield fewer DE genes than DESeq2 and edgeR, while the Wilcoxon rank-sum test detects none. 
    (\textbf{b}) MA plots comparing DE genes retained by Nullstrap-DESeq2 vs. DESeq2 (left) and Nullstrap-edgeR vs. edgeR (right); Nullstrap-DE methods prioritize genes with stronger fold changes and higher average expression. 
    (\textbf{c}) Heatmaps of normalized expression for DE genes identified by Nullstrap-edgeR, edgeR, and edgeR only sets. Nullstrap-edgeR DE genes show clearer separation between treated and untreated samples, with hierarchical clustering correctly recovering treatment groups. 
    (\textbf{d}) KEGG pathway enrichment for DE genes upregulated under dexamethasone treatment. KEGG term colors indicate biological relevance to glucocorticoid signaling and airway smooth muscle function: red = highly relevant, purple = relevant, black = not relevant. Nullstrap-DE methods preferentially enrich for biologically meaningful pathways, while DESeq2, edgeR, and their respective ``only'' sets tend to identify more general or unrelated pathways.
    }
    \label{fig:case3}
\end{figure}
}

{\captionsetup{font=small} 
\begin{figure}[htbp]
    \centering
    \includegraphics[width=\textwidth]{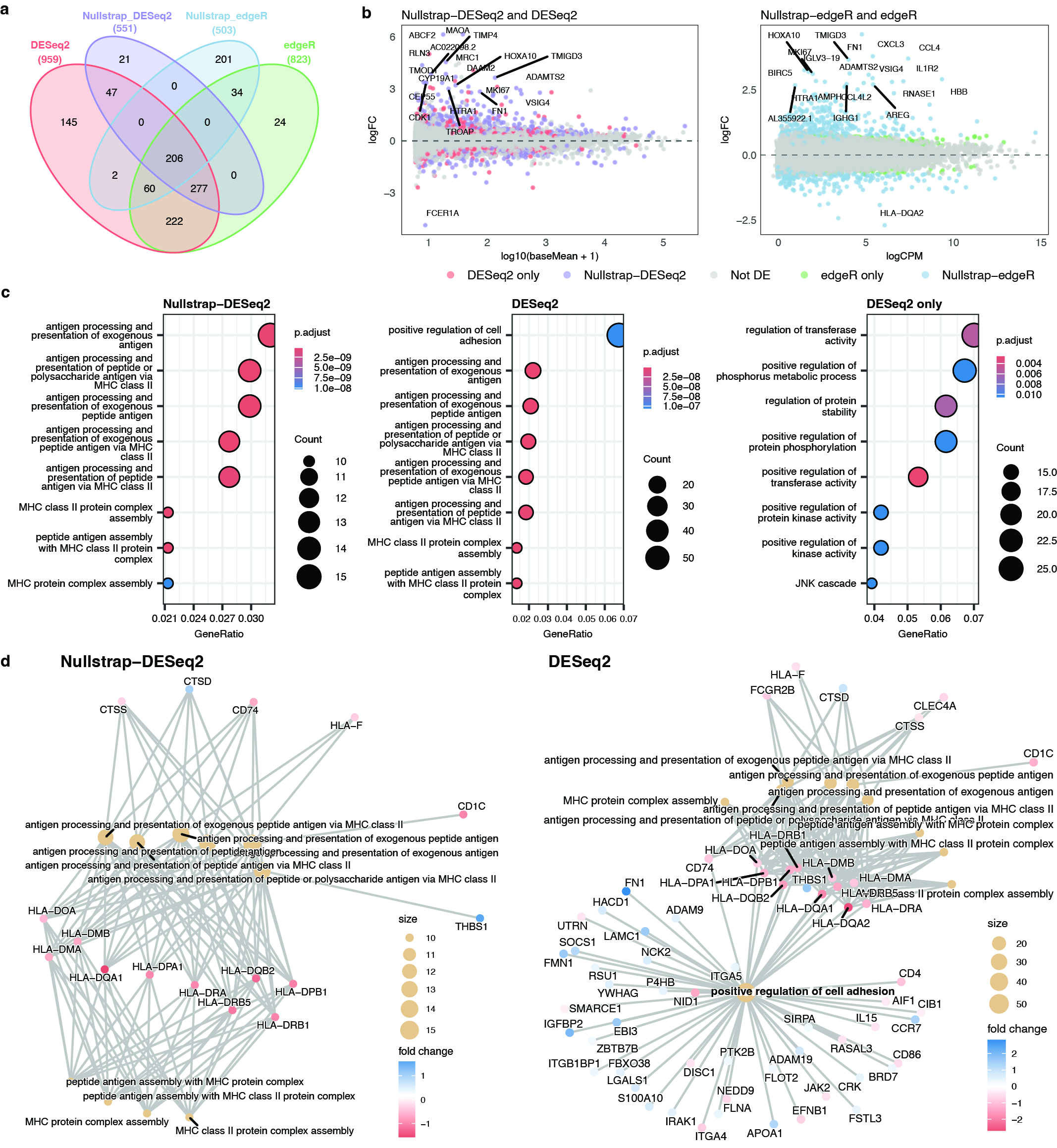}
    \caption{
        \textbf{Comparison of DE methods on pseudobulk scRNA-seq data from COVID-19 patient monocytes.}
   (\textbf{a}) Venn diagram of DE genes (FDR $<$ 0.1) showing Nullstrap-DE methods yield fewer DE genes than their parent methods. (Note: Nullstrap-DESeq2 detects a small set of DE genes not found by DESeq2; these genes are automatically filtered out by DESeq2’s built-in independent filtering step prior to multiple testing correction, but are retained in Nullstrap-DESeq2 because they still have valid Wald statistics and nominal $p$-values.)
    (\textbf{b}) MA plots indicate that Nullstrap-DE prioritizes DE genes with stronger fold changes and higher expression.
    (\textbf{c}) GO enrichment of DE genes in monocytes from mild/moderate vs. severe/critical COVID-19 patients. Nullstrap-DESeq2 recovers immune-specific processes specific to COVID-19 severity (e.g., antigen presentation via MHC class II, peptide–MHC assembly), whereas DESeq2 prioritizes broader categories like cell adhesion, and DESeq2 only genes enrich for less specific terms. 
    (\textbf{d}) Bipartite gene–concept networks show Nullstrap-DESeq2 captures a coherent immune module (e.g., \textit{HLA-DRA}, \textit{HLA-DPB1}, \textit{HLA-DQA1}), while DESeq2 yields a more diffuse network dominated by general adhesion-related genes (\textit{ITGA5, LAMC1, FN1, NID1}).
  }
    \label{fig:case4}
\end{figure}
}

%
%
%
\printbibliography[title={References}]
\end{refsection}

\clearpage

\title{Supplementary Material for ``Nullstrap-DE: A General Framework for Calibrating FDR and
Preserving Power in Differential Expression Methods, with
Adaptivity to DESeq2 and edgeR''}
\titlerunning{Nullstrap-DE: A General Framework for Calibrating FDR and Preserving Power in DE Methods}
%
\author{Chenxin Flora Jiang\inst{1}$^\dagger$\orcidID{0009-0005-7369-4116} \and
Changhu Wang\inst{2}$^\dagger$\orcidID{0009-0002-8567-0961} \and
Jingyi Jessica Li\inst{1,2}\textsuperscript{*}\orcidID{0000-0002-9288-5648}}
\authorrunning{C. F. Jiang et al.}
%
\institute{Department of Statistics and Data Science, University of California, Los Angeles, CA 90095, USA \and
Biostatistics Program, Public Health Science Division, Fred Hutchinson Cancer Center, Seattle, WA 98109, USA 
\\[6pt]
$^\dagger$\ These authors contributed equally to this work.\\[3pt]
\textsuperscript{*}\ Contact: \email{lijy03@fredhutch.org}; \email{jli@stat.ucla.edu}
}
\maketitle              

\begingroup
\setlength{\parindent}{2em}
\makeatletter\@afterindenttrue\makeatother 

\setcounter{section}{0}
\renewcommand\thesection{S\arabic{section}}
\renewcommand\thesubsection{S\arabic{section}.\arabic{subsection}}

\begin{refsection}
\section{Supplementary Methods}

\subsection{Theoretical Guarantees of Nullstrap-DE}\label{supp:theory}

In this section, we provide theoretical results of Nullstrap-DE for FDR control. For simplicity, we consider the case with two experimental conditions (\(K = 2\)), so that the condition effect vector \(\bm{\beta}_j\) reduces to a scalar coefficient \(\beta_j \in \mathbb{R}\) for each gene \(j\).  Let \(\mathcal{S}_0\) and \(\mathcal{S}_1\) denote the sets of ground-truth
non-DE and DE genes, respectively. In this setting, the Wald test statistics 
are equivalent to 
$\widehat{T}_j = {|\hat{\beta}_j|}/{{\mathrm{se}(\hat{\beta}_j)}}$ and $\widetilde{T}_j = {|\tilde{\beta}_j|}/{{\mathrm{se}(\tilde{\beta}_j)}}.
$

We assume in Assumption \ref{assump:estimation} that the estimation procedure $\mathcal{E}(\cdot, \cdot,\cdot)$ is accurate enough. In particular, the original-data estimate $\hat\beta_j$ and the synthetic-null-data estimate $\tilde\beta_j$ both have non-zero and non-infinity standard errors with high probability, and they are both close to their respective true parameters $\beta_j$ and $\beta_{0j} = 0$ with high probability.

\begin{assumption}[Estimation accuracy]\label{assump:estimation}

Assume that the standard errors $\mathrm{se}(\hat{\beta}_j)$ and $\mathrm{se}(\tilde{\beta}_j)$ are uniformly bounded away from zero and infinity with high probability. Specifically, there exist constants $L > l > 0$ such that
\[
\mathbb{P}\!\left(
L\ge \mathrm{se}(\hat{\beta}_j),\ \mathrm{se}(\tilde{\beta}_j) \ge l
\ \text{for all } j = 1,\dots,p
\right) \ge 1 - p^{-1}.
\]

Further, suppose there exists a deterministic sequence $\gamma_{n,p} = O(\log p / \sqrt{n})$ with $\gamma_{n,p} \to 0$ such that
\[
\mathbb{P}\!\left(
\max_{j \in \{1,\dots,p\}}|\hat{{\beta}}_j - {\beta}_j| \le \gamma_{n,p}
\right) \ge 1 - p^{-2},
\qquad
\mathbb{P}\!\left(
\max_{j \in \{1,\dots,p\}}|\tilde{{\beta}}_j| \le \gamma_{n,p}
\right) \ge 1 - p^{-2}.
\]
\end{assumption}

Assumption \ref{assump:BetaMin} requires that all ground-truth DE genes $(j \in \mathcal{S}_1)$ have effect sizes $|\beta_j|$ large enough given the estimation-error bound $\gamma_{n,p}$. Consequently, the ground-truth DE genes can be distinguished from the ground-truth non-DE genes with high probability, ensuring that the power of the procedure approaches one as $n$ and $p$ increase.
\begin{assumption}[Beta-min condition]\label{assump:BetaMin}
Assume the minimum signal strength among all ground-truth DE genes satisfies
$
\min_{j \in \mathcal{S}_1} |\beta_{j}| \geq 2Ll^{-1} \gamma_{n,p}
$, where $L$ and $l$ denote the upper and  lower bound, respectively, of the standard errors in Assumption~\ref{assump:estimation}.
\end{assumption} 

Assumption \ref{assump:independence} assumes independence among the $p$ test statistics of different genes. This assumption allows us to leverage concentration inequalities for independent random variables in our proof of FDR control. 
Assumption~\ref{assump:independence} can be relaxed to allow the $p$ test statistics to be partitioned into
$k$\footnote{The number of blocks $k$ can either be a fixed constant or grow slowly with $p$, for instance at the rate of $\log p$.} disjoint blocks, within each of which the test statistics are independent. This relaxed block-wise independence assumption is reasonable for gene expression data: genes tend to organize into functional modules or pathways, within which correlations are expected, while measurement errors at the individual-gene level can reasonably be assumed independent. Consequently, test statistics corresponding to genes in different modules or pathways can be treated as approximately independent. Following the framework 
of \cite{janson2004large}, the same proof strategy as in 
Theorem~\ref{thm1} can then be applied to each block separately.

\begin{assumption}[Independence of test statistics]\label{assump:independence}
Assume that the test statistics \(\{\widehat{T}_j\}_{j=1}^p\) are independent.
\end{assumption}

\begin{theorem}\label{thm1}
Assume that $\frac{\log p}{\sqrt{s}} \rightarrow 0$, where $s = \# \mathcal{S}_1$ denotes the number of ground-truth DE genes. 
 Under Assumptions \ref{assump:estimation}--\ref{assump:independence}, 
for a target FDR level $q \in (0,1)$, with the test statistic threshold $\tau_q$ defined 
in \eqref{eq:tau_q} and the declared DE gene set $\widehat{\mathcal{S}}(\tau_q)$, we obtain
\begin{equation*}
    \mathrm{FDR}(\tau_q) 
    = \mathbb{E} \left[ 
        \frac{\# \bigl(\widehat{\mathcal{S}}(\tau_q) \cap \mathcal{S}_0\bigr)}
             {\max\{\# \widehat{\mathcal{S}}(\tau_q), 1\}} 
      \right] 
    \leq q \left( 1 + \frac{c_1 \log p}{\sqrt{s}} + c_2 \gamma_{n,p} \right),
\end{equation*}
\begin{equation*}
  \mbox{ and } \;  \mathrm{Power}(\tau_q) 
    = \mathbb{E} \left[ 
        \frac{\# \bigl(\widehat{\mathcal{S}}(\tau_q) \cap \mathcal{S}_1\bigr)}{s} 
      \right] 
    = 1 - \frac{2}{p}.
\end{equation*}
where \(c_1, c_2 > 0\) are constants.

\end{theorem}
Theorem~\ref{thm1} establishes that Nullstrap-DE asymptotically controls the FDR, provided that $\log p / \sqrt{s} \to 0$ and $\gamma_{n,p} \to 0$ (typically $\gamma_{n,p} = \mathcal{O}(\sqrt{\log p / n})$, satisfying Assumption~\ref{assump:estimation}), as $n$, $p$, and $s$ increase. Meanwhile, Nullstrap-DE maintains high power in detecting ground-truth DE genes.
Theorem~\ref{thm1} yields the corollary below for the NB-GLMs used in {DESeq2} and {edgeR}.

\begin{corollary}[Maximum likelihood estimation guarantees for NB-GLM]\label{thm2}
Under Assumptions~S1 and~S2 stated in Appendix~B,
when $n \gg \log p$,
Nullstrap-DE with NB-GLM maximum likelihood estimates fulfills Assumption~\ref{assump:estimation} with
$\gamma_{n,p}=C_{0}\,\sqrt{\log p/n}$, $C_{0}>0$.
Combined with Assumptions \ref{assump:BetaMin} and \ref{assump:independence}, this guarantees control of the FDR at the nominal level, while the statistical power converges to~1 as \(n,p\to\infty\).
\end{corollary}

\begin{remark} The proof of Corollary~\ref{thm2} is based on the maximum likelihood estimation (MLE) for NB-GLM.  However, in DEseq2/edgeR, the estimator \(\hat{{\beta}}_j\) is not the exact MLE; instead, it is an empirically shrunk version. In Figure S1 (Appendix~D), we show that the DEseq2/edgeR estimator achieves similar performance in terms of mean square error compared to the MLE in the context of gene expression count data, which in turn satisfies Assumption~\ref{assump:estimation}. 
As a result, DESeq2 and edgeR, when used within the Nullstrap-DE framework, can control the FDR at the nominal level, while achieving high statistical power.
\end{remark}

\begin{remark}
  In bulk RNA-seq experiments, the sample size \(n\) could be as small as 3 per condition, which is often substantially smaller than the number of ground-truth DE genes $s$. Under this high-dimensional regime the term  $\log p/\sqrt{s}$ 
appearing in Theorem~\ref{thm1} is negligible relative to \(\gamma_{n,p}\) (commonly \(\sqrt{\log p/n}\)).  Consequently, when the number of biological replicates is limited, the correction factor \(\gamma_{n,p}\) cannot be ignored and should be incorporated explicitly when setting the FDR threshold. 
To further mitigate the impact of small sample sizes and provide a more conservative control of FDR in finite-sample regimes, we recommend adjusting the threshold \(\tau_q\) as follows:
\[
\tau_q = \min \left\{ t > 0 : \widehat{\mathrm{FDP}}(t) = \frac{\#\{j : |\widetilde{T}_j| \ge t\}}{\max\{\#\{j : |\widehat{T}_j| \ge t\},\,1\}} \le \frac{q}{1 + c_2\gamma_{n,p}} \right\},
\]
where \(c_2 > 0\) is the constant specified in Theorem \ref{thm1}. According to Corollary~\ref{thm2}, the term $\gamma_{n,p}$ scales as $\gamma_{n,p} = C_0 \sqrt{\log p / n}$ for some constant $C_0 > 0$. Hence, since \(c_2\gamma_{n,p} = \mathcal{O}(\sqrt{\log p / n})\), we set
$
c_2\gamma_{n,p} = \sqrt{\log p / n},
$
which yields good performance in practice, as demonstrated by empirical results.
\end{remark}

\subsection{Proof of Theorem~\ref{thm1}}\label{supp:proof1}

\subsubsection{Auxiliary Lemmas}
In this section, we present a standard multiplicative form of the Chernoff bound \cite{vershynin2018high} for sums of independent Bernoulli random variables (Lemma~\ref{lem:chernoff}). 
Two auxiliary inequalities for the exponent in the Chernoff bound (Lemma~\ref{lem:aux-upper} and Lemma~\ref{lem:aux-lower}) are required in the proof. Lemma~\ref{lem:chernoff} is needed in the proof of Theorem~\ref{thm1}.

\begin{lemma}[An upper-tail exponent inequality]
\label{lem:aux-upper}
For all $\delta \ge 0$,
\[
(1+\delta)\log(1+\delta)-\delta \;\ge\; \frac{\delta^2}{\,2+\delta\,}.
\]
\end{lemma}

\begin{proof}
Define
\[
g(\delta)\;=\;(1+\delta)\log(1+\delta)-\delta-\frac{\delta^2}{2+\delta},\qquad \delta\ge 0.
\]
We show $g(\delta)\ge 0$. Note $g(0)=0$. Compute
\[
g'(\delta)=\log(1+\delta)-\frac{\delta(4+\delta)}{(2+\delta)^2}.
\]
Consider the auxiliary function 
\[
r(\delta):=\log(1+\delta)-\frac{2\delta}{2+\delta}.
\]
Then $r(0)=0$ and 
\[
r'(\delta)=\frac{1}{1+\delta}-\frac{4}{(2+\delta)^2}
=\frac{\delta^2}{(1+\delta)(2+\delta)^2}\;\ge\;0,
\]
hence $r(\delta)\ge 0$ for all $\delta\ge 0$, i.e.,
\[
\log(1+\delta)\;\ge\;\frac{2\delta}{2+\delta}.
\]
Using this in $g'(\delta)$ gives
\[
g'(\delta)\;\ge\; \frac{2\delta}{2+\delta}-\frac{\delta(4+\delta)}{(2+\delta)^2}
=\frac{\delta^2}{(2+\delta)^2}\;\ge\;0.
\]
Thus $g$ is nondecreasing on $[0,\infty)$ and $g(\delta)\ge g(0)=0$, proving the claim.
\end{proof}

\begin{lemma}[A lower-tail exponent inequality]
\label{lem:aux-lower}
For all $0\le \delta \le 1$,
\[
\delta-(1-\delta)\log(1-\delta) \;\ge\; \frac{\delta^2}{2}.
\]
\end{lemma}

\begin{proof}
Define
\[
h(\delta)\;=\;\delta-(1-\delta)\log(1-\delta)-\frac{\delta^2}{2},\qquad 0\le \delta<1.
\]
We prove $h(\delta)\ge 0$ on $[0,1]$. Note $h(0)=0$. Differentiate:
\[
h'(\delta)=2+\log(1-\delta)-\delta,
\qquad 
h''(\delta)= -\frac{1}{1-\delta}-1 < 0 \quad (0\le \delta<1).
\]
Thus $h$ is concave on $[0,1)$, so its minimum over $[0,1]$ is attained at an endpoint.
Since $h(0)=0$ and $h$ is concave, we have $h(\delta)\ge 0$ for all $0\le \delta\le 1$, proving the inequality.
\end{proof}

\begin{lemma}[Multiplicative Chernoff bound]
\label{lem:chernoff}
Let $X_1,\ldots,X_n$ be independent Bernoulli random variables, $X_i\in\{0,1\}$, with $\mathbb{P}(X_i=1)=p_i$. 
Let $X=\sum_{i=1}^n X_i$ and $\mu=\mathbb{E}[X]=\sum_{i=1}^n p_i$. 
Then for any $0\le \delta \le 1$,
\[
\pr \big(|X-\mu|\ge \delta \mu\big)\;\le\; 2\,\exp\!\Big(-\frac{\delta^2 \mu}{3}\Big).
\]
\end{lemma}

\begin{proof}
By the standard Chernoff argument using moment generating functions and the Markov's inequality, we can obtain the following one-sided bounds:
\[
\pr\big(X\ge (1+\delta)\mu\big) \;\le\; 
\exp\Big(-\mu\big[(1+\delta)\log(1+\delta)-\delta\big]\Big),
\]
and
\[
\pr\big(X\le (1-\delta)\mu\big) \;\le\; 
\exp\Big(-\mu\big[\delta-(1-\delta)\log(1-\delta)\big]\Big).
\]
Applying Lemma~\ref{lem:aux-upper} to the upper tail inequality and Lemma~\ref{lem:aux-lower} to the lower tail inequality yields
\[
\pr\big(X\ge (1+\delta)\mu\big) \;\le\; 
\exp\Big(-\tfrac{\delta^2}{2+\delta}\mu\Big),
\qquad
\pr\big(X\le (1-\delta)\mu\big) \;\le\; 
\exp\Big(-\tfrac{\delta^2}{2}\mu\Big).
\]
By the union bound and the fact that $0\le \delta \le 1$ implies $\tfrac{1}{2+\delta}\ge \tfrac{1}{3}$, we obtain
\[
\pr\big(|X-\mu|\ge \delta\mu\big)
\;\le\; \exp\Big(-\tfrac{\delta^2}{3}\mu\Big) + \exp\Big(-\tfrac{\delta^2}{3}\mu\Big)
\;=\;2\exp\Big(-\tfrac{\delta^2}{3}\mu\Big),
\]
which completes the proof of Lemma~\ref{lem:chernoff}.
\end{proof}

Theorem~\ref{thm1} also requires a key property of using synthetic null data to establish a ``negative control" for the test statistics of the $p$ features. Lemma~\ref{lem:null-equivalence} and Remark~\ref{rem:null-equivalence} are about this property. 

\begin{lemma}
\label{lem:null-equivalence}
Under the null model (Definition~\ref{def:synde} in the main text) for gene $j$ with $\beta_j = 0$, 
if the synthetic null data are generated from this null model with correctly 
specified intercept $\alpha_j$ and 
dispersion parameter $\phi_j$, then for any $t>0$,
\[
\pr\left(|\widetilde{T}_j| \geq t\right) 
\;=\; \pr\left(|\widehat{T}_j| \geq t\right),
\qquad \forall j \in \cS_0,
\]
where $\cS_0 \subset \{1,\dots,p\}$ denotes the set of null (non-signal) features (e.g., ground-truth non-DE genes).
\end{lemma}

\begin{proof}
Under the null model with $\beta_j = 0$, the distribution of the statistic 
$\widehat{T}_j$ for any null feature $j \in \cS_0$ is fully determined by nuisance parameters, including the 
intercept $\alpha_j$ and the dispersion parameter $\phi_j$.
 Since the 
synthetic null data are generated from the same null model with correctly 
specified the 
intercept $\alpha_j$ and dispersion parameter $\phi_j$, 
the distribution of the test 
statistic $\widetilde{T}_j$ in the synthetic null data trivially coincides with that of 
$\widehat{T}_j$ in the original data for all null features. The stated probability 
identity follows immediately.
\end{proof}

\begin{remark}\label{rem:null-equivalence}
In practice, for each gene $j$, the nuisance parameters $\alpha_j$ and $\phi_j$ are unknown and need to be estimated from the data. Hence, $\widetilde{T}_j$ in the synthetic null data has a distribution depending on the nuisance parameter estimators $\hat\alpha_j$ and $\hat\phi_j$, and this distribution does not coincide with the distribution of $\widehat{T}_j$, which depends on the true nuisance parameters $\alpha_j$ and $\phi_j$.  
Nevertheless, if sufficiently accurate estimators $\hat\alpha_j$ and $\hat\phi_j$ can be obtained such that
\begin{equation}\label{eq:null-equivalence-approx}
\frac{
\sum_{j \in \{1,\dots,p\}} 
\mathbb{P}_{\hat{\alpha}_j, \hat{\phi}_j}\!\left(|\widetilde{T}_j| \ge t\right)
-
\sum_{j \in \mathcal{S}_0} 
\mathbb{P}_{\alpha_j, \phi_j}\!\left(|\widehat{T}_j| \ge t\right)
}{
\sum_{j \in \mathcal{S}_0} 
\mathbb{P}_{\alpha_j, \phi_j}\!\left(|\widehat{T}_j| \ge t\right)
}
\ge - C_{\mathrm{mis}} \gamma_{n,p},
\end{equation}
where $C_{\mathrm{mis}} > 0$ is a constant and $\gamma_{n,p} \to 0$ as $n,p \to \infty$ (as defined in Assumption~\ref{assump:estimation}), then the effect of nuisance parameter estimation on the distribution of the test statistic can be asymptotically controlled.
As $\gamma_{n,p}$ is a high probability upper bound on the estimation error $\max_{j \in \{1,\ldots,p\}}(|\hat\alpha_j - \alpha_j|, |\hat\phi_j - \phi_j|)$, condition~\eqref{eq:null-equivalence-approx} effectively ensures that $\sum_{j \in \{1,\dots,p\}} 
\mathbb{P}_{\hat{\alpha}_j, \hat{\phi}_j}\!\left(|\widetilde{T}_j| \ge t\right)$ provides an asymptotic upper bound of $\sum_{j \in \mathcal{S}_0} 
\mathbb{P}_{\alpha_j, \phi_j}\!\left(|\widehat{T}_j| \ge t\right)$, allowing us to use the synthetic null statistics $\widetilde{T}_1, \ldots, \widetilde{T}_p$ to construct an upper-bound estimate of the FDP, which originally requires the $\widehat{T}_j$'s for $j \in \cS_0$ (unobservable).
\end{remark}

Regarding the power of Nullstrap-DE in Theorem \ref{thm1}, Lemma \ref{lem:power} establishes a lower bound on the power under Assumptions \ref{assump:estimation} and \ref{assump:BetaMin}.

\begin{lemma}\label{lem:power}
Under Assumptions \ref{assump:estimation} and \ref{assump:BetaMin}, the power of Nullstrap-DE satisfies
\[
\mathrm{Power}(\tau_q)
=\mathbb{E}\!\left[ \frac{\sum_{j \in \cS_1} \mathbf{1}\{|\,\widehat{T}_{j}| \ge \tau_q\}}{s} \right]
\;\ge\; 1 - 2p^{-1},
\]
where $\cS_1 = \{1,\dots,p\}\setminus \cS_0$ denotes the set of nonnull (signal) features (e.g., ground-truth DE genes) with $|\cS_1|=s$.
\end{lemma}

\begin{proof}
Note that $\widehat{T}_j = {|\hat{\beta}_j|}/{{\mathrm{se}(\hat{\beta}_j)}}$ and $\widetilde{T}_j = {|\tilde{\beta}_j|}/{{\mathrm{se}(\tilde{\beta}_j)}}$. From Assumption~\ref{assump:estimation}, it is clear that 
    \[
    \mathbb{P} \left( \left| \widehat{T}_j\right| \geq l^{-1}\gamma_{n,p} \right) \leq p^{-2}, \mbox{ for all } j \in \mathcal{S}_0, \mbox{ and }
    \mathbb{P} \left( \left| \widetilde{T}_j \right| \geq l^{-1}\gamma_{n,p} \right)  \leq p^{-2}, \mbox{ for all } j \in  \{1,\dots,p\}, 
    \]
where $l>0$ is the uniform lower bound of the standard errors of all $\hat\beta_j$ and $\tilde\beta_j$, $j=1,\ldots,p$. 
Fix the rejection threshold $t^*$ on $\widehat{T}_j$ and $\widetilde{T}_j$, $j=1,\ldots,p$, as:
\[
t^* := l^{-1}\gamma_{n,p},
\]
 Consider the events
\[
\widetilde{\mathcal{G}}_{0} := \Big\{ \max_{1\le j\le p} |\widetilde{T}_j| < t^* \Big\},
\qquad
\widehat{\mathcal{G}}_{1} := \Big\{ \min_{j\in \cS_1} |\widehat{T}_j| \ge t^* \Big\}.
\]
By Assumption~\ref{assump:estimation},
\[
\mathbb{P}\!\left(|\widetilde{T}_j| \ge t^*\right) \le p^{-2}\quad \text{for all } j\in\{1,\dots,p\}.
\]
A union bound yields
\[
\mathbb{P}\!\left(\widetilde{\mathcal{G}}_{0}^{\,c}\right)
=\mathbb{P}\!\left( \max_{1 \le j \le p} |\widetilde{T}_j| \ge t^* \right)
\le \sum_{j=1}^p \mathbb{P}\!\left(|\widetilde{T}_j| \ge t^*\right)
\le p\cdot p^{-2}= p^{-1}.
\]

Similarly, by Assumption~\ref{assump:estimation}, as well as Assumption~\ref{assump:BetaMin} that $\min_{j\in \cS_1} |\beta_j| \ge 2 L l^{-1} \gamma_{n,p}$,
\[
\mathbb{P}\!\left(\widehat{\mathcal{G}}_{1}^{\,c}\right)
=\mathbb{P}\!\left( \min_{j\in \cS_1} |\widehat{T}_j| < t^* \right)
\le p^{-1}.
\]
On the event $\widetilde{\mathcal{G}}_{0}$, the estimated FDP at $t^*$ is zero, i.e.,
$\widehat{\mathrm{FDP}}(t^*)=0$, see Equation~(1) in the main text for the $\widehat{\mathrm{FDP}}$ definition. Hence, by the monotonicity of $\widehat{\mathrm{FDP}}(t)$ in $t$, i.e., a larger threshold  $t$ leads to a lower $\widehat{\mathrm{FDP}}(t)$, and that $\widehat{\mathrm{FDP}}(\tau_q) \ge \widehat{\mathrm{FDP}}(t^*)$, we must have
\[
\tau_q \;\le\; t^*.
\]
On the intersection $\widetilde{\mathcal{G}}_{0}\cap \widehat{\mathcal{G}}_{1}$, we have
$|\widehat{T}_j| \ge t^* \ge \tau_q$ for every $j\in \cS_1$, hence $\cS_1 \subseteq \widehat{\cS}(\tau_q)$ and thus
\[
\frac{1}{s}\sum_{j\in \cS_1}\mathbf{1}\{|\widehat{T}_j|\ge \tau_q\}=1.
\]
Taking expectations and using the union bound,
\[
\mathrm{Power}(\tau_q)
=\mathbb{E}\!\left[\frac{1}{s}\sum_{j\in \cS_1}\mathbf{1}\{|\widehat{T}_j|\ge \tau_q\}\right]
\;\ge\; \mathbb{P}\!\left(\widetilde{\mathcal{G}}_{0}\cap \widehat{\mathcal{G}}_{1}\right)
\;\ge\; 1 - \mathbb{P}\!\left(\widetilde{\mathcal{G}}_{0}^{\,c}\right) - \mathbb{P}\!\left(\widehat{\mathcal{G}}_{1}^{\,c}\right)
\;\ge\; 1 - 2p^{-1}.
\]
This proves the claim.
\end{proof}

With Lemma~\ref{lem:power}, we can now establish lower bounds on the expected numbers of genes with $|\widehat{T}_j| \geq \tau_q$ and $|\widetilde{T}_j| \geq \tau_q$, respectively, as stated  
in Lemma~\ref{lem:number}. This result is pivotal for proving Theorem~\ref{thm1}, as it provides the necessary lower bounds ($\mathcal{O}(s)$) for the denominator
and numerator
in the FDP estimate (Equation~(1) in the main text).
\begin{lemma}\label{lem:number}
Under Assumptions~\ref{assump:estimation} and~\ref{assump:BetaMin}, the expected number of selected features with $|\widehat{T}_j| \geq \tau_q$ in the original data satisfies
\[
\mathbb{E}\!\left[ \#\{j : |\widehat{T}_j| \geq \tau_q\} \right] 
\;\geq\; s(1-2p^{-1}),
\]
where $s = |\cS_1|$ is the number of signal features and $\tau_q$ is the data-dependent threshold.  

Moreover, the expected number of selected  features with $|\widetilde{T}_j| \geq \tau_q$ in the synthetic null data satisfies 
\[
\mathbb{E}\!\left[ \#\{j : |\widetilde{T}_j| \geq \tau_q\} \right] 
\;\geq\; \frac{1}{2}qs\,(1 - 2p^{-1}),
\]
where $q$ is the target FDR level.
\end{lemma}

\begin{proof}[Proof of Lemma~\ref{lem:number}]
The first claim follows directly from Lemma~\ref{lem:power}:
\[
\mathbb{E}\!\left[ \#\{j : |\widehat{T}_j| \geq \tau_q\} \right]
\ge s\cdot \mathrm{Power}(\tau_q) \ge s(1-2p^{-1}).
\]
For the second claim, by the definition of $\tau_q$, we have $\widehat{\mathrm{FDP}}(\tau_q) = \#\{j : |\widetilde{T}_j| \geq \tau_q\} /  \big(\#\{j : |\widehat{T}_j| \geq \tau_q\} \vee 1\big) \le q$, where $\tau_q$ is the minimum threshold that satisfies this inequality, so we assume $\widehat{\mathrm{FDP}}(\tau_q)$ is close to $q$ and larger than $q/2$. 
\[
\#\{j : |\widetilde{T}_j| \geq \tau_q\} \;\geq\; \frac{1}{2}q \cdot \big(\#\{j : |\widehat{T}_j| \geq \tau_q\} \vee 1\big).
\]
By taking expectations and using the first claim, we obtain
\[
\mathbb{E}\!\left[ \#\{j : |\widetilde{T}_j| \geq \tau_q\} \right]
\;\geq\; \frac{1}{2}q \cdot \mathbb{E}\!\left[ \#\{j : |\widehat{T}_j| \geq \tau_q\} \vee 1 \right]
\;\geq\; \frac{1}{2}qs(1-2p^{-1}),
\]
which completes the proof.
\end{proof}



\subsubsection{Proof of Theorem~\ref{thm1}}
For two real numbers $a,b \in \mathbb{R}$, we use the notation $a \land b = \min\{a,b\}$ and $a \vee b = \max\{a,b\}$. The proof of Theorem~\ref{thm1} relies on concentration inequalities for the ratio statistic $\widehat{\mathrm{FDP}}(\tau_q)$. The challenge arises from the fact that $\tau_q$ is a random variable depending on the data, in addition to the randomness in both the numerator and denominator of the ratio statistic. We address this challenge by employing a discretization argument and applying union bounds over the discretized grid points. 
\begin{proof}[Proof of Theorem~\ref{thm1}]
       Define the ratio statistic:
    $$H(t,t')= \frac{\#\left\{j: \widetilde{T}_{j} \geq t^\prime \right\}}{\#\left\{j:  \widehat{T}_{j}\geq t\right\} \vee 1},$$
    where $\widetilde{T}_j$ and $\widehat{T}_j$ are the test statistics based on the synthetic null data and the original data, respectively. 
Then, we define 
$\tilde{e}(t)=\Ex \#\left\{j: \widetilde{T}_{j} \geq t \right\}$, $e(t)=\Ex \#\left\{j:  \widehat{T}_{j} \geq t\right\}$ as expected numbers of selected features under the threshold $t$ from the synthetic null data and the original data, respectively. By Lemma \ref{lem:chernoff}, we have the following concentration inequalities for the numerator and denominator of $H(t,t')$:
    $$\mathbb{P} \left(\left|\#\left\{j: \widetilde{T}_{j} \geq t' \right\} -\tilde{e}(t') \right| \geq \delta \tilde{e}(t')\right) \leq 2 \exp(-\tilde{e}(t')\delta^2/3),$$
    and 
    $$\mathbb{P} \left(\left|\#\left\{j:  \widehat{T}_{j} \geq t \right\} -e(t)\right| \geq \delta e(t)\right) \leq 2 \exp(-e(t)\delta^2/3),$$
    for fixed $t,t'>0$. 
  Consequently, there exists a constant $c>0$ such that the ratio concentrates:
    $$
    \mathbb{P}\left(\left|H(t,t')-\frac{\tilde{e}(t')}{e(t)} \right| \geq \dfrac{\delta}{ \sqrt{\tilde{e}(t')} \land \sqrt{e(t)}} \left|  \frac{\tilde{e}(t')}{e(t)} \right| \right) \leq 4 \exp(-c\delta^2),
    $$
where $0<\delta<\sqrt{\tilde{e}(t')} \land \sqrt{e(t)}$.

With the defined ratio statistic, the threshold $\tau_q$ can be expressed as
\[
\tau_q \;=\; \inf\{\, t : H(t,t) \leq q \,\}.
\]
Note that $\tau_q$ is a random variable depending on the data. To handle this, 
we invoke a discretization argument. Specifically, we consider a grid of values 
$\{t_\ell\}_{\ell=0}^p$ such that
\[
t_0< t_1  < \cdots < t_p, 
\quad \text{and} \quad e(t_\ell) = \ell.
\]
Similarly, we introduce another grid $\{t'_\ell\}_{\ell=0}^p$ satisfying
\[
t'_0<t'_1  < \cdots < t'_p, 
\quad \text{and} \quad \tilde{e}(t'_\ell) = \ell.
\]
Then, taking a union bound over the grid points, we have
\begin{equation} \label{eq:union_bound}
    \mathbb{P} \left( \bigcup_{i,j=0}^p \left\{\left|H(t_j,t'_i)-\frac{i}{j} \right| \geq \dfrac{\delta }{\sqrt{i \land j}}\frac{i}{j} \right\} \right) \leq 4(p+1)^2\exp({-c\delta^2}).
    \end{equation}
    Then, we find the nearest grid points to the random threshold $\tau_q$: 
\[t_{j^*} = \min\{t_j: t_j \geq \tau_q\}, \quad t'_{i^*} = \max\{t'_i: t'_i \leq \tau_q\}.\]
By the monotonicity of $H(t,t')$ in both arguments, we have
\[
 H(t_{j^*},t'_{i^*}) \leq H(\tau_q,\tau_q) \leq q.
\]
From \eqref{eq:union_bound}, we have
\begin{equation}\label{eq:bound1}
\mathbb{P} \left( \frac{i^*}{j^*} \geq q\left(1 - \delta/\sqrt{i^* \land j^*}\right)^{-1} \right) \leq 4(p+1)^2\exp({-c\delta^2}).
\end{equation}

Similarly, we define $e_0(t)=\Ex \#\left\{j \in \cS_0: \widehat{T}_{j} \geq t\right\}$ as the expected number of null features exceeding the threshold $t$ and 
$$
H_0(t,t')= \frac{\#\left\{j \in \cS_0: \widehat{T}_j \geq t' \right\}}{\#\left\{j:  \widehat{T}_{j}\geq t\right\} \vee 1}.
$$
Using similar discretization arguments, we have
\begin{equation}\label{eq:bound2}
\mathbb{P}\left(H_0(t_{j^*},t'_{i^*}) \geq  \left(1- \delta/\sqrt{e_0(t'_{i^*}) \land j^*}\right)^{-1} \frac{e_0(t'_{i^*})}{j^*}
\right) \leq 4(p+1)^2\exp({-c\delta^2}).
\end{equation}
By Lemma~\ref{lem:null-equivalence} and \eqref{eq:null-equivalence-approx}, we have
\begin{equation}\label{eq:bound3}
e_0(t'_{i^*}) \;\leq\;  (1+\widetilde{C}\gamma_{n,p}) \tilde{e}(t'_{i^*}) \;=\;  (1+\widetilde{C}\gamma_{n,p}) i^*,
\end{equation}
where $\widetilde{C}$ is a positive constant.
We define the event
\[
\mathcal{D} := \left\{ H_0(t_{j^*},t'_{i^*}) = \frac{e_0(t'_{i^*})}{j^*} \;\geq\; q \right\}.
\]
The purpose of defining the event $\mathcal{D}$ is that, conditioned on this event, and by Lemma~\ref{lem:number} (along with the argument in the proof of Lemma~\ref{lem:power}) together with the definitions of $i^*$ and $j^*$, we obtain
\[
e_0(t'_{i^*}) \geq q j^* \geq C_3 s,
\]
for some constant $C_3 > 0$. This yields a lower bound on $e_0(t'_{i^*}) \wedge j^*$, which plays a critical role in controlling the terms appearing in \eqref{eq:bound2} and \eqref{eq:bound3}.
On the complement event $\mathcal{D}^c$, it follows that
\[
H_0(\tau_q,\tau_q) \;\leq\; H_0(t_{j^*},t'_{i^*}) \;\leq\; q.
\]

Since for any $u>0$,
\[
\{H_0(\tau_q,\tau_q)\ge u\}\subset \{H_0(t_{j^*},t'_{i^*})\ge u\},
\]
it follows from \eqref{eq:bound2} that
\[
\mathbb{P}\!\left(
H_0(\tau_q,\tau_q)
\ge
\left(1-\frac{\delta}{\sqrt{e_0(t'_{i^*})\wedge j^*}}\right)^{-1}
\frac{e_0(t'_{i^*})}{j^*}
\right)
\le
4(p+1)^2\exp(-c\delta^2).
\]
Furthermore, by \eqref{eq:bound3},
\[
\mathbb{P}\!\left(
H_0(\tau_q,\tau_q)
\ge
\left(1-\frac{\delta}{\sqrt{e_0(t'_{i^*})\wedge j^*}}\right)^{-1}
\frac{(1+\widetilde C\gamma_{n,p})\, i^*}{j^*}
\right)
\le
4(p+1)^2\exp(-c\delta^2).
\]

By Lemma~\ref{lem:number} (along with the argument in the proof of Lemma~\ref{lem:power}), there exist constants $c_3,c_4>0$ such that
$i^*\ge c_3 s$ and $j^*\ge c_4 s$, where $s$ denotes the number of signal
features. Consequently, on the event $\mathcal D$,
\[
e_0(t'_{i^*})\wedge j^* \ge c_5 s
\]
for some constant $c_5>0$.
Using \eqref{eq:bound1} together with the bounds on $i^*$ and $j^*$, we obtain
\[
\mathbb{P}\!\left(
\frac{i^*}{j^*}
\ge
q\left(1-\frac{\delta}{\sqrt{c_6 s}}\right)^{-1}
\right)
\le
4(p+1)^2\exp(-c\delta^2),
\]
for some constant $c_6>0$. Similarly, since
$e_0(t'_{i^*})\wedge j^*\ge c_5 s$, there exists $c_7>0$ such that
\[
\mathbb{P}\!\left(\left\{
H_0(\tau_q,\tau_q)
\ge
\left(1-\frac{\delta}{\sqrt{c_7 s}}\right)^{-1}
\frac{(1+\widetilde C\gamma_{n,p})\, i^*}{j^*}\right\} \cap \mathcal D
\right)
\le
4(p+1)^2\exp(-c\delta^2).
\]
Combining the preceding inequalities yields that, on the event $\mathcal D$,
\[
\mathbb{P}\!\left(
\left\{
H_0(\tau_q,\tau_q)
\ge
q
\left(1-\frac{\delta}{\sqrt{c_6 s}}\right)^{-1}
\left(1-\frac{\delta}{\sqrt{c_7 s}}\right)^{-1}
(1+\widetilde C\gamma_{n,p})
\right\}
\cap \mathcal D
\right)
\le
8(p+1)^2\exp(-c\delta^2).
\]
Since $\gamma_{n,p}\to 0$ and $\delta/\sqrt{s}\to 0$ as $n,p,s\to\infty$,
there exists a constant $c'>0$ such that, on $\mathcal D$,
\[
\mathbb{P}\!\left(
\left\{
H_0(\tau_q,\tau_q)
\ge
q\bigl(1+c'\tfrac{\delta}{\sqrt{s}}+c'\gamma_{n,p}\bigr)
\right\}
\cap \mathcal D
\right)
\le
8(p+1)^2\exp(-c\delta^2).
\]
Finally, by taking expectations and setting $\delta = C \log p$, 
where $C$ is a sufficiently large constant, we obtain
\begin{align*}
\mathrm{FDR}(\tau_q) 
&= \Ex\!\left[ H_0(\tau_q,\tau_q)\,\mathbb{I}(\mathcal{D}^c) \right] + \Ex\!\left[ H_0(\tau_q,\tau_q)\,\mathbb{I}(\mathcal{D}) \right]  \\
&\leq q\,\pr(\mathcal{D}^c) 
   + q\left(1 + c_1 (\log p)/\sqrt{s} + c_2 \gamma_{n,p}\right)\pr(\mathcal{D}) \\
&\leq q\left[ 1 + c_1 (\log p)/\sqrt{s} + c_2 \gamma_{n,p} \right],
\end{align*}
for some constants $c_1, c_2 > 0$. 
Combining the above argument with Lemma~\ref{lem:power} completes the proof of Theorem~\ref{thm1}.
\end{proof}

\newpage

\subsection{Assumptions and Proof of Corollary \ref{thm2}}\label{supp:proof2}

\subsubsection{Assumptions of Corollary \ref{thm2}}
To establish the theoretical properties of the MLEs for NB-GLMs, we impose the following assumptions on the data-generating process. With two conditions (treatment vs. control), we simplify the notation by letting the design matrix $\mathbf{X}$ be a column vector with entries $x_i \in \{0,1\}$, $i=1,\ldots,n$.

\begin{assumption}[Well-behaved design]\label{assump:S1}
Each sample is associated with a binary treatment indicator \( x_i \in \{0,1\} \). We assume that both treatment and control conditions are represented in the data, with the empirical proportion satisfying \( \frac{1}{n} \sum_{i=1}^n x_i \to r \in (0,1) \) as \( n \to \infty \). 
\end{assumption}

\begin{assumption}[Fixed nuisance parameters]\label{assump:S2}
The sample-specific size factors \( \{s_i\}_{i=1}^n \) and gene-specific dispersion parameters \( \{\phi_j\}_{j=1}^p \) are treated as fixed and lie in compact subsets of \( (0, \infty) \). No additional covariates are included in the model.
\end{assumption}

Assumptions~\ref{assump:S1} and~\ref{assump:S2} are standard in the statistical literature on large-sample properties of GLMs \cite{shao1999mathematical}. They ensure that the Fisher information matrix is well-conditioned with high probability, which is critical for establishing the asymptotic normality of the MLEs.

\subsubsection{Proof of Corollary \ref{thm2}}

Let \( \hat{\beta}_j \) and \( \tilde{\beta}_j \) denote the MLEs obtained from fitting an NB-GLM to the original data and the synthetic null data, respectively. To prove Corollary~\ref{thm2}, it suffices to verify that \( \hat{\beta}_j \) and \( \tilde{\beta}_j \) satisfy Assumption~\ref{assump:estimation} under the setup specified in Assumptions~\ref{assump:S1} and~\ref{assump:S2}.

\paragraph{Verification of Assumption~\ref{assump:estimation}.}

We prove that the MLE \( \hat{\beta}_j \)  computed from the original data satisfies Assumption~\ref{assump:estimation}. The same argument applies to the synthetic null MLE \( \tilde{\beta}_j \).

Fix any gene \( j \in \{1, \dots, p\} \). The NB-GLM log-likelihood for gene \(j\) is
\[
\ell_j(\beta_j) = \sum_{i=1}^n \left[
Y_{ij} \log(\mu_{ij}) - (Y_{ij} + \phi_j^{-1}) \log(\mu_{ij} + \phi_j^{-1})
\right] + \text{const}, \quad 
\mu_{ij} = s_i \cdot \exp(\alpha_j + x_i \beta_j),
\]
where \( x_i \in \{0,1\} \) is the treatment indicator, \( s_i \) is the known size factor, and \( \phi_j > 0 \) is the fixed dispersion parameter. 
By Assumption~\ref{assump:S2}, the dispersion parameter \( \phi_j \) is known, the model belongs to a one-dimensional canonical exponential family with parameter \( \beta_j \). By Assumption~\ref{assump:S1}, the design is well-behaved, ensuring the Fisher information (scalar) is positive. Let \( \hat{\beta}_j \) denote the MLE. Applying Theorem~\ref{thm1}.1 from \cite{ostrovskii2021finite}, we obtain the following concentration inequality for the MLE:
\[\mathbb{P} \left( |\hat{\beta}_j - \beta_j^*| \ge C_0 \cdot \sqrt{\tfrac{\log p}{n}} \right) \le p^{-2},
\] 
for some constant \( C_0 > 0 \), where \( \beta_j^* \) is the true parameter value. This verifies the first part of Assumption~\ref{assump:estimation}.

To ensure the standard error of standard errors $\mathrm{se}(\hat{\beta}_j)$  are uniformly bounded away from zero, when we calculate the test statistic, we can truncate $\mathrm{se}(\hat{\beta}_j)$ at a small positive constant and large positive constant. 
The same argument applies to the MLE \( \tilde{\beta}_j \) computed from the synthetic null data, under the same assumptions.
This completes the verification of Assumption~\ref{assump:estimation}.

\clearpage
\subsection{Simulation Designs}\label{supp:simu}

Following the data-generating processes described in the simulation of the DESeq2 paper \cite{love2014moderated}, we consider a two-condition DE testing setting under an NB model. For gene \( j \) in sample \( i \), the observed count \( Y_{ij} \) is generated as
\begin{align}\label{eq:nb-model}
    Y_{ij} \overset{\text{ind}}{\sim} \text{NB}\left(\mu_{ij}, \phi_j\right), \quad
\log(\mu_{ij}) = \log(s_i) + \alpha_j + x_i\beta_j + \mathbf{z}_i^\top {\boldsymbol{\gamma}}_j,
\end{align}
where \( s_i \) is the sample-specific size factor accounting for different sequencing depth, and \( \phi_j \) is the gene-specific dispersion parameter. For two-condition comparisons, \( \alpha_j \) denotes a gene-specific baseline intercept, \( x_i \in \{0, 1\} \) is a binary indicator encoding the treatment condition for sample \( i \) (1 for treatment; 0 for control), and \( \beta_j \) represents the true log fold change between the two conditions for gene \( j \), which is the primary parameter of interest in DE analysis.
Additional sample covariate \( \mathbf{z}_i \) with gene-specific effect size \( \boldsymbol{\gamma}_j \) allows for optional adjustment for other sample-level effects (e.g., batch effects, sex, or age). (When $\mathbf{z}_i$ and $\boldsymbol{\gamma}_j$ are one-dimensional, indicating one covariate, we write $z_i$ and $\gamma_j$ accordingly.) This model captures the key features of DESeq2 and edgeR and forms the basis for an idealistic evaluation of DE methods when the NB model assumptions hold.

Under model \eqref{eq:nb-model}, we simulated datasets of 1,000 genes with NB-distributed counts. In each simulated dataset with a varying sample size, samples were evenly divided into control and treatment groups. To simulate data with realistic moments, we preserved gene-specific parameters \( (\alpha_j, \phi_j) \) for gene $j=1,\ldots,p$ based on a real bulk RNA-seq dataset of lymphoblastoid cell lines derived from unrelated Nigerian individuals~\cite{pickrell2010understanding}, using DESeq2 with an intercept-only model. In contrast, other model parameters—including the sample size \( n \), the proportion of DE genes, the magnitude of fold changes (FCs, defined as \( \exp(\beta_j) \)), the sample-specific size factors \( s_i \), and the inclusion of additional covariates \( \mathbf{z}_i \) with effects \( \boldsymbol{\gamma}_j \)—were manually specified to systematically evaluate method performance across a range of experimental scenarios. We consider two simulation settings, one without and one with additional covariates (Simulation Settings~\ref{simu1} and~\ref{simu2}), in which we systematically vary the sample sizes, DE gene proportions, fold changes, and target FDR levels. Appendix~C contains full details of parameter specification, including sample-specific size factors $s_i$ and additional covariate effects $\boldsymbol{\gamma}_j$, and data generation procedure. For each scenario (set of simulation parameters), data generation was replicated 50 times, and performance metrics were averaged across replicates.

\begin{simsetting}\textbf{(Without additional covariates: varying DE gene proportions, sample sizes, fold changes, and target FDR levels)}\label{simu1}
In this setting, additional covariates are not included, while four simulation parameters are varied: the DE gene proportion \( \in \{0.1, 0.15, 0.2\} \), the sample size \( n \in \{6, 8, 10, 12, 14, 16, 20, 24\} \), the fold change \( \mathrm{FC} \in \{2,2.5,3\} \), and the target FDR level \( q \in \{0.05, 0.1, 0.2, 0.3, 0.4\} \). 
\end{simsetting}

Fig.~\ref{fig:simu1}a,b summarizes the results under Simulation Setting~\ref{simu1}, with the DE gene proportion set to \(0.1\), and the other three simulation parameters varied. Supplementary Fig.~\ref{fig:simu1_de=0.2} and \ref{fig:simu1_de=0.15} summarize the results for the other two DE proportions. Specifically, Fig.~\ref{fig:simu1}a shows methods' performance across sample sizes \( n \in \{6, 8, 10, 12, 14, 16, 20, 24\} \) and fold changes \( \mathrm{FC} \in \{2,2.5,3\} \), with the target FDR level fixed at \(q = 0.1\). Fig.~\ref{fig:simu1}b examines the effect of varying the target FDR level \( q \in \{0.05, 0.1, 0.2, 0.3, 0.4\} \) across selected sample sizes \( n \in \{6, 10, 16\} \), with the fold change fixed at \(\mathrm{FC} = 3\).

Across all settings, both Nullstrap-DESeq2 and Nullstrap-edgeR consistently control the FDR at the target level across all simulation settings, while achieving comparable or higher statistical power than the other four methods. In contrast, the original DESeq2 and edgeR exhibit increasing FDR inflation as the sample size, fold change, or target FDR level grows, consistent with previous findings~\cite{li2022exaggerated}. Between the two Wilcoxon rank-sum tests, Wilcoxon\_raw (applied to raw counts) effectively controls the FDR, but Wilcoxon\_norm (applied to normalized counts) fails to do so in most settings, possibly due to distortions introduced by the normalization process.

In terms of statistical power, Nullstrap-DESeq2 and Nullstrap-edgeR perform comparably to DESeq2 and edgeR, and substantially outperform Wilcoxon\_raw and Wilcoxon\_norm, particularly at smaller sample sizes. For example, in Fig.~\ref{fig:simu1}a, at \(n = 8\) and \(\mathrm{FC} = 2.5\), both Nullstrap-DE methods achieve a power of approximately 0.75, slightly lower than those of DESeq2 and edgeR, but markedly higher than the near-zero power of the two Wilcoxon rank-sum tests. In Fig.~\ref{fig:simu1}b, at \(n = 6\), Nullstrap-DESeq2 and Nullstrap-edgeR maintain strong power as the target FDR level increases, whereas Wilcoxon\_raw and Wilcoxon\_norm consistently yield near-zero power.

These results highlight the effectiveness of Nullstrap-DESeq2 and Nullstrap-edgeR in controlling FDR while maintaining high power, especially in small-sample scenarios.

\begin{simsetting}\textbf{(With additional covariates: varying DE gene proportions, sample sizes, fold changes, and target FDR levels)}\label{simu2} 
In this setting, an additional binary covariate \( z_i \) representing a potential confounder, such as sex or batch, is introduced and designed to be correlated with the treatment indicator \( x_i \) to simulate confounding effects. Four simulation parameters are varied: the DE gene proportion \( \in \{0.1, 0.15, 0.2\} \), the sample size \( n \in \{6, 8, 10, 12, 14, 16, 20, 24\} \), the fold change \( \mathrm{FC} \in \{2.5,3,3.5\} \), and the target FDR level \( q \in \{0.05, 0.1, 0.2, 0.3, 0.4\} \). 
\end{simsetting}

Fig.~\ref{fig:simu1}c,d summarizes the results under Simulation Setting~\ref{simu2}, with the DE gene proportion set to \(0.2\), and the other three simulation parameters varied. Supplementary Fig.~\ref{fig:simu2_de=0.15} and \ref{fig:simu2_de=0.1} summarize the results for the other two DE proportions. Specifically, Fig.~\ref{fig:simu1}c shows methods' performance across sample sizes \( n \in \{6, 8, 10, 12, 14, 16, 20, 24\} \) and fold changes \( \mathrm{FC} \in \{2.5,3,3.5\} \), with the target FDR level fixed at \(q = 0.1\). Fig.~\ref{fig:simu1}d examines the effect of varying the target FDR level \( q \in \{0.05, 0.1, 0.2, 0.3, 0.4\} \) across selected sample sizes \( n \in \{8, 16, 24\} \), with the fold change fixed at \(\mathrm{FC} = 3\).

Overall, both Nullstrap-DESeq2 and Nullstrap-edgeR maintain effective FDR control and high power across all conditions. In contrast, DESeq2 and edgeR exhibit substantial FDR inflation as the sample size, fold change, or target FDR level increases. Both Wilcoxon\_raw and Wilcoxon\_norm consistently fail to control the FDR across nearly all scenarios due to the presence of additional covariates, and their power remains markedly lower than the other approaches, especially in small-sample regimes.

To evaluate the robustness of each method under model misspecification, we introduce two additional simulation settings based on alternative data-generating distributions: Poisson and zero-inflated NB. Both settings follow the same semi-synthetic design used in Simulation Setting~\ref{simu1}, with gene-specific base intercepts \( \alpha_j \) and dispersions \( \phi_j \) estimated from real data. We omit additional covariates in these settings to focus on the effect of model misspecification.

\begin{simsetting}\textbf{(Poisson: no additional covariates, varying sample sizes, fold changes, and target FDR levels)}\label{simu3}
The counts are generated from a Poisson distribution. The DE gene proportion is fixed at \(0.2\), and other simulation parameters are varied: the sample size \( n \in \{6, 8, 10, 12, 14, 16, 20, 24\} \), fold change \( \mathrm{FC} \in \{2,2.5,3,3.5\} \), and target FDR level \( q \in \{0.05, 0.1, 0.2, 0.3, 0.4\} \).
\end{simsetting}

\begin{simsetting}\textbf{(Zero-inflated NB: no additional covariates, varying sample sizes, fold changes, and target FDR levels)}\label{simu4}
The counts are generated from a zero-inflation NB distribution with zero proportion \(=0.1\). The DE gene proportion is fixed at \(0.2\), and other simulation parameters are varied: the sample size \( n \in \{6, 8, 10, 12, 14, 16, 20, 24\} \), fold change \( \mathrm{FC} \in \{3,3.5,4\} \), and target FDR level \( q \in \{0.05, 0.1, 0.2, 0.3, 0.4\} \).
\end{simsetting}

The simulation results for Simulation Settings~\ref{simu3} and~\ref{simu4} are presented in Supplementary Fig.~\ref{fig:simu3_de=0.2}, \ref{fig:simu4_de=0.2}. Both Nullstrap-DESeq2 and Nullstrap-edgeR consistently maintain valid FDR control across all scenarios, though with slightly reduced power under zero-inflated NB due to excess zeros. In contrast, DESeq2 and edgeR exhibit substantial FDR inflation as sample size, fold change, or target FDR level increases. The Wilcoxon\_norm method fails to control the FDR in nearly all settings, and both Wilcoxon rank-sum tests demonstrate notably low power, particularly in small-sample scenarios.

\clearpage
\section{Details of Simulation Design}

Following the data-generating process described in the DESeq2 paper~\cite{love2014moderated}, we consider a two-condition DE testing setting under an NB-GLM. For gene \( j \) in sample \( i \), the observed count \( Y_{ij} \) is generated as
\begin{align}\label{eq:nb-model}
    Y_{ij} \overset{\text{ind}}{\sim} \text{NB}\left(\mu_{ij}, \phi_j\right), \quad
\log(\mu_{ij}) = \log(s_i) + \alpha_j + x_i\beta_j + z_i \gamma_j,
\end{align}
where \( x_i \in \{0, 1\} \) is a binary indicator encoding the treatment condition for sample \( i \) (1 for treatment, 0 for control; samples are evenly divided into control and treatment groups), and \( \beta_j \) denotes the true log fold change between the two conditions for gene \( j \), which is the primary parameter of interest in DE analysis. The gene-specific dispersion parameter \( \phi_j \) and baseline intercept \( \alpha_j \) are estimated from original data, while the sample-specific size factors \( s_i \) and an additional covariate \( z_i \) with gene-specific effect \( \gamma_j \) are manually specified. The full specification of parameters and the data generation procedure is as follows:

\begin{enumerate}
    \item Based on a real bulk RNA-seq dataset of lymphoblastoid cell lines from unrelated Nigerian individuals~\cite{pickrell2010understanding}, we fitted DESeq2 using an intercept-only model to estimate gene-specific intercepts \( \alpha_j \), and we extracted the mean–dispersion relationship as defined by \texttt{dds@dispersionFunction} in the DESeq2 package.
    \item Given a specified DE gene proportion, genes are randomly assigned to either the truly DE group or the non-DE group. For truly DE genes, the true log fold change for gene \( j \) is set to \( \beta_j = \log(\mathrm{FC}_j) \), where \( \mathrm{FC}_j \) is the fold change specified in the simulation scenario; for non-DE genes, $\beta_j=0$.

    \item Sample-specific size factors \( s_i \) are independently drawn from the uniform distribution \( \mathrm{Uniform}(0.9, 1.1) \), mimicking moderate variability in sequencing depth.

    \item If an additional covariate \( z_i \) is included, it is generated as a binary variable (e.g., representing sex or batch) that is deliberately correlated with the treatment indicator \( x_i \) to simulate confounding. Specifically, for a two-condition comparison with balanced group sizes, we simulate 80\% group-wise imbalance between the covariate and treatment. That is, within each treatment group, 80\% of the samples share the same covariate value (e.g., ``female'' in group A and ``male'' in group B), while the remaining 20\% are assigned the opposite value. The covariate values are then permuted within each group to avoid perfect separation while maintaining the overall imbalance. The additional covariate \( z_i \) is assumed to affect 20\% of genes, whose gene-specific effect sizes \( \gamma_j \) are independently drawn from \( \mathrm{Uniform}(2, 3) \), while the remaining 80\% of genes are unaffected (\( \gamma_j = 0 \)).

    \item The mean parameter of the NB distribution is calculated as 
    \[
    \mu_{ij} = \exp\left(\log(s_i) + \alpha_j + x_i\beta_j + z_i \gamma_j\right),
    \]
    and the dispersion parameter \( \phi_j \) is determined from the empirical mean–dispersion relationship using \( \phi_j = \texttt{dds@dispersionFunction}(\mu_{ij}) \).
\end{enumerate}

\clearpage
\section{Supplementary Figures}
\renewcommand{\thefigure}{S\arabic{figure}}
\renewcommand{\thetable}{S\arabic{table}}
\setcounter{figure}{0}
\setcounter{table}{0}

\begin{figure}[htbp]
    \centering
    \includegraphics[width=0.8\linewidth]{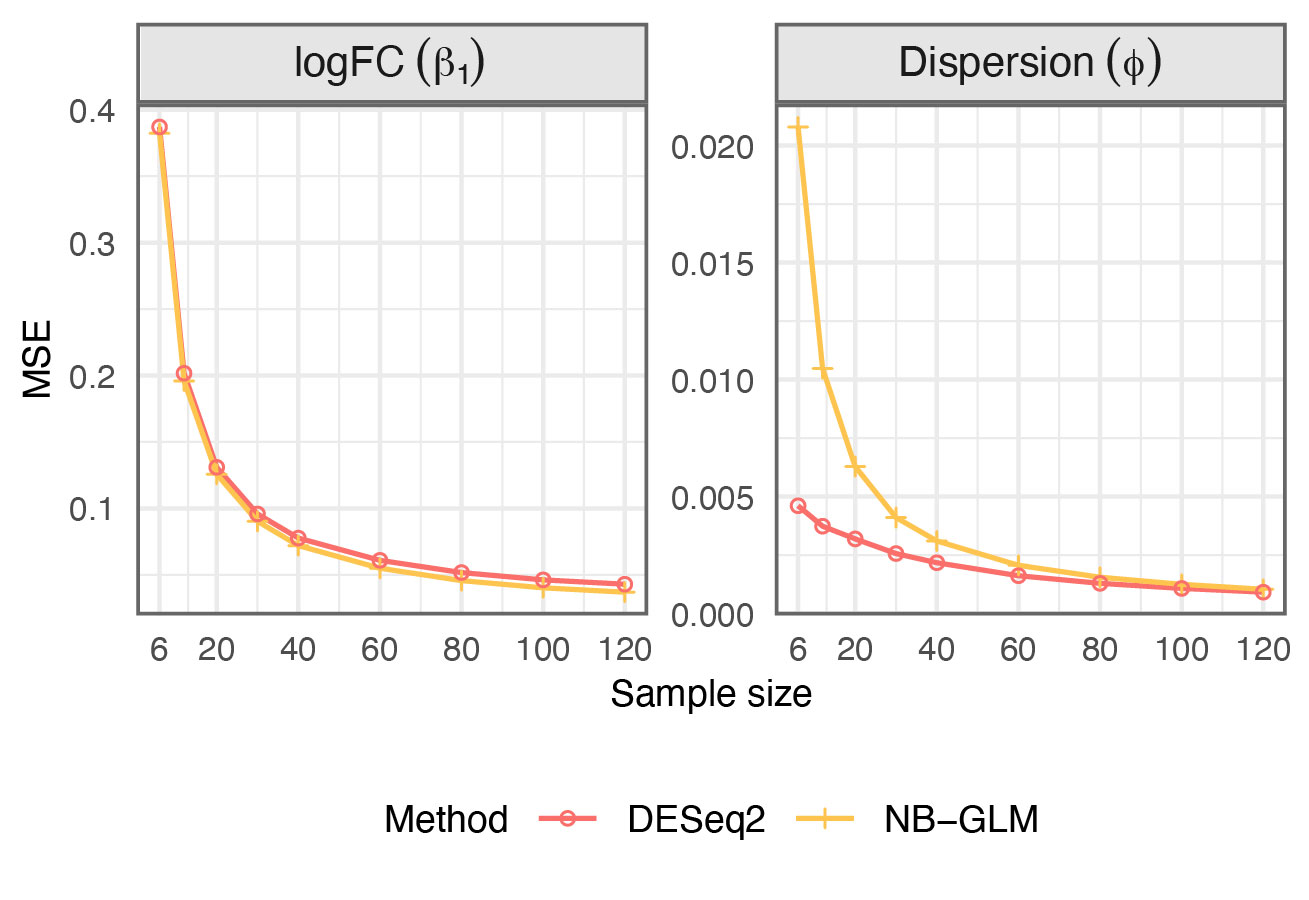}
    \caption{
    Mean squared errors (MSEs) of gene-specific log fold change (\( \beta_{1j} = \beta_1 \)) and dispersion (\( \phi_j = \phi \)) estimates obtained using DESeq2 and NB-GLM-based MLE. Results are based on numerical simulations where the mean–dispersion relationship is drawn from a real RNA-seq dataset. Each point indicates the average MSE across 50 simulated datasets with varying sample sizes. While both methods yield decreasing MSE with increasing sample size, DESeq2 achieves notably lower MSE for dispersion estimation, especially in small-sample settings. For log fold change estimation, both methods show comparable performance across sample sizes.
    }
    \label{fig:theory_nb_glm}
\end{figure}


\begin{figure}[htbp]
    \centering
    \includegraphics[width=0.9\textwidth]{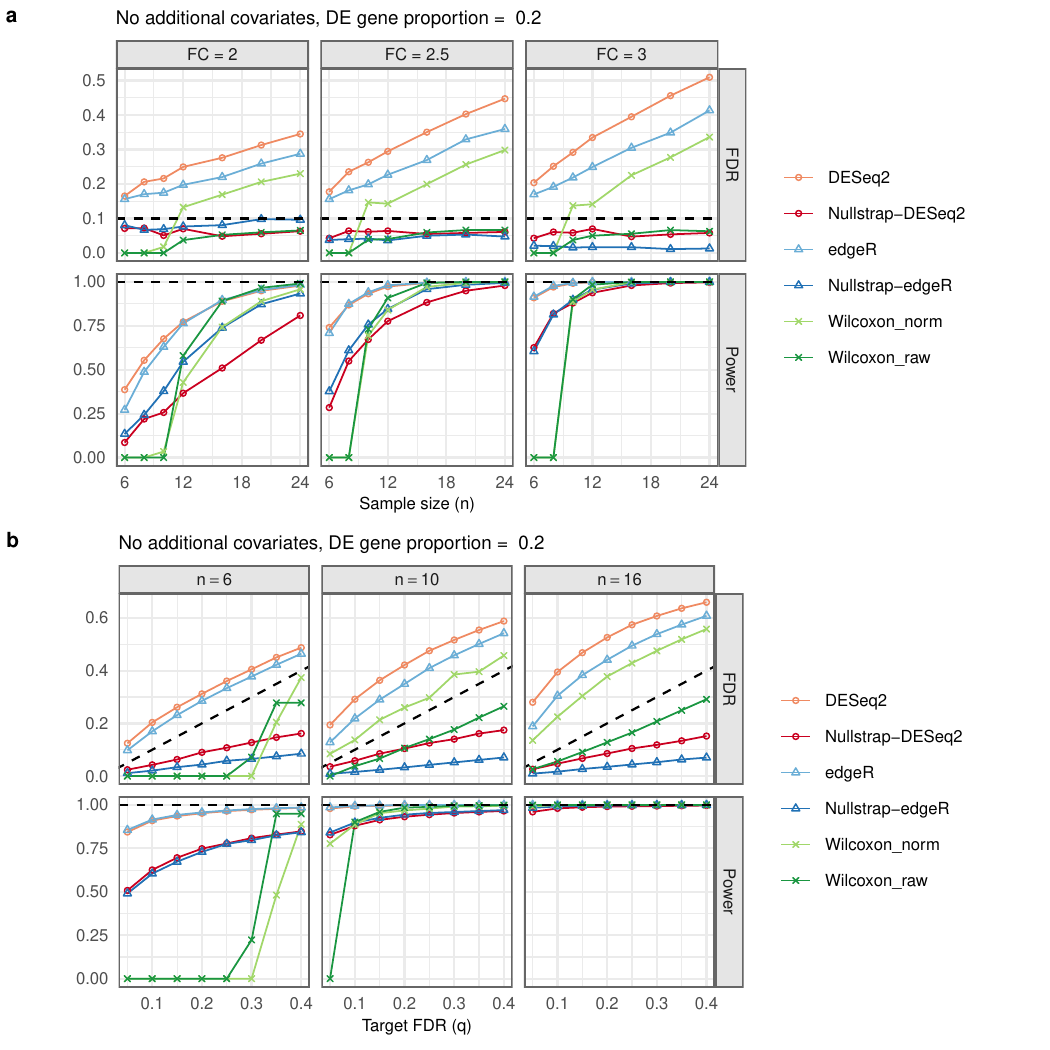}
    \caption{Simulation setting 1 (no additional covariates): \textbf{(a)} Empirical FDR and power versus sample size ($n$) under different fold changes (FC), with DE proportion $= 0.2$ and target FDR level $q = 0.1$. \textbf{(b)} Empirical FDR and power versus target FDR level ($q$) under different sample sizes ($n$), with DE proportion $= 0.2$ and fixed fold change $\mathrm{FC} = 3$.}
    \label{fig:simu1_de=0.2}
\end{figure}

\begin{figure}[htbp]
    \centering
    \includegraphics[width=0.9\textwidth]{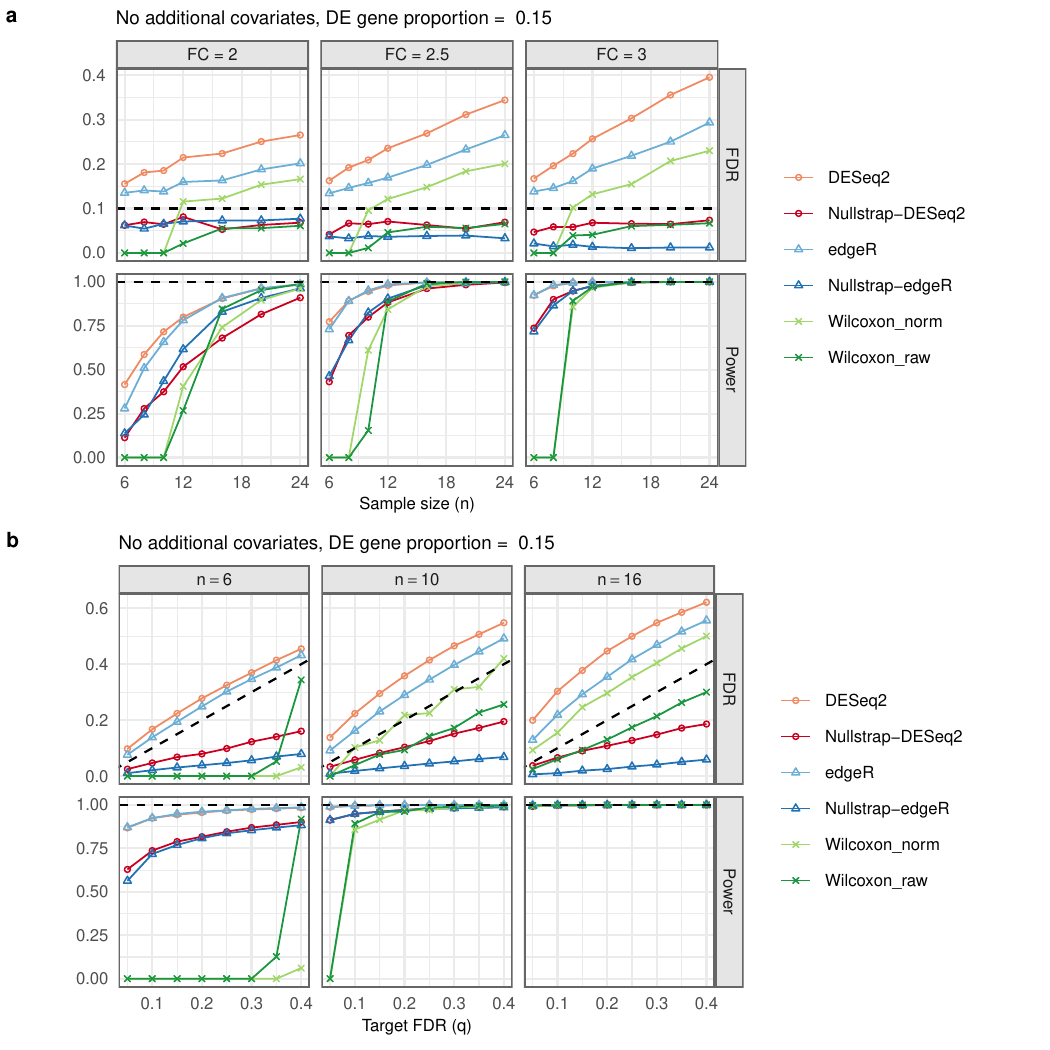}
    \caption{Simulation setting 1 (no additional covariates): \textbf{(a)} Empirical FDR and power versus sample size ($n$) under different fold changes (FC), with DE proportion $= 0.15$ and target FDR level $q = 0.1$. \textbf{(b)} Empirical FDR and power versus target FDR level ($q$) under different sample sizes ($n$), with DE proportion $= 0.15$ and fixed fold change $\mathrm{FC} = 3$.}
    \label{fig:simu1_de=0.15}
\end{figure}


\begin{figure}[htbp]
    \centering
    \includegraphics[width=0.9\textwidth]{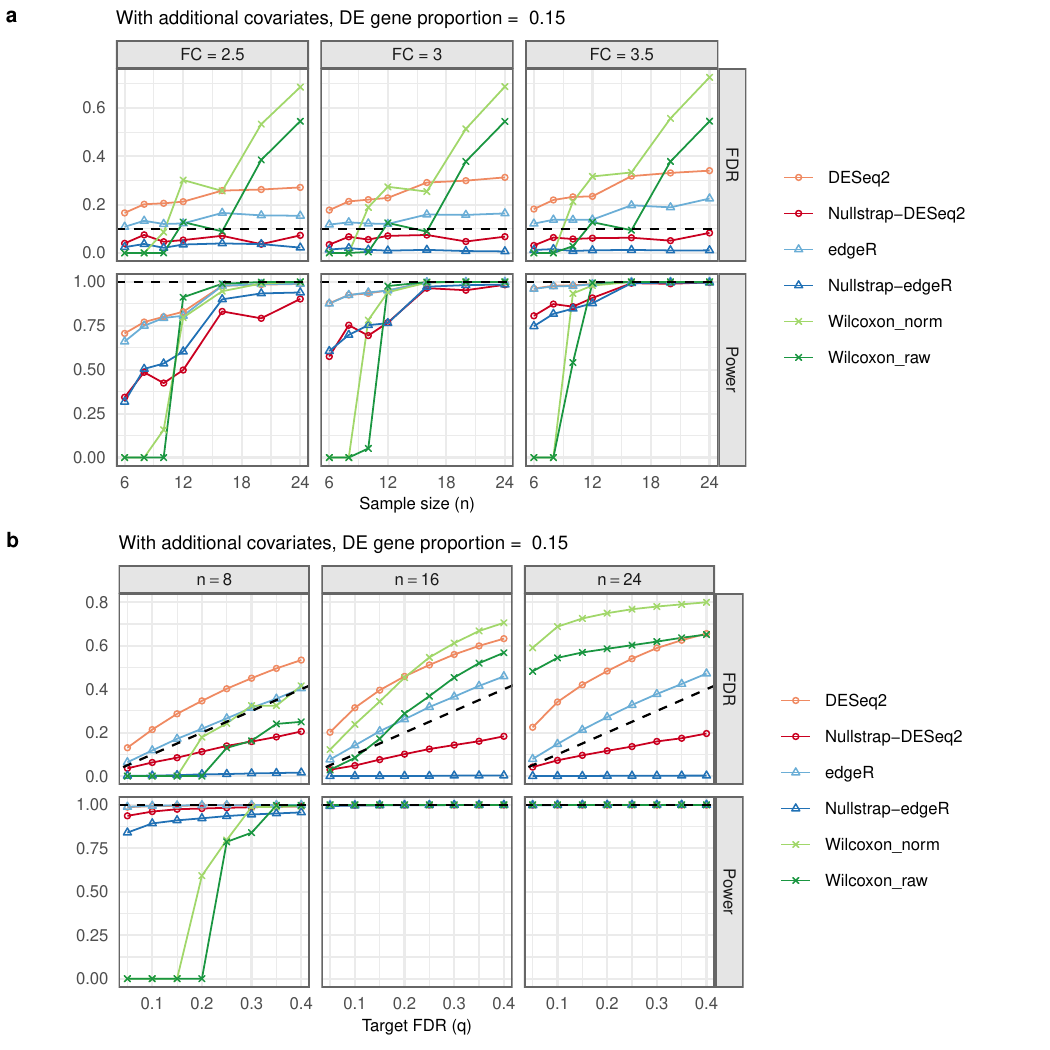}
    \caption{Simulation setting 2 (with additional covariates): \textbf{(a)} Empirical FDR and power versus sample size ($n$) under different fold changes (FC), with DE proportion $= 0.15$ and target FDR level $q = 0.1$. \textbf{(b)} Empirical FDR and power versus target FDR level ($q$) under different sample sizes ($n$), with DE proportion $= 0.15$ and fixed fold change $\mathrm{FC} = 3$.}
    \label{fig:simu2_de=0.15}
\end{figure}

\begin{figure}[htbp]
    \centering
    \includegraphics[width=0.9\textwidth]{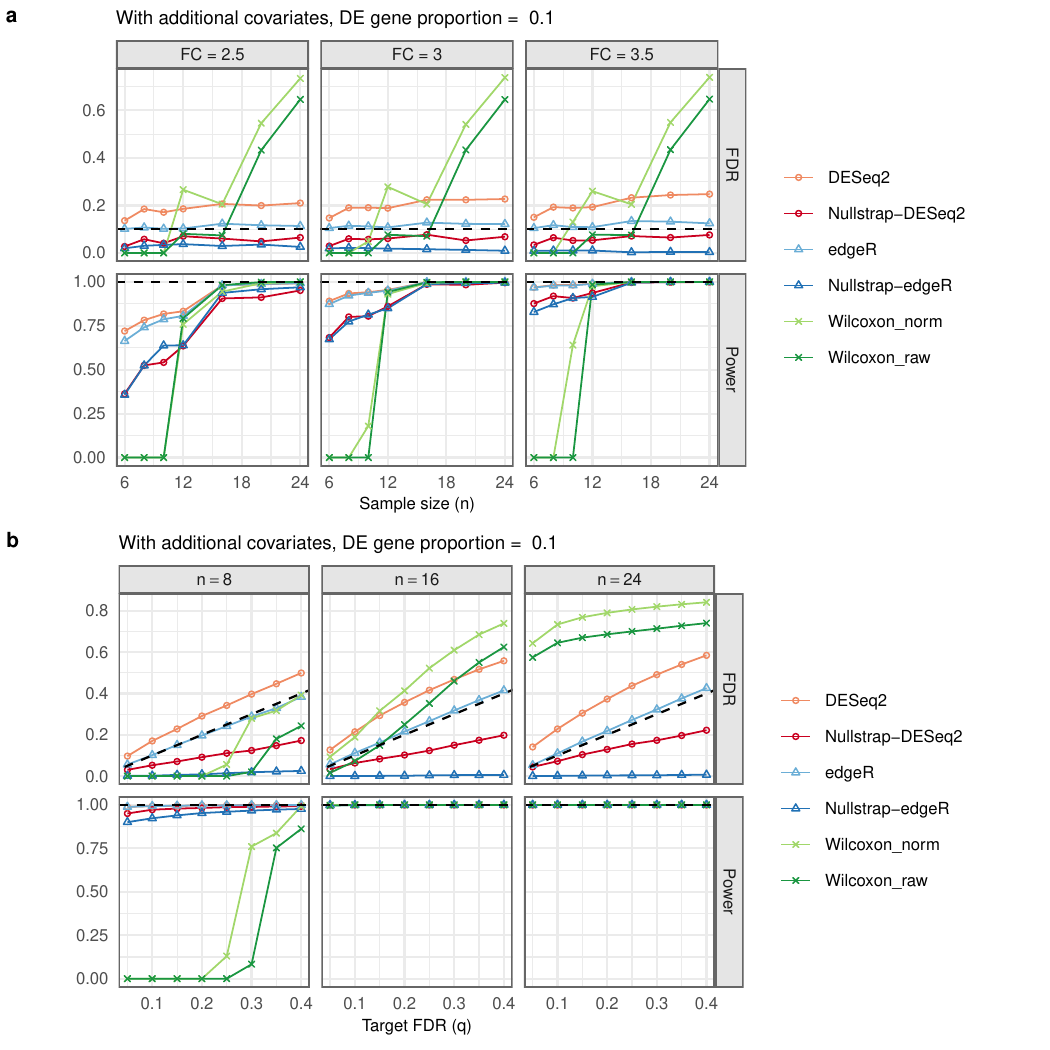}
     \caption{Simulation setting 2 (with additional covariates): \textbf{(a)} Empirical FDR and power versus sample size ($n$) under different fold changes (FC), with DE proportion $= 0.1$ and target FDR level $q = 0.1$. \textbf{(b)} Empirical FDR and power versus target FDR level ($q$) under different sample sizes ($n$), with DE proportion $= 0.1$ and fixed fold change $\mathrm{FC} = 3$.}
    \label{fig:simu2_de=0.1}
\end{figure}


\begin{figure}[htbp]
    \centering
    \includegraphics[width=0.9\textwidth]{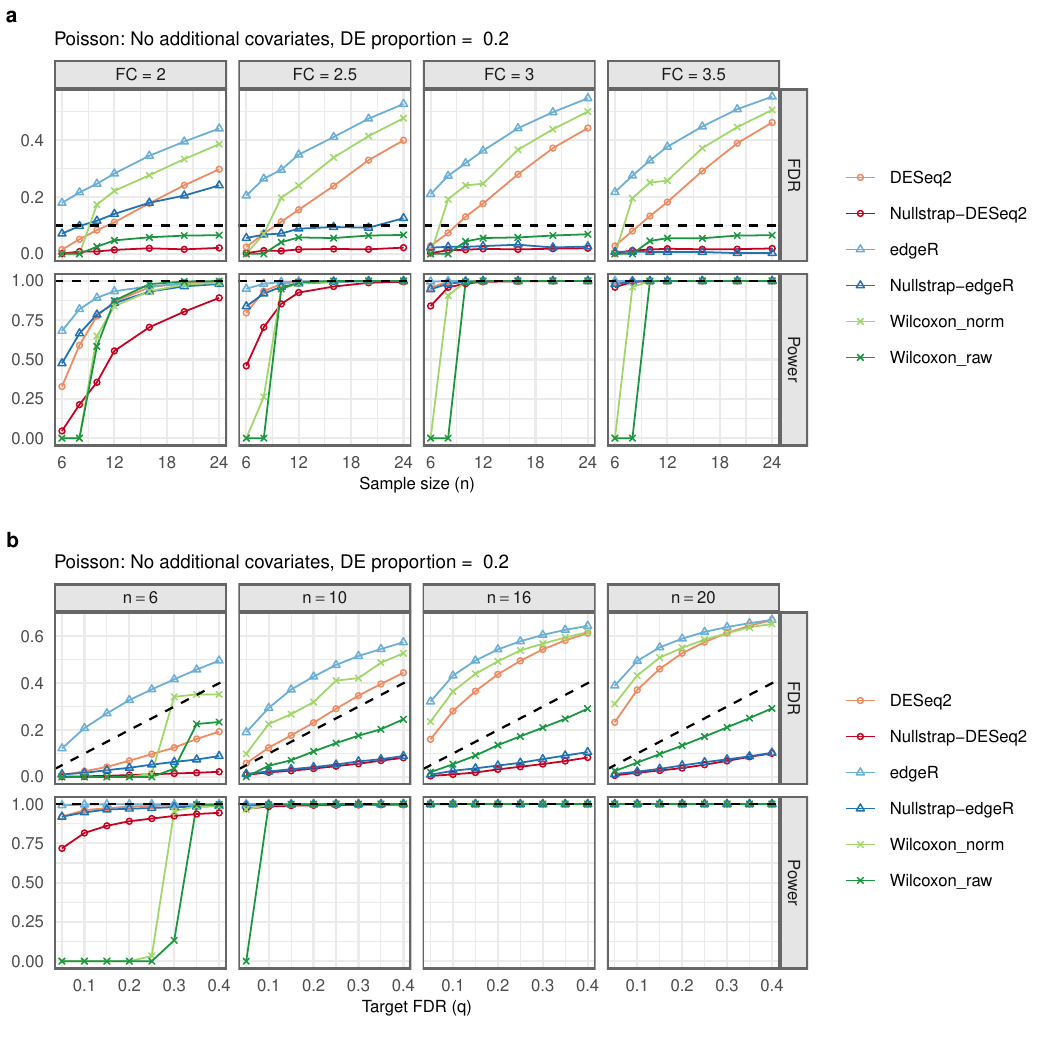}
    \caption{Simulation setting 3 (Poisson, no additional covariates, DE proportion = 0.2): \textbf{(a)} Empirical FDR and power versus sample size ($n$) under different fold changes (FC), with target FDR level $q = 0.1$. \textbf{(b)} Empirical FDR and power versus target FDR level ($q$) under different sample sizes ($n$) fixed fold change $\mathrm{FC} = 3$.}
    \label{fig:simu3_de=0.2}
\end{figure}

\begin{figure}[htbp]
    \centering
    \includegraphics[width=0.9\textwidth]{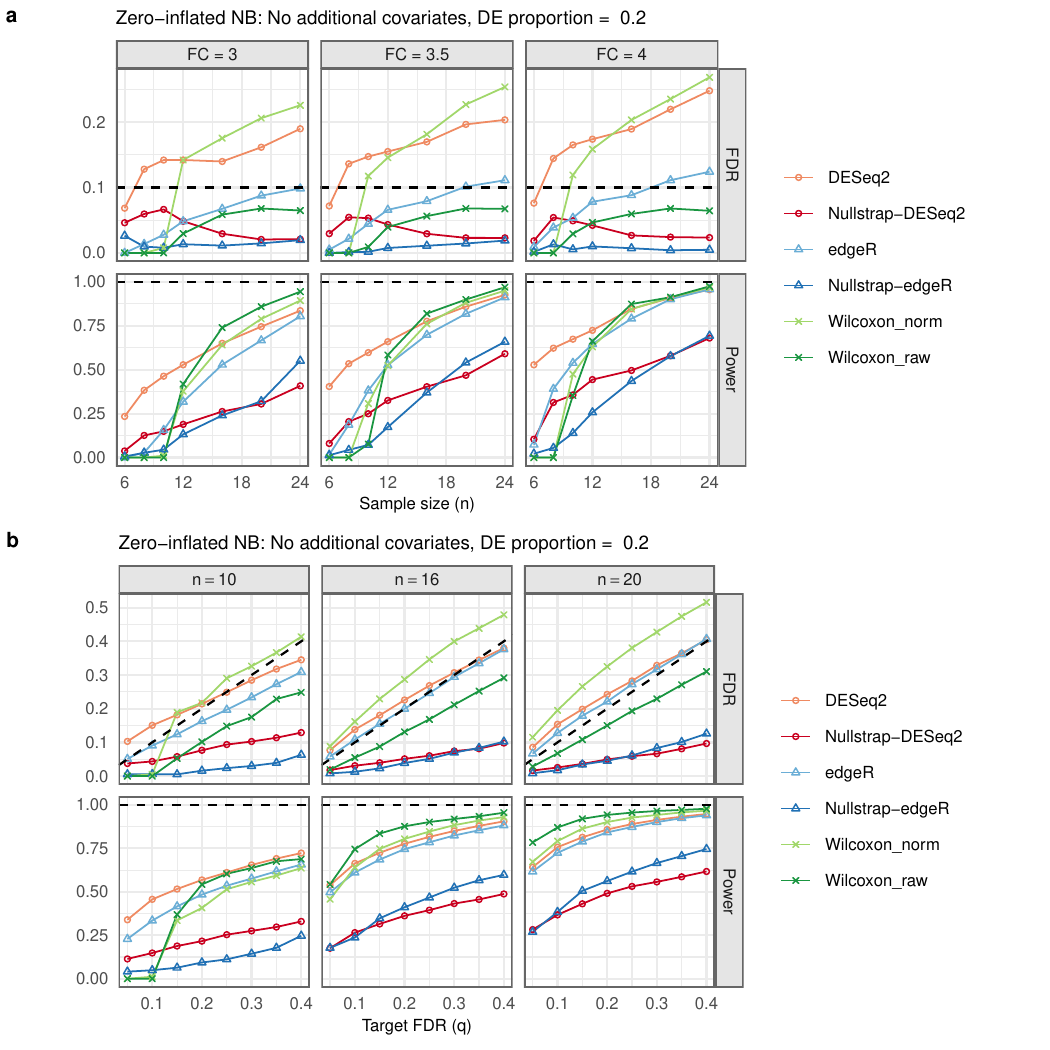}
    \caption{Simulation setting 4 (Zero-inflated negative binomial, no additional covariates, DE proportion = 0.2): \textbf{(a)} Empirical FDR and power versus sample size ($n$) under different fold changes (FC), with target FDR level $q = 0.1$. \textbf{(b)} Empirical FDR and power versus target FDR level ($q$) under different sample sizes ($n$), with fixed fold change $\mathrm{FC} = 3$.}
    \label{fig:simu4_de=0.2}
\end{figure}


{\captionsetup{font=small} 
\begin{figure}[htbp]
    \centering
    \includegraphics[width=\textwidth]{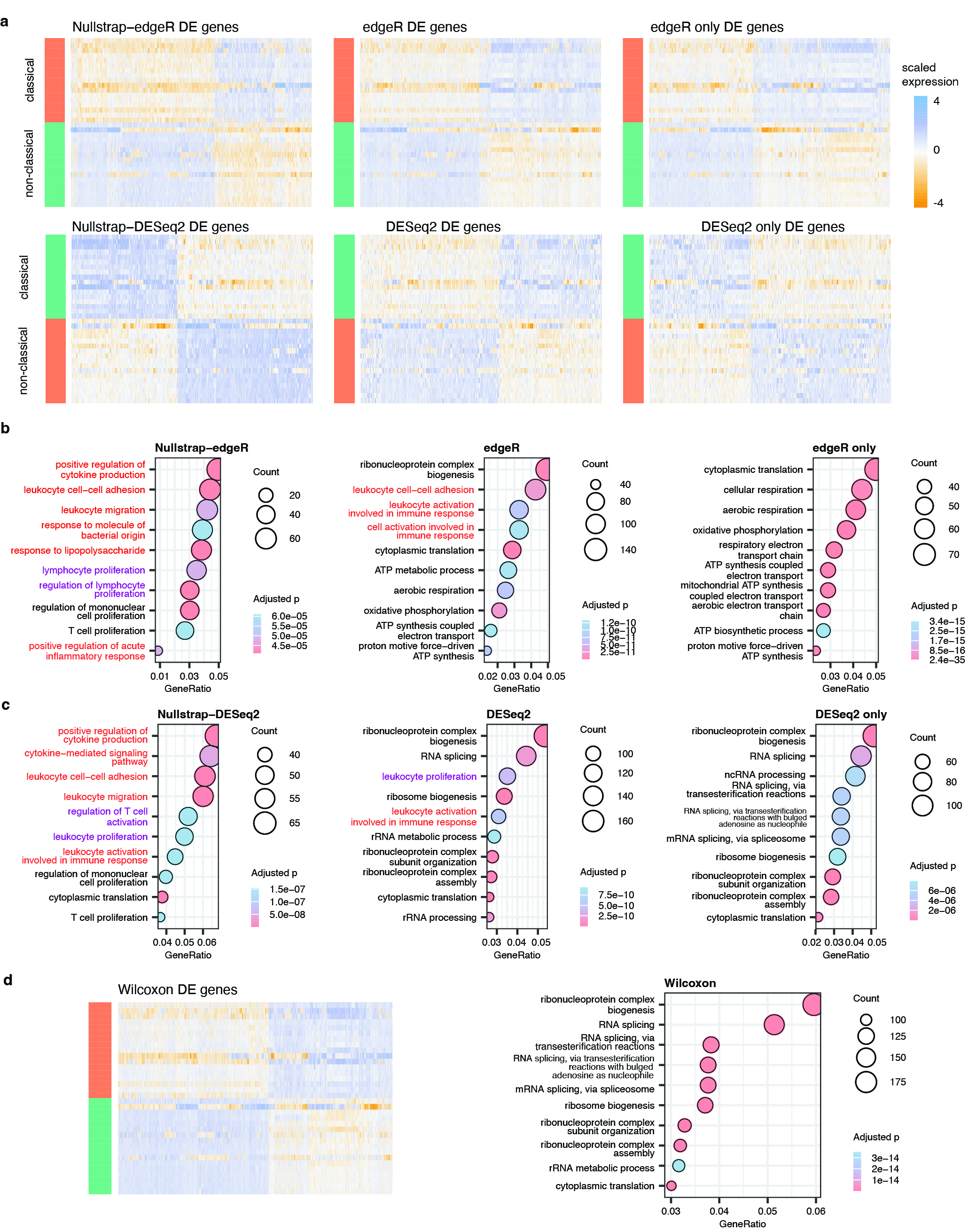}
    \caption{\textbf{Additional comparison of DE methods in human monocyte RNA-seq data.}
(\textbf{a}) Heatmaps of scaled expression for DE genes identified by Nullstrap-edgeR/DESeq2, edgeR/DESeq2, and edgeR/DESeq2 only sets. Nullstrap-edgeR/DESeq2 DE genes show clearer separation between classical and non-classical monocytes. 
(\textbf{b}) GO enrichment analysis for DE genes from Nullstrap-edgeR, edgeR, and edgeR only sets. Nullstrap-edgeR primarily identifies immune-related functions, whereas edgeR and edgeR only include many general cellular processes.  
(\textbf{c}) GO enrichment analysis for DE genes from Nullstrap-DESeq2, DESeq2, and DESeq2-only sets. Nullstrap-DESeq2 enriches for immune-specific functions, while DESeq2 and DESeq2 only capture broader transcriptional programs.  
(\textbf{d}) Heatmap and GO enrichment for DE genes from the Wilcoxon test. Many Wilcoxon DE genes are associated with general cellular functions, with limited specificity to monocyte subsets.
}
    \label{fig:case2_supp1}
\end{figure}
}

\newpage

\printbibliography[title={References}]

\end{refsection}

\end{document}


\if1\blind
{
    \title{Nullstrap-DE: A General Framework for Calibrating FDR and Preserving Power in Differential Expression Methods, with Applications to DESeq2 and edgeR}
  \author{ 
  Chenxin Flora Jiang \thanks{Equal contribution} \hspace{.2cm}\\
   Department of Statistics and Data Science \\ University of California, Los Angeles \\
  Changhu Wang \footnotemark[1] \hspace{.2cm}\\
   Jingyi Jessica Li\hspace{.1cm} \thanks{Correspondence should be addressed to Jingyi Jessica Li (jli@stat.ucla.edu, lijy03@fredhutch.org)}\\
 Department of Statistics and Data Science \\ University of California, Los Angeles \\
 Biostatistics
Program, Public Health Science
Division \\ Fred Hutchinson Cancer
Center}
  \date{}
  \maketitle
} \fi

\if0\blind
{
  \bigskip
  \bigskip
  \bigskip
  \begin{center}
    {\LARGE\bf Title}
\end{center}
  \medskip
} \fi

\maketitle

\def\spacingset#1{\renewcommand{\baselinestretch}%
{#1}\small\normalsize} \spacingset{1}


\newpage

\spacingset{1.9}

\setcounter{figure}{0}
\renewcommand{\thefigure}{S\arabic{figure}}
\setcounter{table}{0}
\renewcommand{\thetable}{S\arabic{table}}

\appendix
\section{Proof of Theorem 1}

\subsection{Auxiliary Lemmas}
In this section we present a standard multiplicative form of the Chernoff bound for sums of independent Bernoulli random variables. 
Two auxiliary inequalities (Lemma~\ref{lem:aux-upper} and Lemma~\ref{lem:aux-lower}) are required in the proof.
\begin{lemma}[An upper-tail exponent inequality]
\label{lem:aux-upper}
For all $\delta \ge 0$,
\[
(1+\delta)\log(1+\delta)-\delta \;\ge\; \frac{\delta^2}{\,2+\delta\,}.
\]
\end{lemma}

\begin{proof}
Define
\[
g(\delta)\;=\;(1+\delta)\log(1+\delta)-\delta-\frac{\delta^2}{2+\delta},\qquad \delta\ge 0.
\]
We show $g(\delta)\ge 0$. Note $g(0)=0$. Compute
\[
g'(\delta)=\log(1+\delta)-\frac{\delta(4+\delta)}{(2+\delta)^2}.
\]
Consider the auxiliary function 
\[
r(\delta):=\log(1+\delta)-\frac{2\delta}{2+\delta}.
\]
Then $r(0)=0$ and 
\[
r'(\delta)=\frac{1}{1+\delta}-\frac{4}{(2+\delta)^2}
=\frac{\delta^2}{(1+\delta)(2+\delta)^2}\;\ge\;0,
\]
hence $r(\delta)\ge 0$ for all $\delta\ge 0$, i.e.,
\[
\log(1+\delta)\;\ge\;\frac{2\delta}{2+\delta}.
\]
Using this in $g'(\delta)$ gives
\[
g'(\delta)\;\ge\; \frac{2\delta}{2+\delta}-\frac{\delta(4+\delta)}{(2+\delta)^2}
=\frac{\delta^2}{(2+\delta)^2}\;\ge\;0.
\]
Thus $g$ is nondecreasing on $[0,\infty)$ and $g(\delta)\ge g(0)=0$, proving the claim.
\end{proof}

\begin{lemma}[A lower-tail exponent inequality]
\label{lem:aux-lower}
For all $0\le \delta \le 1$,
\[
\delta-(1-\delta)\log(1-\delta) \;\ge\; \frac{\delta^2}{2}.
\]
\end{lemma}

\begin{proof}
Define
\[
h(\delta)\;=\;\delta-(1-\delta)\log(1-\delta)-\frac{\delta^2}{2},\qquad 0\le \delta<1.
\]
We prove $h(\delta)\ge 0$ on $[0,1]$. Note $h(0)=0$. Differentiate:
\[
h'(\delta)=2+\log(1-\delta)-\delta,
\qquad 
h''(\delta)= -\frac{1}{1-\delta}-1 < 0 \quad (0\le \delta<1).
\]
Thus $h$ is concave on $[0,1)$, so its minimum over $[0,1]$ is attained at an endpoint.
Since $h(0)=0$ and $h$ is concave, we have $h(\delta)\ge 0$ for all $0\le \delta\le 1$, proving the inequality.
\end{proof}

\begin{lemma}[Multiplicative Chernoff bound]
\label{lem:chernoff}
Let $X_1,\ldots,X_n$ be independent Bernoulli random variables, $X_i\in\{0,1\}$, with $\mathbb{P}(X_i=1)=p_i$. 
Let $X=\sum_{i=1}^n X_i$ and $\mu=\mathbb{E}[X]=\sum_{i=1}^n p_i$. 
Then for any $0\le \delta \le 1$,
\[
\pr \big(|X-\mu|\ge \delta \mu\big)\;\le\; 2\,\exp\!\Big(-\frac{\delta^2 \mu}{3}\Big).
\]
\end{lemma}

\begin{proof}
By the standard Chernoff argument using moment generating functions and Markov's inequality, we can obtain the following one-sided bounds:
\[
\pr\big(X\ge (1+\delta)\mu\big) \;\le\; 
\exp\Big(-\mu\big[(1+\delta)\log(1+\delta)-\delta\big]\Big),
\]
and
\[
\pr\big(X\le (1-\delta)\mu\big) \;\le\; 
\exp\Big(-\mu\big[\delta-(1-\delta)\log(1-\delta)\big]\Big).
\]
Applying Lemma~\ref{lem:aux-upper} to the upper tail inequality and Lemma~\ref{lem:aux-lower} to the lower tail inequality yields
\[
\pr\big(X\ge (1+\delta)\mu\big) \;\le\; 
\exp\Big(-\tfrac{\delta^2}{2+\delta}\mu\Big),
\qquad
\pr\big(X\le (1-\delta)\mu\big) \;\le\; 
\exp\Big(-\tfrac{\delta^2}{2}\mu\Big).
\]
By the union bound and the fact that $0\le \delta \le 1$ implies $\tfrac{1}{2+\delta}\ge \tfrac{1}{3}$, we obtain
\[
\pr\big(|X-\mu|\ge \delta\mu\big)
\;\le\; \exp\Big(-\tfrac{\delta^2}{3}\mu\Big) + \exp\Big(-\tfrac{\delta^2}{3}\mu\Big)
\;=\;2\exp\Big(-\tfrac{\delta^2}{3}\mu\Big),
\]
which completes the proof of Lemma~\ref{lem:chernoff}.
\end{proof}

\begin{lemma}
\label{lem:null-equivalence}
Under the null model with $\beta = 0$, the distribution of the test statistic 
$\hat{T}_j$ for any null feature $j \in \cS_0$ depends solely on the intercept 
$\alpha$. 
If the synthetic null data are generated from this null model with correctly 
specified $\alpha$, then for any $t>0$,
\[
\pr\left(|\tilde{T}_j| \geq t\right) 
\;=\; \pr\left(|\hat{T}_j| \geq t\right),
\qquad \forall j \in \cS_0.
\]
\end{lemma}

\begin{proof}
Under the null model with $\beta = 0$, the distribution of the statistic 
$\hat{T}_j$ for any null feature $j \in \cS_0$ is fully determined by the 
intercept $\alpha$.
 Since the 
synthetic null data are generated from the same null model with correctly 
specified $\alpha$, 
the induced distribution of the test 
statistic $\tilde{T}_j$ in the synthetic data coincides with that of 
$\hat{T}_j$ in the original data for all null features. The stated probability 
identity follows immediately.
\end{proof}

When the null model is misspecified, the distributions of $\hat{T}_j$ and 
$\tilde{T}_j$ may not coincide. However, if we can obtain accurate estimators 
of the nuisance parameters $\alpha$, then for null features 
the distributions of $\hat{T}_j$ and $\tilde{T}_j$ will be close. We assume 
that the estimation error of $\alpha$ can be uniformly bounded 
over all null features by a sequence $\gamma_{n,p} \to 0$ (as defined in 
Assumption~1) as $n,p \to \infty$. Specifically, let $\bm{\alpha} = (\alpha_1, \dots, \alpha_p)$ and $\hat{\bm{\alpha}} = (\hat{\alpha}_1, \dots, \hat{\alpha}_p)$ denote the true and estimated intercepts for all $p$ features, respectively. We assume that  
\begin{equation}\label{eq:null-equivalence-approx}
G(\bm{\alpha}, \hat{\bm{\alpha}}) =  
\frac{\sum_{j \in  \{1,\dots,p\}}\Pr\left(|\tilde{T}_j| \geq t\right) - \sum_{j \in \mathcal{S}_0}\Pr\left(|\hat{T}_j| \geq t\right)}{\sum_{j \in  \{1,\dots,p\}}\Pr\left(|\hat{T}_j| \geq t\right) + \sum_{j \in  \mathcal{S}_0}\Pr\left(|\tilde{T}_j| \geq t\right)}  
\geq - \gamma_{n,p},
\end{equation}
which serves as a relaxed version of Lemma~\ref{lem:null-equivalence}, accounting for the estimation error of nuisance parameters. Note that $G(\bm{\alpha}, \bm{\alpha}) \geq 0$ and $G(\bm{\alpha}, \bm{\alpha}) \in [-1, 1]$. Therefore, condition~\eqref{eq:null-equivalence-approx} effectively requires that the estimation error of the nuisance parameters is sufficiently small, such that $G(\bm{\alpha}, \hat{\bm{\alpha}})$ remains close to $G(\bm{\alpha}, \bm{\alpha})$.

From Assumption 1, it is clear that 
\begin{itemize}
    \item For the real-data estimator,
    \[
    \mathbb{P} \left( \left| \hat{T}_j\right| \geq l^{-1}\gamma_{n,p} \right) \leq p^{-2}, \mbox{ for all } j \in \mathcal{S}_0.
    \]
    
    \item For the synthetic-null-data estimator,
    \[
    \mathbb{P} \left( \left| \tilde{T}_j \right| \geq l^{-1}\gamma_{n,p} \right)  \leq p^{-2}, \mbox{ for all } j \in  \{1,\dots,p\}. 
    \]
\end{itemize}
because the standard errors are uniformly bounded away from zero and infinity. 
Let $\cS_1 \subset \{1,\dots,p\}$ denote the set of nonnull (signal) features with $|\cS_1|=s$, and let $\cS_0=\{1,\dots,p\}\setminus \cS_1$ denote the null set.
\begin{lemma}\label{lem:power}
Under Assumptions~1 and~2, the power of the proposed method satisfies
\[
\mathrm{Power}
=\mathbb{E}\!\left[ \frac{\sum_{j \in \cS_1} \mathbf{1}\{|\,\hat{T}_{j}| \ge \tau_q\}}{s} \right]
\;\ge\; 1 - 2p^{-1}.
\]
\end{lemma}

\begin{proof}[Proof of Lemma~\ref{lem:power}]

Fix the threshold
\[
t^* := \ell^{-1}\gamma_{n,p},
\]
where $\ell>0$ is the constant from Assumption~1. Consider the events
\[
\mathcal{G}_{\mathrm{syn}} := \Big\{ \max_{1\le j\le p} |\tilde{T}_j| < t^* \Big\},
\qquad
\mathcal{G}_{\mathrm{sig}} := \Big\{ \min_{j\in \cS_1} |\hat{T}_j| \ge t^* \Big\}.
\]

By Assumption~1,
\[
\mathbb{P}\!\left(|\tilde{T}_j| \ge t^*\right) \le p^{-2}\quad \text{for all } j\in\{1,\dots,p\}.
\]
A union bound yields
\[
\mathbb{P}\!\left(\mathcal{G}_{\mathrm{syn}}^{\,c}\right)
=\mathbb{P}\!\left( \max_{j} |\tilde{T}_j| \ge t^* \right)
\le \sum_{j=1}^p \mathbb{P}\!\left(|\tilde{T}_j| \ge t^*\right)
\le p\cdot p^{-2}= p^{-1}.
\]

By Assumption~2,
\[
\mathbb{P}\!\left(\mathcal{G}_{\mathrm{sig}}^{\,c}\right)
=\mathbb{P}\!\left( \min_{j\in \cS} |\hat{T}_j| < t^* \right)
\le p^{-1}.
\]

On the event $\mathcal{G}_{\mathrm{syn}}$, the estimated FDP at $t^*$ is zero, i.e.,
$\widehat{\mathrm{FDP}}(t^*)=0$, so by the construction of the selection rule (monotonic in $t$) we must have
\[
\tau_q \;\le\; t^*.
\]

On the intersection $\mathcal{G}_{\mathrm{syn}}\cap \mathcal{G}_{\mathrm{sig}}$, we have
$|\hat{T}_j| \ge t^* \ge \tau_q$ for every $j\in \cS_1$, hence $\cS_1 \subseteq \widehat{\cS}(\tau_q)$ and thus
\[
\frac{1}{s}\sum_{j\in \cS}\mathbf{1}\{|\hat{T}_j|\ge \tau_q\}=1.
\]
Taking expectations and using the union bound,
\[
\mathrm{Power}
=\mathbb{E}\!\left[\frac{1}{s}\sum_{j\in \cS}\mathbf{1}\{|\hat{T}_j|\ge \tau_q\}\right]
\;\ge\; \mathbb{P}\!\left(\mathcal{G}_{\mathrm{syn}}\cap \mathcal{G}_{\mathrm{sig}}\right)
\;\ge\; 1 - \mathbb{P}\!\left(\mathcal{G}_{\mathrm{syn}}^{\,c}\right) - \mathbb{P}\!\left(\mathcal{G}_{\mathrm{sig}}^{\,c}\right)
\;\ge\; 1 - 2p^{-1}.
\]
This proves the claim.
\end{proof}

\begin{lemma}\label{lem:number}
Under Assumptions~1 and~2, the expected number of selected nonnull features satisfies
\[
\mathbb{E}\!\left[ \#\{j : |\hat{T}_j| \geq \tau_q\} \right] 
\;\geq\; s - \frac{2s}{p},
\]
where $s$ is the number of nonnull features and $\tau_q$ is the data-dependent threshold.  

Moreover, the expected number of selected synthetic-null features satisfies
\[
\mathbb{E}\!\left[ \#\{j : |\tilde{T}_j| \geq \tau_q\} \right] 
\;\geq\; \frac{1}{2}qs\,(1 - 2p^{-1}),
\]
where $q_0$ denotes the nominal number of null exceedances at level $q$.
\end{lemma}

\begin{proof}[Proof of Lemma~\ref{lem:number}]
The first claim follows directly from Lemma~\ref{lem:power}:
\[
\mathbb{E}\!\left[ \#\{j : |\hat{T}_j| \geq \tau_q\} \right]
=s\cdot \mathrm{Power} \ge s(1-2m^{-1})=s-\frac{2s}{p}.
\]
For the second claim, by the definition of $\tau_q$ we have
\[
\#\{j : |\tilde{T}_j| \geq \tau_q\} \;\geq\; \frac{1}{2}q \cdot \big(\#\{j : |\hat{T}_j| \geq \tau_q\} \vee 1\big).
\]
By taking expectations and using the first claim, we obtain
\[
\mathbb{E}\!\left[ \#\{j : |\tilde{T}_j| \geq \tau_q\} \right]
\;\geq\; \frac{1}{2}q \cdot \mathbb{E}\!\left[ \#\{j : |\hat{T}_j| \geq \tau_q\} \vee 1 \right]
\;\geq\; \frac{1}{2}qs(1-2p^{-1}),
\]
which completes the proof.
\end{proof}



\subsection{Proof of Theorem 1}
\begin{proof}[Proof of Theorem 1]

       Define the ratio statistic:
    $$H(t,t')= \frac{\#\left\{j: \tilde{T}_{j} \geq t^\prime \right\}}{\#\left\{j:  \hat{T}_{j}\geq t\right\} \vee 1},$$
    where $\tilde{T}_j$ and $T_j$ are the test statistics based on the synthetic null data and the original data, respectively. 
Then, we define 
$\tilde{e}(t)=\Ex \#\left\{j: \tilde{T}_{j} \geq t \right\}$, $e(t)=\Ex \#\left\{j:  \hat{T}_{j} \geq t\right\}$ as expected counts. By Lemma \ref{lem:chernoff}, we have the following concentration inequalities for the numerator and denominator of $H(t,t')$:
    $$\mathbb{P} \left(\left|\#\left\{j: \tilde{T}_{j} \geq t' \right\} -\tilde{e}(t') \right| \geq \delta \tilde{e}(t')\right) \leq 2 \exp(-\tilde{e}(t')\delta^2/3),$$
    and 
    $$\mathbb{P} \left(\left|\#\left\{j:  \hat{T}_{j} \geq t \right\} -e(t)\right| \geq \delta e(t)\right) \leq 2 \exp(-e(t)\delta^2/3),$$
    for fixed $t,t'>0$. 
  Consequently, there exists a constant $c>0$ such that the ratio concentrates:
    $$
    \mathbb{P}\left(\left|H(t,t')-\frac{\tilde{e}(t')}{e(t)} \right| \geq \dfrac{\delta}{ \sqrt{\tilde{e}(t')} \land \sqrt{e(t)}} \left|  \frac{\tilde{e}(t')}{e(t)} \right| \right) \leq 4 \exp(-c\delta^2),
    $$
where $0<\delta<\sqrt{\tilde{e}(t')} \land \sqrt{e(t)}$.

With the defined ratio statistic, the threshold $\tau_q$ can be expressed as
\[
\tau_q \;=\; \inf\{\, t : H(t,t) \leq q \,\}.
\]
Note that $\tau_q$ is a random variable depending on the data. To handle this, 
we invoke a discretization argument. Specifically, we consider a grid of values 
$\{t_\ell\}_{\ell=1}^p$ such that
\[
t_1 < t_2 < \cdots < t_p, 
\quad \text{and} \quad e(t_\ell) = \ell.
\]
Similarly, we introduce another grid $\{t'_\ell\}_{\ell=1}^p$ satisfying
\[
t'_1 < t'_2 < \cdots < t'_p, 
\quad \text{and} \quad \tilde{e}(t'_\ell) = \ell.
\]
Then, taking a union bound over the grid points, we have
\begin{equation} \label{eq:union_bound}
    \mathbb{P} \left( \bigcup_{i,j=0}^p \left\{\left|H(t_j,t'_i)-\frac{i}{j} \right| \geq \dfrac{\delta }{\sqrt{i \land j}}\frac{i}{j} \right\} \right) \leq 4(p+1)^2\exp({-c\delta^2}).
    \end{equation}
    Then, we find the nearest grid points to the random threshold $\tau_q$: 
\[t_{j^*} = \min\{t_j: t_j \geq \tau_q\}, \quad t'_{i^*} = \max\{t'_i: t'_i \leq \tau_q\}.\]
By the monotonicity of $H(t,t')$ in both arguments, we have
\[
 H(t_{j^*},t'_{i^*}) \leq H(\tau_q,\tau_q) \leq q.
\]
From \eqref{eq:union_bound}, we have
\begin{equation}\label{eq:bound1}
\mathbb{P} \left( \frac{i^*}{j^*} \geq q\left(1 - \delta/\sqrt{i^* \land j^*}\right)^{-1} \right) \leq 4(p+1)^2\exp({-c\delta^2}).
\end{equation}
\
Similarly, we define $e_0(t)=\Ex \#\left\{j \in S_0: {T}_{j} \geq t\right\}$ as the expected count of null features exceeding the threshold $t$ and 
$$
H_0(t,t')= \frac{\#\left\{j \in S_0: \hat{T}_j \geq t' \right\}}{\#\left\{j:  \hat{T}_{j}\geq t\right\} \vee 1}.
$$
Using similar discretization arguments, we have
\begin{equation}\label{eq:bound2}
\mathbb{P}\left(H_0(t_{j^*},t'_{i^*}) \geq  \left(1- \delta/\sqrt{e_0(t'_{i^*}) \land j^*}\right)^{-1} \frac{e_0(t'_{i^*})}{j^*}
\right) \leq 4(p+1)^2\exp({-c\delta^2}).
\end{equation}
By Lemma~\ref{lem:null-equivalence}, we have
\begin{equation}\label{eq:bound3}
e_0(t'_{i^*}) \;\leq\; e(t'_{i^*}) \;=\; i^*.
\end{equation}

We define the event
\[
\mathcal{D} := \left\{ \frac{e_0(t'_{i^*})}{j^*} \;\geq\; q \right\}.
\]
On the complement event $\mathcal{D}^c$, it follows that
\[
H_0(\tau_q,\tau_q) \;\leq\; H_0(t_{j^*},t'_{i^*}) \;\leq\; q.
\]
Moreover, by Lemma~\ref{lem:number} and the monotonicity of $H(t,t')$ in both arguments, 
combining \eqref{eq:bound1}, \eqref{eq:bound2}, and \eqref{eq:bound3}, 
we obtain that on the event $\mathcal{D}$, there exists a constant $c'>0$ such that
\[
\pr\!\left( \left\{ H_0(\tau_q,\tau_q) \;\geq\; q\left(1 + c'\tfrac{\delta}{\sqrt{s}} + c'\gamma_{n,p}\right)\right\} \cap \mathcal{D} \right) 
\;\leq\; 8(p+1)^2 \exp({-c\delta^2}).
\]
Finally, by taking expectations and setting $\delta = C \log p$, 
where $C$ is a sufficiently large constant, we obtain
\begin{align*}
\mathrm{FDR} 
&= \Ex\!\left[ H_0(\tau_q,\tau_q)\,\mathbb{I}(\mathcal{D}) \right] 
   + \Ex\!\left[ H_0(\tau_q,\tau_q)\,\mathbb{I}(\mathcal{D}^c) \right] \\
&\leq q\,\pr(\mathcal{D}^c) 
   + q\left(1 + c_1 (\log m)/\sqrt{s} + c_2 \gamma_{n,p}\right)\pr(\mathcal{D}) \\
&\leq q\left[ 1 + c_1 (\log m)/\sqrt{s} + c_2 \gamma_{n,p} \right],
\end{align*}
for some constants $c_1, c_2 > 0$. 
Combining the above argument with Lemma~\ref{lem:power} completes the proof of Theorem~1.
\end{proof}

\section{Assumptions and Proof of Corollary 1}

\subsection{Assumptions of Corollary 1}

\begin{assumptionS}[Well-behaved design]\label{assump:S1}
Each sample is associated with a binary treatment indicator \( x_i \in \{0,1\} \). We assume that both treatment groups are represented in the data, with the empirical proportion satisfying \( \frac{1}{n} \sum_{i=1}^n x_i \to p \in (0,1) \) as \( n \to \infty \). 
\end{assumptionS}

\begin{assumptionS}[Fixed nuisance parameters]\label{assump:S2}
The sample-specific size factors \( \{s_i\}_{i=1}^n \) and gene-specific dispersion parameters \( \{\phi_j\}_{j=1}^m \) are treated as fixed and lie in compact subsets of \( (0, \infty) \). No additional covariates are included in the model.
\end{assumptionS}

\subsection{Proof of Corollary 1}

Let \( \hat{\beta}_j \) and \( \tilde{\beta}_j \) denote the MLEs obtained from fitting a negative binomial GLM to the real data and the synthetic null data, respectively. To prove Corollary~1, it suffices to verify that \( \hat{\beta}_j \) and \( \tilde{\beta}_j \) satisfy Assumptions~1–3 under the setup specified in Assumptions~\ref{assump:S1} and~\ref{assump:S2}.

\paragraph{Verification of Assumption~1.}

We prove that the MLE \( \hat{\beta}_j \)  computed from the real data satisfies Assumption~1. The same argument applies to the synthetic null MLE \( \tilde{\beta}_j \).

Fix any gene \( j \in \{1, \dots, m\} \). The negative binomial log-likelihood for gene \(j\) is
\[
\ell_j(\beta_j) = \sum_{i=1}^n \left[
Y_{ij} \log(\mu_{ij}) - (Y_{ij} + \phi_j^{-1}) \log(\mu_{ij} + \phi_j^{-1})
\right] + \text{const}, \quad 
\mu_{ij} = s_i \cdot \exp(\alpha_j + x_i \beta_j),
\]
where \( x_i \in \{0,1\} \) is the treatment indicator, \( s_i \) is the known size factor, and \( \phi_j > 0 \) is the fixed dispersion parameter. Let \( \hat{\beta}_j \) denote the MLE obtained by solving the score equation:
\[
\frac{d}{d\beta_j} \ell_j(\hat{\beta}_j) = 0.
\]

By a Taylor expansion of the score function around the true parameter value \( \beta_j^* \), 
\[
\hat{\beta}_j - \beta_j^* = - \left[ \frac{d^2}{d\beta_j^2} \ell_j(\tilde{\beta}_j) \right]^{-1} \cdot 
\left[ \frac{d}{d\beta_j} \ell_j(\beta_j^*) \right],
\]
for some \( \bar{\beta}_j \) between \( \hat{\beta}_j \) and \( \beta_j^* \). Therefore,
\begin{equation}\label{eq:assump1_pf1}
|\hat{\beta}_j - \beta_j^*| 
\le \left| \left[ \frac{d^2}{d\beta_j^2} \ell_j(\bar{\beta}_j) \right]^{-1} \right| \cdot 
\left| \frac{d}{d\beta_j} \ell_j(\beta_j^*) \right|.
\end{equation}

The second derivative is given by the observed information:
\[
\frac{d^2}{d\beta_j^2} \ell_j(\beta) 
= \sum_{i=1}^n H_{ij}(\beta), \quad 
H_{ij}(\beta) := -\frac{d^2}{d\beta^2} \log f(Y_{ij}; \mu_{ij}, \phi_j).
\]
Under Assumption~\ref{assump:S1} and Assumption~\ref{assump:S2}, the observed information satisfies
\[
\sum_{i=1}^n H_{ij}(\beta) \ge n \cdot \lambda_{\min}
\]
with high probability, for some constant \( \lambda_{\min} > 0 \). Thus,
\begin{equation}\label{eq:assump1_pf2}
\left| \left[ \frac{d^2}{d\beta_j^2} \ell_j(\bar{\beta}_j) \right]^{-1} \right| 
\le \frac{1}{n \lambda_{\min}}.
\end{equation}

Substituting \eqref{eq:assump1_pf2} into \eqref{eq:assump1_pf1}, for some constant \( C := 1/\lambda_{\min} \), we have
\begin{equation}\label{eq:assump1_pf3}
|\hat{\beta}_j - \beta_j^*| \le C \cdot \frac{1}{n} \left| \frac{d}{d\beta_j} \ell_j(\beta_j^*) \right|.
\end{equation}

We now bound the score function. Define
\[
\frac{d}{d\beta_j} \ell_j(\beta_j^*) = \sum_{i=1}^n U_{ij}, \quad 
U_{ij} := \frac{\partial \log f(Y_{ij}; \mu_{ij}, \phi_j)}{\partial \mu_{ij}} \cdot \mu_{ij} \cdot x_i.
\]
Each \( U_{ij} \in \mathbb{R} \) depends on \( Y_{ij} \sim \mathrm{NB}(\mu_{ij}, \phi_j) \), the indicator \( x_i \in \{0,1\} \), and fixed quantities \( s_i, \phi_j, \beta_j^* \). Under Assumptions~\ref{assump:S1} and~\ref{assump:S2}, \( U_{ij} \) are uniformly sub-exponential.

By Bernstein’s inequality for sub-exponential random variables, there exists a constant \( C_1 > 0 \) such that
\[
\mathbb{P} \left( \left| \sum_{i=1}^n U_{ij} \right| \ge \sqrt{n \log m} \right) 
\le 2 \exp(- C_1 \log m) = 2 m^{-C_1}.
\]
Choosing \( C_1 \ge 2 \), we obtain
\begin{equation}\label{eq:assump1_pf4}
\mathbb{P} \left( \left| \frac{d}{d\beta_j} \ell_j(\beta_j^*) \right| \ge \sqrt{n \log m} \right) 
\le 2 m^{-2}.
\end{equation}

Substituting this into equation~\eqref{eq:assump1_pf3}, we have
\[
\mathbb{P} \left( |\hat{\beta}_j - \beta_j^*| \ge C_0 \cdot \sqrt{\tfrac{\log m}{n}} \right) \le 2 m^{-2}
\]
for some constant \( C_0 > 0 \).

Let $\gamma_{n,m}=C_{0}\,\sqrt{\log(m)/n}, C_{0}>0$, then
\[
\mathbb{P} \left( |\hat{\beta}_j - \beta_j^*| \ge \gamma_{n,m} \right) \le 2 m^{-2}, \mbox{ for all } j \in [m].
\]

The same argument applies to the MLE \( \tilde{\beta}_j \) computed from the synthetic null data, under the same assumptions.

This completes the verification of Assumption~1.


By Theorem~1, since \( \hat{\beta}_j \) and \( \tilde{\beta}_j \) satisfy Assumptions~1, 2, and~3, the Nullstrap-DE procedure based on NB-GLM MLEs guarantees control of the false discovery rate at the nominal level. Moreover, the statistical power converges to one as \( n, m \to \infty \).

\clearpage
\section{Details of Simulation Design}

Following the data-generating process described in the DESeq2 paper~\citep{love2014moderated}, we consider a two-condition DE testing setting under a negative binomial (NB) model. For gene \( j \) in sample \( i \), the observed count \( Y_{ij} \) is generated as
\begin{align}\label{eq:nb-model}
    Y_{ij} \overset{\text{ind}}{\sim} \text{NB}\left(\mu_{ij}, \phi_j\right), \quad
\log(\mu_{ij}) = \log(s_i) + \alpha_j + x_i\beta_j + z_i^\top \gamma_j,
\end{align}
where \( x_i \in \{0, 1\} \) is a binary indicator encoding the treatment condition for sample \( i \) (1 for treatment, 0 for control; samples are evenly divided into control and treatment groups), and \( \beta_j \) denotes the true log fold change between the two conditions for gene \( j \), which is the primary parameter of interest in DE analysis. The gene-specific dispersion parameter \( \phi_j \) and baseline intercept \( \alpha_j \) are estimated from real data, while the sample-specific size factors \( s_i \) and additional covariates \( z_i \) with gene-specific effects \( \gamma_j \) are manually specified. The full specification of parameters and the data generation procedure are as follows:

\begin{enumerate}
    \item Based on a real bulk RNA-seq dataset of lymphoblastoid cell lines from unrelated Nigerian individuals~\citep{pickrell2010understanding}, we fitted DESeq2 using an intercept-only model to estimate gene-specific intercepts \( \alpha_j \), and we extracted the mean–dispersion relationship as defined by \texttt{dds@dispersionFunction} in the DESeq2 package.
    \item Given a specified DE gene proportion, genes are randomly assigned to either the truly DE group or the non-DE group. For truly DE genes, the true log fold change for gene \( j \) is set to \( \beta_j = \log(\mathrm{FC}_j) \), where \( \mathrm{FC}_j \) is the fold change specified in the simulation scenario; for non-DE genes, $\beta_j=0$.

    \item Sample-specific size factors \( s_i \) are independently drawn from the uniform distribution \( \mathrm{Uniform}(0.9, 1.1) \), mimicking moderate variability in sequencing depth.

    \item If an additional covariate \( z_i \) is included, it is generated as a binary variable (e.g., representing sex or batch) that is deliberately correlated with the treatment indicator \( x_i \) to simulate confounding. Specifically, for a two-condition comparison with balanced group sizes, we simulate 80\% group-wise imbalance between the covariate and treatment. That is, within each treatment group, 80\% of the samples share the same covariate value (e.g., ``female'' in group A and ``male'' in group B), while the remaining 20\% are assigned the opposite value. The covariate values are then permuted within each group to avoid perfect separation while maintaining the overall imbalance. The additional covariate \( z_i \) is assumed to affect 20\% of genes, whose gene-specific effect sizes \( \gamma_j \) are independently drawn from \( \mathrm{Uniform}(2, 3) \), while the remaining 80\% of genes are unaffected (\( \gamma_j = 0 \)).

    \item The mean parameter of the NB distribution is calculated as 
    \[
    \mu_{ij} = \exp\left(\log(s_i) + \alpha_j + x_i\beta_j + z_i^\top \gamma_j\right),
    \]
    and the dispersion parameter \( \phi_j \) is determined from the empirical mean–dispersion relationship using \( \phi_j = \texttt{dds@dispersionFunction}(\mu_{ij}) \).
\end{enumerate}

\clearpage

\section{Supplementary Figures}
\begin{figure}[htbp]
    \centering
    \includegraphics[width=\linewidth]{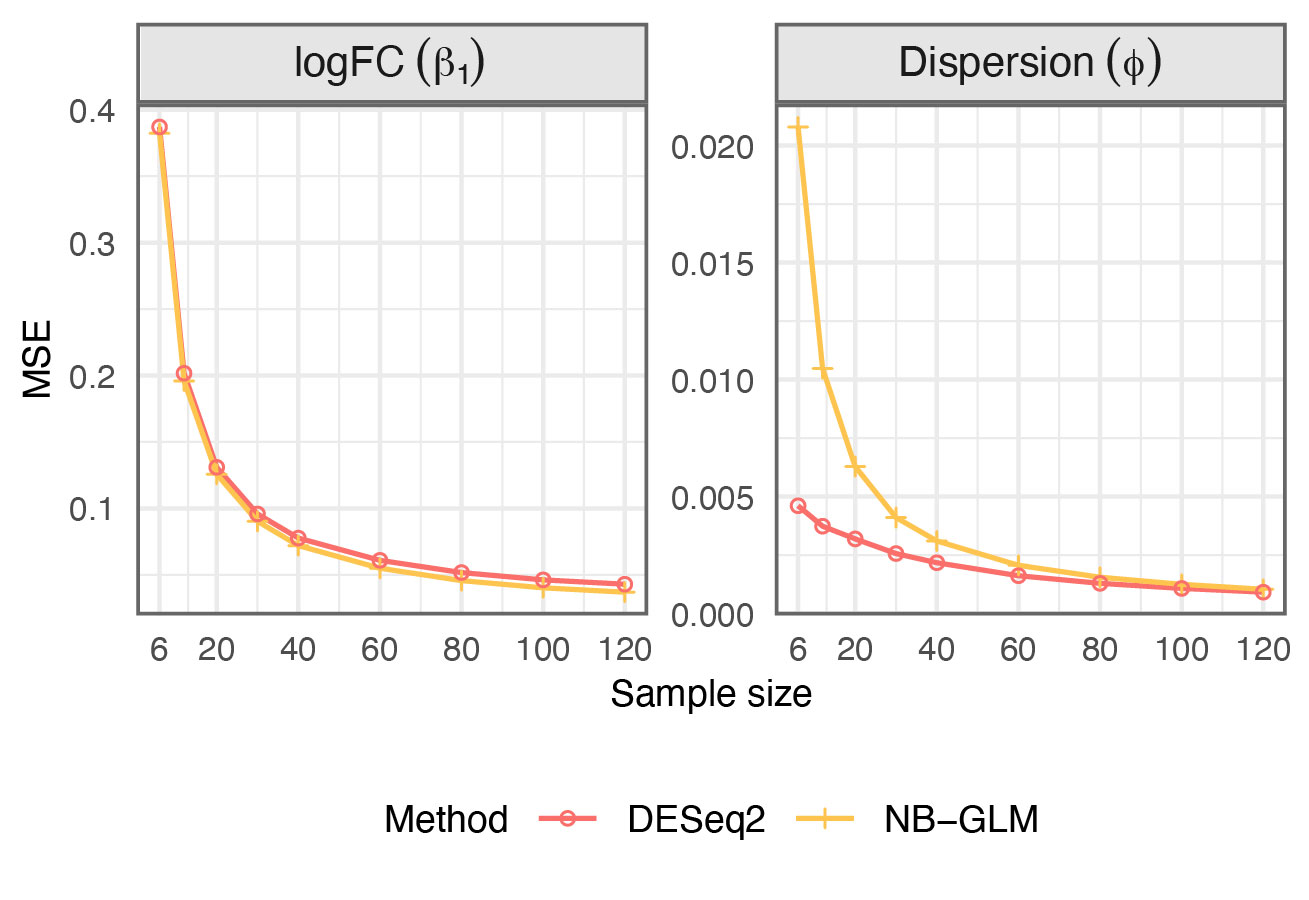}
    \caption{
    Mean squared errors (MSEs) of gene-specific log fold change (\( \beta_{1j} = \beta_1 \)) and dispersion (\( \phi_j = \phi \)) estimates obtained using DESeq2 and NB-GLM-based MLE. Results are based on numerical simulations where the mean–dispersion relationship is drawn from a real RNA-seq dataset. Each point indicates the average MSE across 50 simulated datasets with varying sample sizes. While both methods yield decreasing MSE with increasing sample size, DESeq2 achieves notably lower MSE for dispersion estimation, especially in small-sample settings. For log fold change estimation, both methods show comparable performance across sample sizes.
    }
    \label{fig:theory_nb_glm}
\end{figure}


\begin{figure}[htbp]
    \centering
    \includegraphics[width=0.9\textwidth]{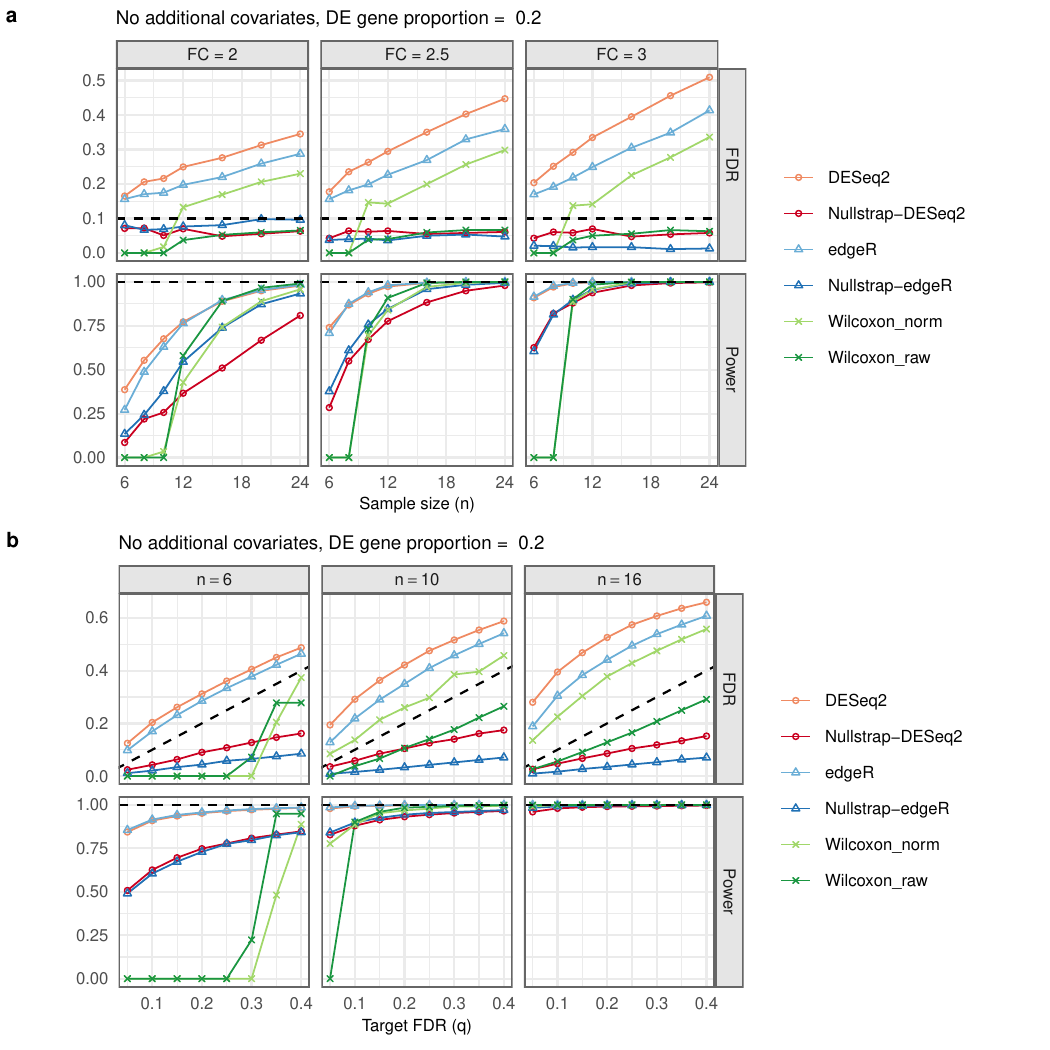}
    \caption{Simulation setting 1 (no additional covariates): \textbf{(a)} Empirical FDR and power versus sample size ($n$) under different fold changes (FC), with DE proportion $= 0.2$ and target FDR level $q = 0.1$. \textbf{(b)} Empirical FDR and power versus target FDR level ($q$) under different sample sizes ($n$), with DE proportion $= 0.2$ and fixed fold change $\mathrm{FC} = 3$.}
    \label{fig:simu1_de=0.2}
\end{figure}

\begin{figure}[htbp]
    \centering
    \includegraphics[width=0.9\textwidth]{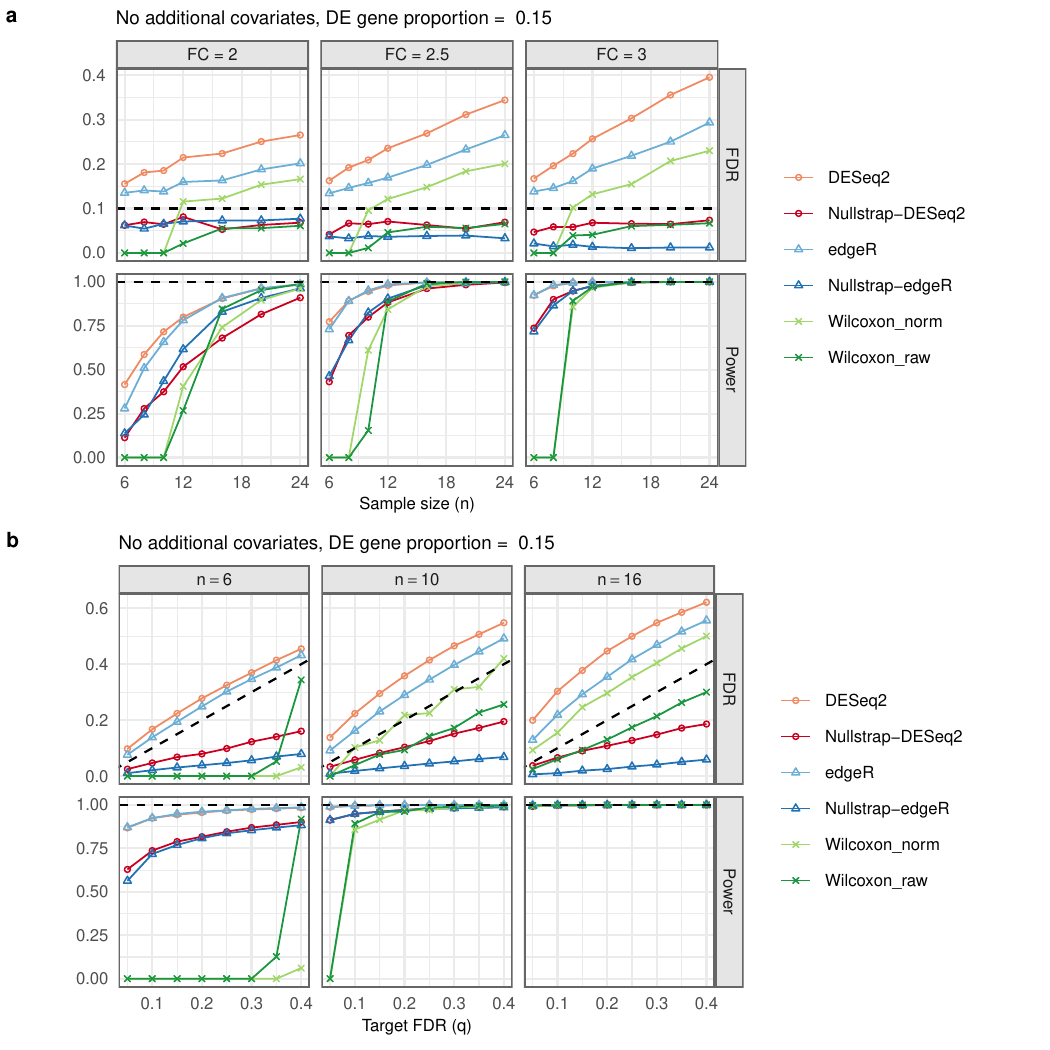}
    \caption{Simulation setting 1 (no additional covariates): \textbf{(a)} Empirical FDR and power versus sample size ($n$) under different fold changes (FC), with DE proportion $= 0.15$ and target FDR level $q = 0.1$. \textbf{(b)} Empirical FDR and power versus target FDR level ($q$) under different sample sizes ($n$), with DE proportion $= 0.15$ and fixed fold change $\mathrm{FC} = 3$.}
    \label{fig:simu1_de=0.15}
\end{figure}


\begin{figure}[htbp]
    \centering
    \includegraphics[width=0.9\textwidth]{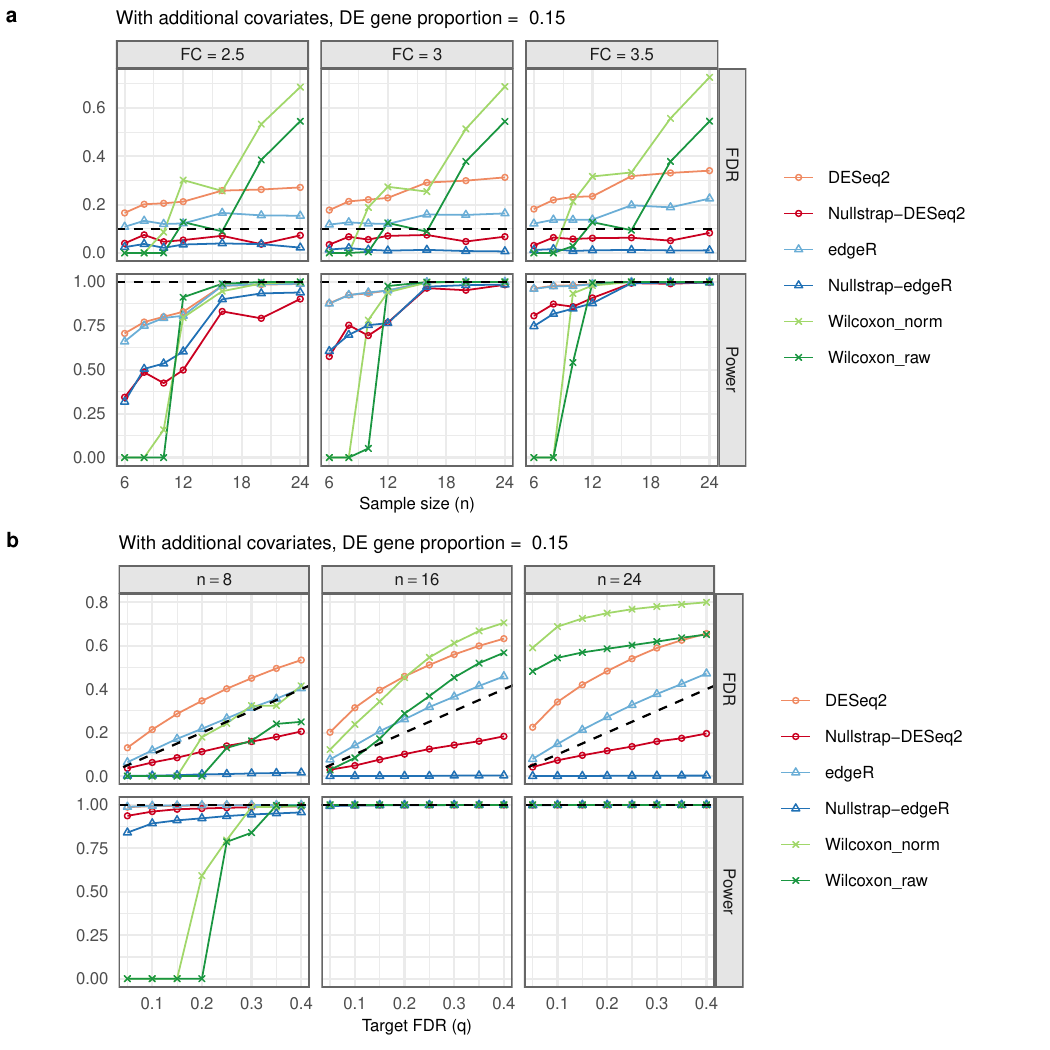}
    \caption{Simulation setting 2 (with additional covariates): \textbf{(a)} Empirical FDR and power versus sample size ($n$) under different fold changes (FC), with DE proportion $= 0.15$ and target FDR level $q = 0.1$. \textbf{(b)} Empirical FDR and power versus target FDR level ($q$) under different sample sizes ($n$), with DE proportion $= 0.15$ and fixed fold change $\mathrm{FC} = 3$.}
    \label{fig:simu2_de=0.15}
\end{figure}

\begin{figure}[htbp]
    \centering
    \includegraphics[width=0.9\textwidth]{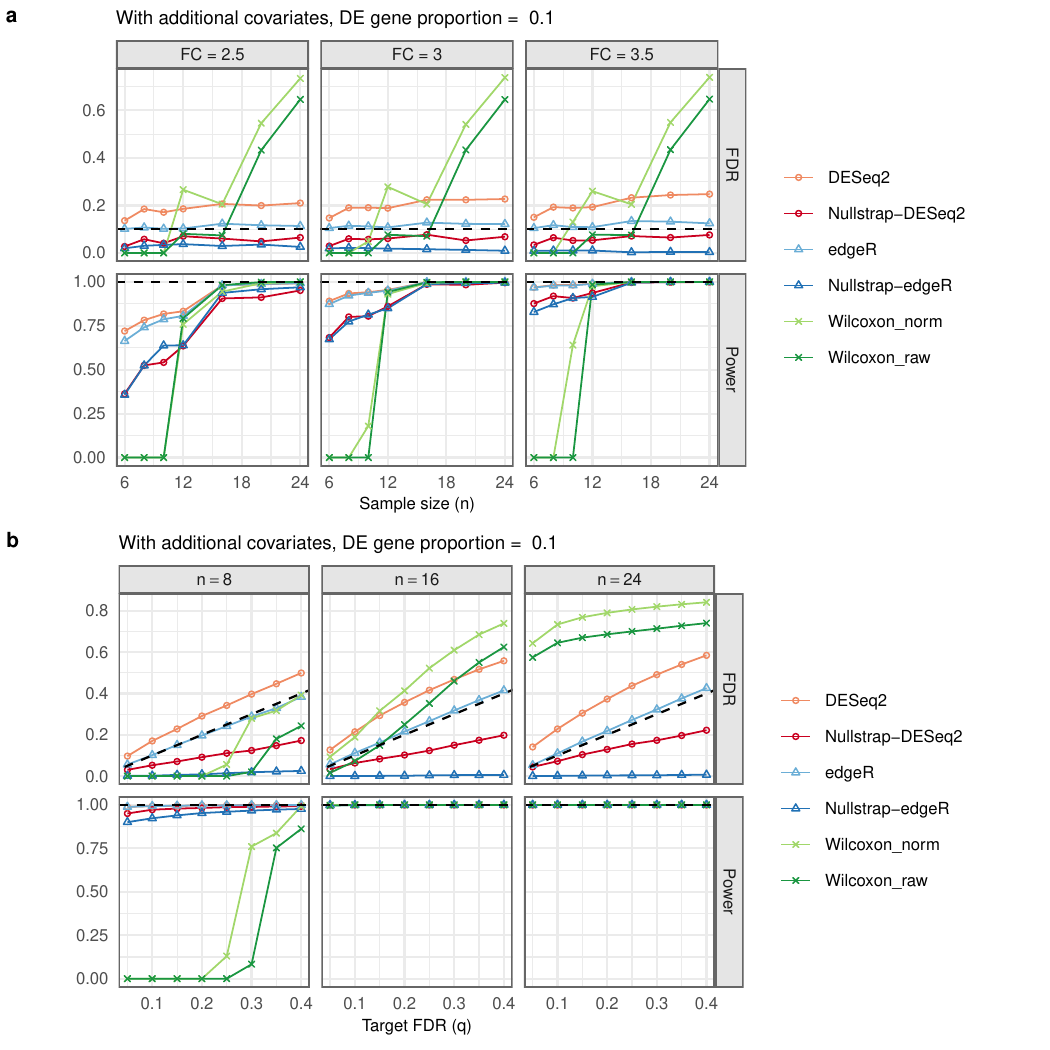}
     \caption{Simulation setting 2 (with additional covariates): \textbf{(a)} Empirical FDR and power versus sample size ($n$) under different fold changes (FC), with DE proportion $= 0.1$ and target FDR level $q = 0.1$. \textbf{(b)} Empirical FDR and power versus target FDR level ($q$) under different sample sizes ($n$), with DE proportion $= 0.1$ and fixed fold change $\mathrm{FC} = 3$.}
    \label{fig:simu2_de=0.1}
\end{figure}


\begin{figure}[htbp]
    \centering
    \includegraphics[width=0.9\textwidth]{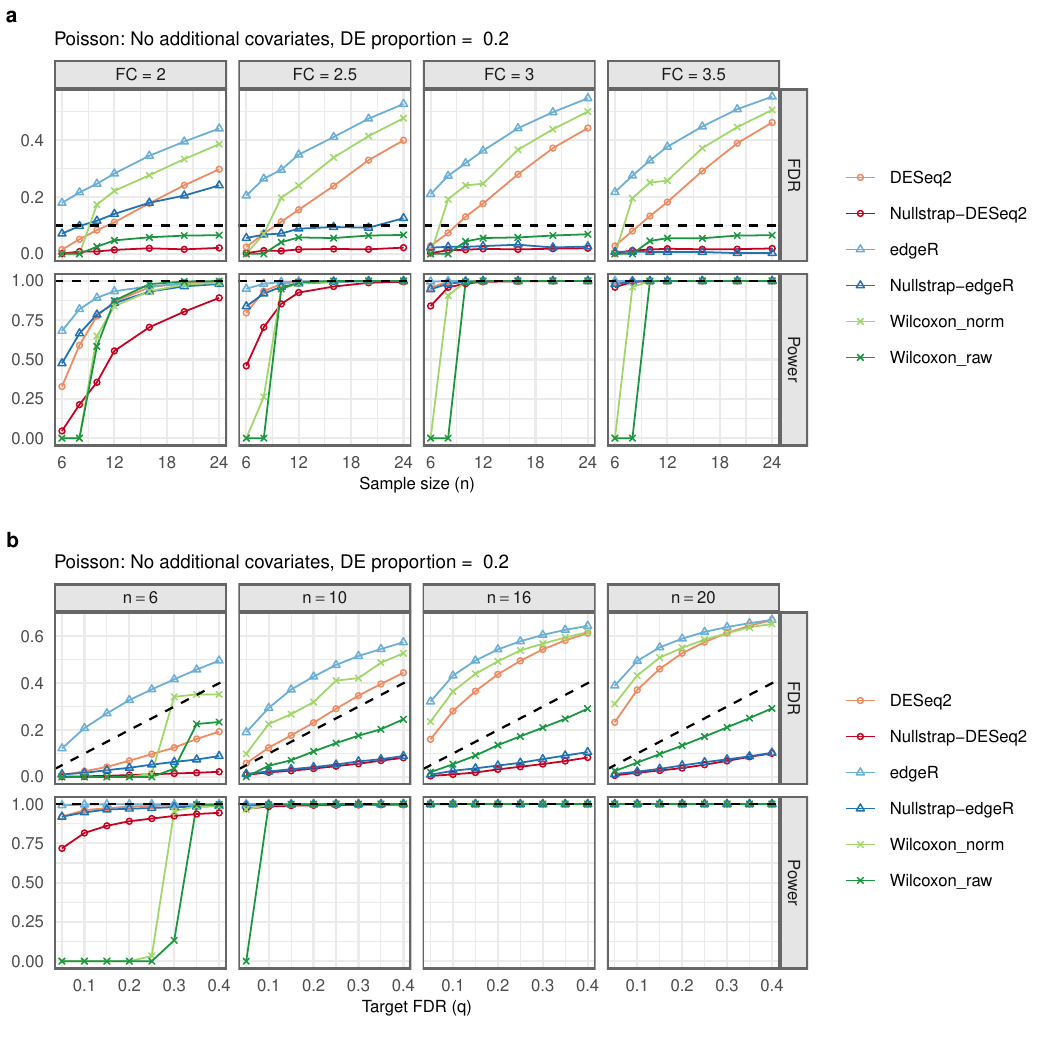}
    \caption{Simulation setting 3 (Poisson, no additional covariates, DE proportion = 0.2): \textbf{(a)} Empirical FDR and power versus sample size ($n$) under different fold changes (FC), with target FDR level $q = 0.1$. \textbf{(b)} Empirical FDR and power versus target FDR level ($q$) under different sample sizes ($n$) fixed fold change $\mathrm{FC} = 3$.}
    \label{fig:simu3_de=0.2}
\end{figure}

\begin{figure}[htbp]
    \centering
    \includegraphics[width=0.9\textwidth]{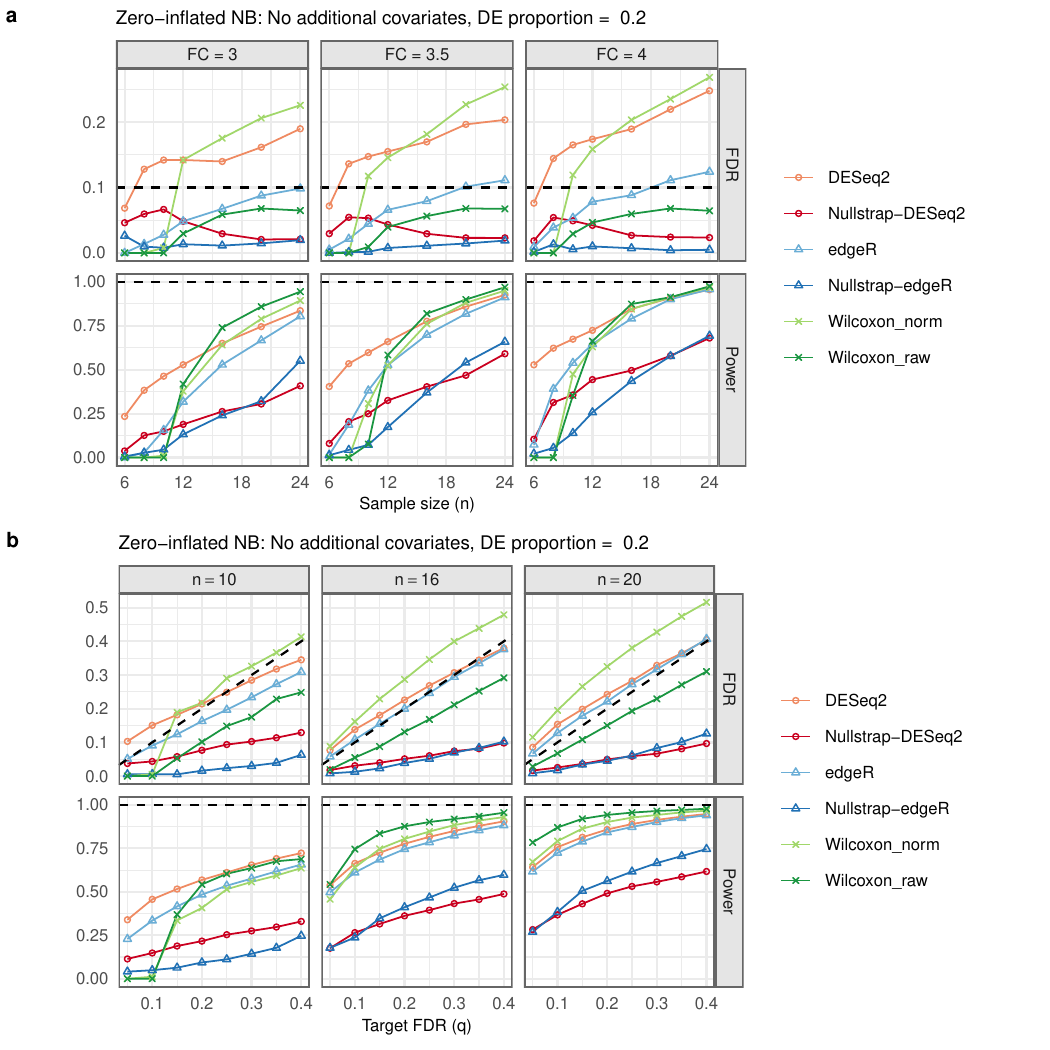}
    \caption{Simulation setting 4 (Zero-inflated negative binomial, no additional covariates, DE proportion = 0.2): \textbf{(a)} Empirical FDR and power versus sample size ($n$) under different fold changes (FC), with target FDR level $q = 0.1$. \textbf{(b)} Empirical FDR and power versus target FDR level ($q$) under different sample sizes ($n$), with fixed fold change $\mathrm{FC} = 3$.}
    \label{fig:simu4_de=0.2}
\end{figure}


\begin{figure}[htbp]
    \centering
    \includegraphics[width=\textwidth]{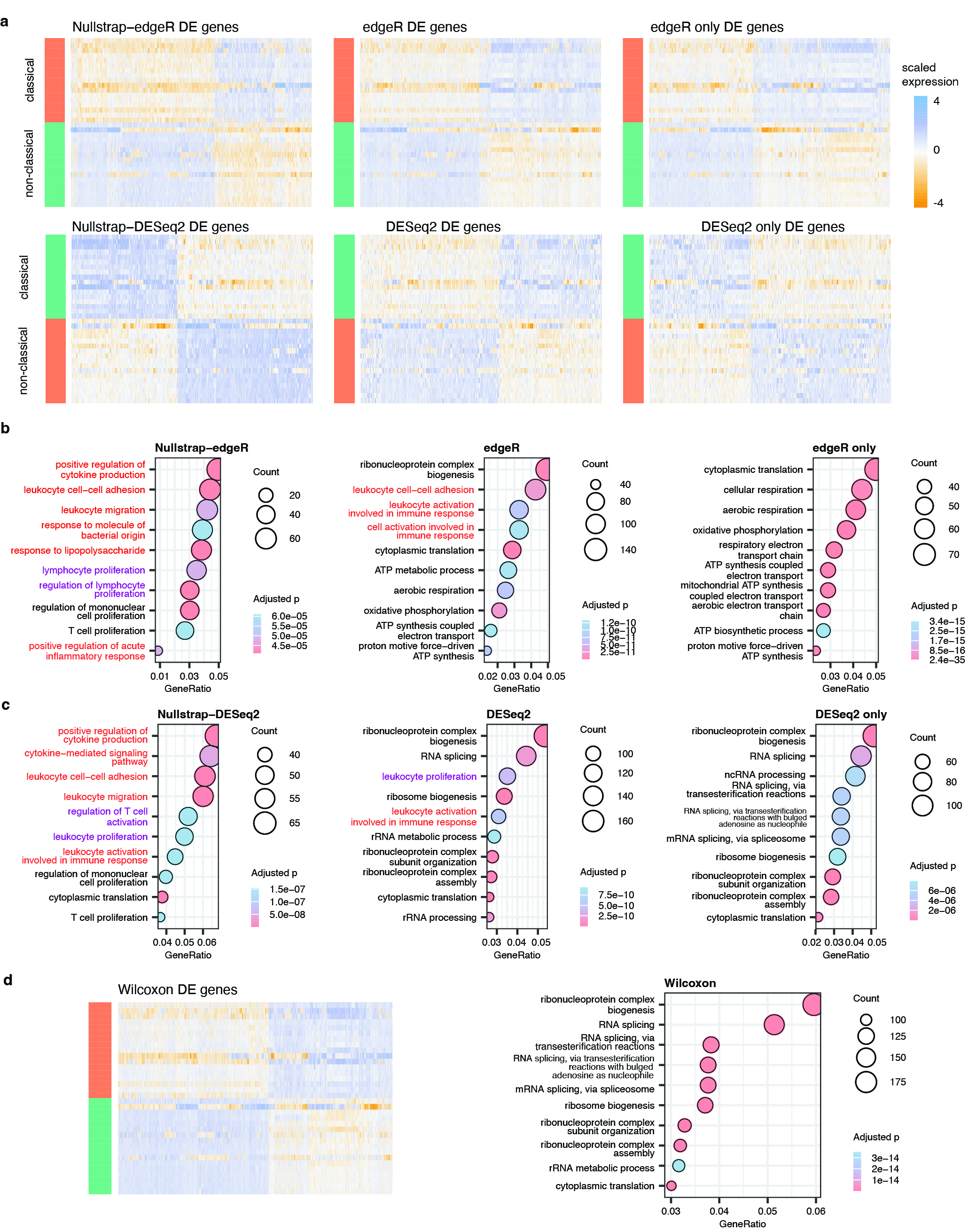}
    \caption{\textbf{Additional comparison of DE methods in human monocyte RNA-seq data.}
(\textbf{a}) Heatmaps of scaled expression for DE genes identified by Nullstrap-DESeq2, DESeq2, and DESeq2 only sets. Nullstrap-DESeq2 DEGs show clearer separation between classical and non-classical monocytes. 
(\textbf{b}) GO enrichment analysis for DE genes from Nullstrap-edgeR, edgeR, and edgeR only sets. Nullstrap-edgeR primarily identifies immune-related functions, whereas edgeR and edgeR only include many general cellular processes.  
(\textbf{c}) GO enrichment analysis for DE genes from Nullstrap-DESeq2, DESeq2, and DESeq2-only sets. Nullstrap-DESeq2 enriches for immune-specific functions, while DESeq2 and DESeq2 only capture broader transcriptional programs.  
(\textbf{d}) Heatmap and GO enrichment for DE genes from the Wilcoxon test. Many Wilcoxon DE genes are associated with general cellular functions, with limited specificity to monocyte subsets.
}
    \label{fig:case2_supp1}
\end{figure}



\clearpage
\bibliographystyle{chicago}

\bibliography{bibliography}